\documentclass{article}

\usepackage[nonatbib, preprint]{neurips_2021}
\usepackage{capt-of}% or \usepackage{caption}
\usepackage{amsmath}
\usepackage{graphics}
\DeclareMathOperator*{\argmin}{argmin}

\usepackage[toc,page]{appendix}

\newtheorem{theorem}{Theorem}
\usepackage{blkarray}

\newtheorem{remark}{Remark}
\usepackage[utf8]{inputenc} % allow utf-8 input
\usepackage[T1]{fontenc}    % use 8-bit T1 fonts
\usepackage{hyperref}       % hyperlinks
\usepackage{url}            % simple URL typesetting
\usepackage{booktabs}       % professional-quality tables
\usepackage{amsfonts}       % blackboard math symbols
\usepackage{nicefrac}       % compact symbols for 1/2, etc.
\usepackage{microtype}      % microtypography
\usepackage{xcolor}         % colors
\usepackage{xcolor}
\usepackage{adjustbox}
\usepackage{graphics}
\usepackage{subfigure}
\usepackage{tabularx,ragged2e,booktabs}
\newcolumntype{L}{>{\RaggedRight\arraybackslash}X}
\usepackage{dsfont}
\usepackage{bbm}
\usepackage{amsmath}
\makeatletter
\def\BState{\State\hskip-\ALG@thistlm}
\makeatother
\usepackage{adjustbox}
\usepackage{graphicx}
\usepackage{cite}
\usepackage{lipsum}
\usepackage{tabu}
\usepackage{bm}
\newcommand{\myMatrix}[1]{\bm{\mathit{#1}}}
\usepackage{xparse}
\usepackage{amsfonts}
\usepackage{amssymb}
\usepackage{textcomp}
\usepackage{mathtools}
\DeclarePairedDelimiter\ceil{\lceil}{\rceil}
\usepackage{resizegather}
\usepackage{multirow}
\usepackage{epsfig}
\usepackage{caption}
\usepackage{latexsym}
\usepackage{mathtools}
\usepackage[ruled,vlined]{algorithm2e}
\title{HeteroSAg: Secure Aggregation with Heterogeneous Quantization in Federated Learning}

\author{
  Ahmed Roushdy Elkordy \\
  ECE Department\\
University of Southern California (USC) \\
  \texttt{aelkordy@usc.edu} \\
   \And
   A. Salman Avestimehr \\
  ECE Department\\
University of Southern California (USC) \\
\texttt{avestime@usc.edu} \\
}

\begin{document}
\maketitle
\begin{abstract}
 Secure model aggregation across many users is a key component of federated learning systems. 
The   state-of-the-art protocols   for secure model aggregation, which are based on additive masking, require all users to quantize their model updates to the same level of quantization. This severely degrades their performance due to lack of adaptation to available communication resources, e.g.,   bandwidth, at different users. As the main contribution of our
paper,  we  propose  \textit{HeteroSAg}, a  scheme   that allows secure model aggregation while using heterogeneous quantization. HeteroSAg    enables the  edge users to adjust  their  quantization  proportional  to  their  available  communication resources,  which  can  provide  a  substantially better  trade-off  between  the  accuracy  of  training  and  the  communication  time.   Our proposed scheme is  based on a grouping  strategy  by partitioning the network into groups, and partitioning  the local model updates of users into segments. Instead of applying  aggregation protocol to the entire local model update vector, it is  applied   on segments with specific coordination between users.  We further  demonstrate how  HeteroSAg can   enable Byzantine robustness while achieving secure aggregation simultaneously. Finally,  we     prove  the convergence guarantees of HeteroSAg under heterogeneous quantization   in the non-Byzantine scenario.
\end{abstract}

\section{Introduction}
Federated learning (FL) is gaining significant interests as it  enables training   machine learning  models   locally at the edge device, e.g.,  mobile phones, instead of sending  raw data to a central server  \cite{cc, kairouz2019advances,FedAvg}.  
The goal in the basic FL framework 
is to learn a  global model $\bm {\theta} \in  \mathbb{R}^{m}$  using the   data stored at the edge device.  This can be represented by minimizing a
global objective function,  
\begin{align}
\argmin_{\bm {\theta}} F(\bm {\theta}) & \text{ such that } F(\bm {\theta}) =\sum_{i=1}^N \frac{n_i}{n} F_i(\bm {\theta}),  \text{and } F_i(\bm {\theta}) = \frac{1}{n_i} \sum_{j=1}^{n_i} f_i(\bm {\theta};x_j,y_j),
\end{align}
where {\boldmath$\bm {\theta}$} is the global  model  to be optimized. Here,  $F_i$ is  the local objective function of user $i$, $f_i(\bm {\theta};x_j,y_j)$ is  the loss of the prediction on example
$(x_j, y_j)$ form user $i$ made with global model $\bm {\theta}$, $n_i$ is the data size at user $i$, and  $n  = \sum_i n_i$.  Without loss of generality, we assume that all users have an equal-sized dataset. 

\begin{figure}[]
\centering
\includegraphics[width=8cm,height=8cm,keepaspectratio]{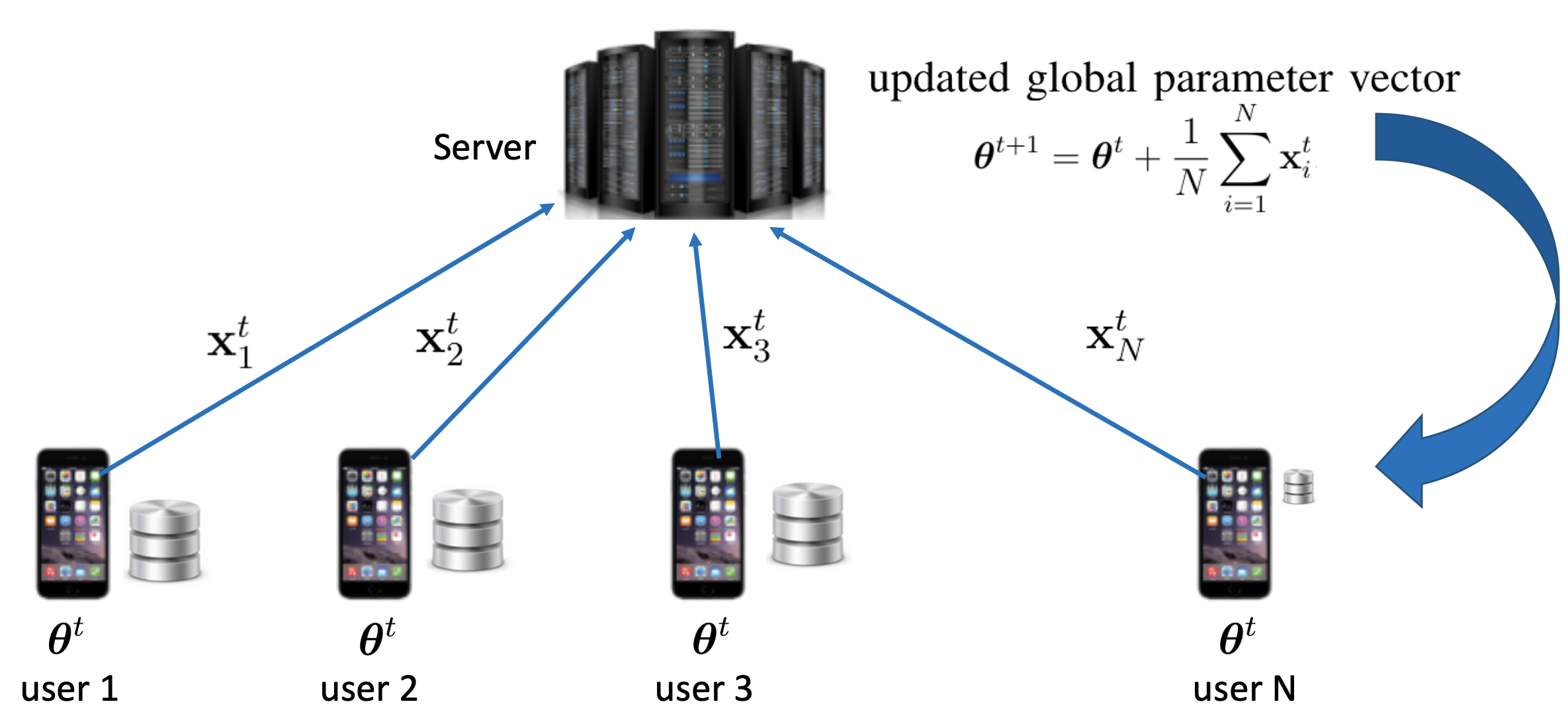}

\caption{ The training process of  federated learning.   }
\label{SystemModel}
\end{figure}

To learn the global model $\bm {\theta}$ that minimizes the objective in   (1),   stochastic gradient descent (SGD) algorithm  can be easily  implemented  distributedly across the $N$ available   devices in the presence of  a central  server  who orchestrates the training  process. The training process in FL by using the distributed SGD is   illustrated in Figure \ref{SystemModel}.
At iteration $t$, the  server sends  the current version of the  global  model   vector,  $\bm {\theta}^{(t)}$, to  the mobile users. User $i$ then   computes its local  model  vector $\bm {\theta}_i^{(t)}$ based on   its local dataset  by using the SGD, 
 %as follows 
 %\begin{equation}
%\label{Agg}
%\bm {\theta}_i^{(t)}=\bm {\theta}^{(t)}-\eta^t g_i(\bm {\theta}^{(t)}),
%\end{equation}
 %where $g_i(\bm {\theta}^{(t)})$ is gradient  computed on the client based on the   global parameter vector $\bm {\theta}^{t}$  with respect to its local dataset.
         so the local model update of each user can be written as $\mathbf{x}^{(t)}_i:=\bm {\theta}_i^{(t)}- \bm{\theta}^{(t)}$. This local model update    could be     a single gradient,  or  could result from  multiple steps of the  SGD  taken on this user’s local dataset. User $i$  sends the  local model update   $\mathbf{x}^{(t)}_i$ to the server.    The local model updates of the $N$ users are then aggregated  by the server. The server then updates the global  model    $\bm{\theta}^{(t + 1)}$ for the next round according to  
 \begin{equation}
\label{Agg1}
\bm {\theta}^{(t+1)}=\bm {\theta}^{(t)}+ \frac{1}{N} \sum_{i=1}^{N} \mathbf{x}^{(t)}_i.
\end{equation}
Although FL   provides many benefits, it still suffers from key challenges such as  communication bottleneck, system failures, malicious users and   users' privacy \cite{kairouz2019advances}.  The  communication bottleneck  in FL  is created by sending a large  model    from each  user to the server  at each iteration of the training.  Researchers have proposed many approaches to    provide  a  communication-efficient FL   system. One of these approaches is to reduce the model size  by performing model   compression  through either         quantization \cite{1-bit, alistarh2016qsgd,wen2017terngrad, rotation, an2016distributed,unknown} or 
 sparsification   \cite{rotation,2017gradient, Strom2015, alistarh2018, ji2020dynamic}.
%In , the authors show that by  allowing every client performs multiple iterations of SGD to compute a model update  instead of communicating after every iteration, the   number of necessary communication rounds will be reduced without harming the convergence speed. 

Another  key challenge for FL   is the Byzantine faults  \cite{Byz} in which  some users may behave arbitrarily due to  software bugs, hardware failure, or even get hacked during training, sending arbitrary or malicious values to the server, thus severely degrading the overall convergence performance. Many  Byzantine robust strategies have been proposed recently for  FL \cite{regatti2020bygars,xie2019zeno,xie2019zenot,krum,By,zhao2020shielding}. These Byzantine robust optimization algorithms combine the gradients received by all workers using robust aggregation rules, to ensure that training is not impacted by malicious users.

Preserving  the privacy of the users  is another  main  consideration for FL system.  There are two approaches to achieve that. First, the training data stays  on the user device, and users locally perform model updates using their individual data. Second,  local models can be securely   aggregated  at the  central server to update the global model. This is achieved through what is known as a secure aggregation (SecAg) protocol \cite{cc},  where users use random masks to mask their local model updates. In this protocol,
each user masks its local update through additive secret sharing using private and pairwise random keys before
sending it to the server. Once the masked models are aggregated at the server, the additional randomness cancels out
and the server learns the aggregate of all user models. At the end of the protocol, the server learns no information
about the individual models beyond the aggregated model, as they are masked by the random keys unknown to
the server. Some other algorithms for secure aggregation  with additive masking have been proposed   \cite{10.1145/3372297.3417885, so2020turboaggregate,kadhe2020fastsecagg,zhao2021information}. 

%\cite{cryptoeprint:2011:157,inproceedings2 , 4568388,aabb, mental, cc,so2020turboaggregate}.    In particular, secure aggregation can be done by using   cryptographic approaches such as homomorphic encryption  that allows aggregations to be performed on encrypted data  \cite{cryptoeprint:2011:157},\cite{inproceedings2}.  The main issue with this approach is that performing computations in the encrypted domain is computationally expensive. Another line of work that achieves secure model aggregation is the   multiparty computation (MPC) \cite{4568388,aabb, mental}. The main limitation of this approach is that it requires  high communication cost. Recently, some works have achieved secure aggregation in a single server setup  by  using additive masking, e.g.,  \cite{cc}, \cite{so2020turboaggregate}, where users use random masks to mask their local model updates. 

In general, the state-of-the-art secure aggregation protocols  with additive masking have some limitations  associated with:
\begin{itemize}
\item (System heterogeneity)  They  require all users to quantize their model updates to the same level of quantization (to guarantee correct decoding Section  \ref{Challenges}),  even if they have different communication resources such as   transmission rates.  Lack of adaptation   to    the  speed of the available  network  (3G, 4G, 5G, Wi-Fi)   and   the fluctuation of the network quality over time  severely  degrades the performance of these protocols. More specifically,  by   making  all users  use a low level quantizer, the communication time will be small, but the test accuracy will decrease. On the other hand,   using   a high level quantizer  will result in increasing the test accuracy at the expense of increasing the  communication time.

 \item    (Robustness) Secure aggregation  protocols make  the adaption of the existing  state-of-the-art  defense  strategies  \cite{krum,By, regatti2020bygars,xie2019zeno,xie2019zenot} against Byzantine users  difficult to implement, as the  server only receive  a   masked model update from  each user, while the success of these  strategies are based on having   users' individual clear  model updates. 
%\footnote{Recently,  the authors in \cite{so2020byzantineresilient} propose a new framework  that achieves secure Byzantine-robust FL   without adapting the  existing   defenses  mechanism.  }

\item (Communication efficiency) The   bandwidth expansion, which  measures   the ratio between the size of the  encoded model in bits to the size of the clear model. This  bandwidth expansion results  from the  additional $\mathcal{O}(\log N)$ bits that should be communicated for each scalar in the model update vector, where $N$ is the total  number of users,  to guarantee correct decoding. Hence, this  expansion   makes them ineffective with aggressive quantization,  specially for large $N$ \cite{cc,kairouz2019advances, bonawitz2019federated}.

\end{itemize}
Overcoming the aforementioned  limitations, specifically the one associated with system heterogeneity,   is a challenging problem as illustrated in detail  in Section \ref{Challenges}. Towards solving these limitations, we propose HeteroSAg. 

\subsection{Main  contributions }\label{sec-1.1}
  \textit{HeteroSAg} has the following   four salient features: 
\begin{enumerate}
\item   HeteroSAg protects the privacy of the local model updates of each individual user in the strong information-theoretic sense by masking the model update of each user such that the mutual information between the masked model and the clear model is zero.
\item  HeteroSAg  allows  using heterogeneous quantization. This enables the edge  users to adjust their quantizations proportional to their available communication resources   which can result in   a substantial     better trade-off between  the  accuracy  of training and the communication time. 

\item HeteroSAg  further enables robustness against Byzantine users, by incorporating distance-based  defense mechanisms such as coordinate-wise median \cite{By}.

\item HeteroSAg reduces   the  bandwidth  expansion. For instance,  we demonstrate  that  for the case of having    $N=2^{10}$ users   using a single bit quantization,    the  bandwidth expansion factor when  using  HeteroSAg    is $ 4\times$, as opposite to   $ 11 \times$ when using SecAg.  
 \end{enumerate}

 We    provide the theoretical   convergence guarantees of HeteroSAg   under heterogeneous quantization  for convex  loss function in the non-Byzantine setting. Furthermore, using neural network with real-world dataset,  we demonstrate  the efficiency of the  heterogeneous quantization given by HeteroSAg. Specifically, we show that we can achieve accuracy close to the baseline  case (no-quantization) with the same communication time  as the case when all users are using 1-bit quantizer. We also show   that  we can achieve   $ \sim {15} \%$ higher test accuracy when compared with    the setting of homogeneous quantization with  $1$-bit quantizer, while the communication time is the same for  both settings.  We  then  experimentally demonstrate the resiliency of HeteroSAg in the presence of Byzantine users under three different attacks  and compare it to the conventional federated   averaging scheme \cite{FedAvg} by using two  different datasets.

%\item We also show that HeteroSAg  can further enable robustness against Byzantine users, by incorporating distance-based  defense mechanisms such as coordinate-wise median or coordinate-wise trimmed mean \cite{By} without  either     extra computation cost at    the users  or     extra communication cost in the system to the original costs of the secure aggregation protocol.  In other words, HeteroSAg   can  be used to achieve communication-efficient   secure Byzantine-robust FL  system. We experimentally validate this result.

\subsection{Related works}\label{sec-1.2}
 The authors in  \cite{unknown}  provide two  heterogeneous quantization  algorithms for distributed  ML in the absence of a central server to  reduce the communication cost, but without any privacy  guarantee for the model updates of the users.  Our work is different from \cite {unknown}, since   our objective is  to provide  a scheme  that  not  only  allows for      heterogeneous    quantization, but also    guarantees the privacy of the users' models by doing    secure model aggregation. We also    consider a network topology where there exists a parameter server. Therefore, the two setups are not comparable. We also highlight that our objective in this paper  is not to design a new quantization scheme, yet to provide a general method that enables using heterogeneous  quantization while doing   secure aggregation. 

\iffalse In the non-Byzantine FL  setting, secure aggregation   can also be done by using   cryptographic approaches such as homomorphic encryption  that allows aggregations to be performed on encrypted data  \cite{cryptoeprint:2011:157},\cite{inproceedings2}.  The main issue with this approach is that performing computations in the encrypted domain is computationally expensive. Another line of work that achieves secure model aggregation is the   multiparty computation (MPC) \cite{4568388,aabb, mental}. The main limitation of this approach is that it requires  high communication cost. Recently, some other works have achieved secure aggregation in a single server setup  by  using additive masking, e.g.,  \cite{10.1145/3372297.3417885, so2020turboaggregate,kadhe2020fastsecagg,zhao2021information}, where users use random masks to mask their local model updates. \fi
 
 In  recent  work,   Byzantine-robust secure aggregation algorithms have been proposed \cite{so2020byzantineresilient,sattler2019robust}. The work in \cite{sattler2019robust}   has been proposed for two  honest (non-colluding) servers who both interact with the mobile users and communicate with each other to
carry out a secure two-party protocol. Unlike this work, the authors in \cite{so2020byzantineresilient} develop  BREA, a single-server Byzantine-resilient secure training framework, to facilitate robust and privacy-preserving training architectures for FL. Our work also  achieves Byzantine-resilient secure aggregation in a single server  by a simple incorporating  of some  state-of-the-art Byzantine robust algorithms which have provable convergence guarantees  such as coordinate-wise median based \cite{By}  without either   extra computation cost to the users  or extra communication cost to the original cost of SecAg. The per-user communication cost   of  HeteroSAg is $\mathcal{O}(m+N)$ as opposite to $\mathcal{O}(N^2+Nm)$ for BREA and $\mathcal{O}(N+Nm)$ for the generalized BREA, where $m$ and $N$ are the model size and the number of users, respectively.  Furthermore, the per-user  computation cost for HeteroSAg is $\mathcal{O}(N^2+ m\log N)$ as opposite to $\mathcal{O}(mN^2+Nm \log^2 N)$ for BREA and the generalized BREA.   We note that  the upper bound on the number of Byzantine nodes for the success of  HeteroSAg    is given by $ B  \leq  \ceil*{0.25 G}-1$, where $G$ the number of groups given by  HeteroSAg,   is less than the upper bound   for  coordinate-wise median based in \cite{By}. However,  our proposal is   initially developed   for enabling secure aggregation while using heterogeneous quantization at different users, while  incorporating defense technique  against Byzantine nodes comes as an extra feature. Additionally,    the   scheme  in   \cite{By} solely   does not provide privacy for the local models of the users. Therefore, we conclude that  HeteroSAg   is the first   scheme  that achieves secure aggregation with heterogeneous quantization while providing  Byzantine-resiliency.

\section{Background}\label{sec-1.2}
 Secure aggregation (SecAg), e.g., \cite{cc},  is a key component in FL  that enables  the distributed  training process while preserving the privacy  of the users.   We summarize SecAg in  the following five steps while considering  $\mathcal {S} \triangleq \mathcal {N}\triangleq \{1, \dots, N\}$, where $N$ is the total  number of   nodes.  We provide this summary      as this protocol   is a key component in  HeteroSAg.

\noindent \textbf{Step 1  (Sharing keys and masks):} 
Users first     establish a secure communication channel between them by using   pairwise keys through a key exchange protocol such as Diffie-Hellman key agreement \cite{dif77}.  All the   communication is forwarded through the  server. Also, each pair of users $i, \; j \in \mathcal {N}$ first agrees on a pairwise random seed $s_{i,j}$ by  using Diffie-Hellman key agreement, such that  $s_{i,j}$ is a function of  the public key   $s_j^{PK}$ of user $j$  and the private key    $s_i^{SK}$ of user $i$.   At the end, each node $i \in \mathcal{N}$ will have this set of agreement keys $\{s_{i,j}\}_{j \in \mathcal{N}/i}$.  Also, according to the key generation in Diffie-Hellman key agreement \cite{dif77}, the public key is symmetric, i.e., $s_{i,j}= s_{j,i}$. Furthermore, the server will have all the set of public keys $s_j^{PK}$ for all $j \in \mathcal{N}$.   
In addition, user $i$ creates a private random seed $b_i$. The role of $b_i$ is to prevent the privacy breaches that may occur if user $i$ is only delayed instead of dropped \footnote{The dropped users  are  those who failed to  send their masked model to the server. In other words, the server will not receive the model  update of  those users for  the current round. 
On the other hand,  the delayed users  are those who send their model updates to the server, but their models   have experienced high delay before receiving by the server. Although,  these users have already  sent  their models to the server,   the server will consider them as dropped users,  because of their high delay. Therefore,  the server  will not include adding the received masked  models  from those users in the model aggregation step} (or declared as
dropped by a malicious server),   in which case the pairwise masks alone are not sufficient for privacy protection.    Further discussion about the rule of $b_i$ is given in Step 5.

\noindent\textbf{Step 2  (Secret sharing):}  User $i$, $i \in \mathcal {N}$, secret shares  the  private key  $s_i^{SK}$  as well as $b_i$  with the other users in the system, via Shamir’s secret sharing \cite{shamir}. To ensures that the local model  is private  against an adversarial server  which  tries to learn information about the local models of the honest users, while  the mobile users are honest and do not collude with the server,  the threshold of the secret share  scheme  should be  $ \ceil[\big]{N/2}+1$.   For the case where users are adversaries,  no matter how we set the threshold value, users on their own learn nothing about other users.

\noindent\textbf{Step 3 (Quantization):}     SecAg  and  cryptographic
protocols require the input  vector   elements to be integers, while using modular operation to transmit these vectors. By considering  the case where  the  model update of each user takes real values, we need to do quantization first  so that we can apply SecAg.     Without loss of generality and for the ease of the analysis, we use 
 the $K$-level quantizer in \cite{an2016distributed}  to quantize the  model  update $\mathbf{x}_i$, for $ i \in \mathcal{S}$. We  assume that  the elements of each     model  $\mathbf{ x}_i$, for $i=1, \dots, N$,  fall in  the range $[r_1,r_2]$.   Let $0 \leq l < K_g$, where $ K_g$ is the number of quantization levels,   be an
integer such that when  $\mathbf{ x}_i(k) \in [ T(l), T(l+1)]$, where  $T(l)=r_1+l \Delta_{K_g}$, and   $ \Delta_{K_g}=\frac{r_2-r_1}{K_g-1}$  is the quantization interval. Then 
\setlength\floatsep{.3\baselineskip plus 3pt minus 2pt}
\begin{equation}\label{quanti}
Q_{K_g}(\mathbf{  x}_i(k)) =\begin{cases} T(l+1)& \text{ with probability } \frac{\mathbf{  x}_i(k)-T(l)}{T(l+1)-T(l)},\\ T(l) & \text{ otherwise.} \end{cases}
\end{equation}
\noindent The output of the quantizer  $\bar {\mathbf{  x}}_i(k)=Q_{K_g}(\mathbf{  x}_i(k))$   takes a discrete  value from this range $\{r_1, r_1+ \Delta_{K_g},r_1+2 \Delta_{K_g}, \dots,  r_2- \Delta_{K_g},  r_2\}$. 
%First we shift each coordinate  $\mathbf{x}_i (k) +t$, and     let $0 \leq l < K_g$  be an
%integer such that when  ($\mathbf{ x}_i(k)+t) \in [ %r_1+l\Delta,r_1+(l+1)\Delta]$, where  %$\Delta=\frac{2t}{K_g-1}$  is the quantization %interval. Then

%\begin{equation}\label{quanti}
%Q_{K_g}(\mathbf{ x}_i(k)) =\begin{cases} \min (r_1+l\Delta,  r_1+(l+1)\Delta)  & \text{ with probability } 1-\frac{ \mathbf{ x}_i(k)-\min (r_1+l\Delta,  r_1+(l+1)\Delta)}{l}\\ \max (r_1+l\Delta,  r_1+(l+1)\Delta)  & \text{ with probability } \frac{ \mathbf{ x}_i(k)-\min (r_1+l\Delta,  r_1+(l+1)\Delta)}{l} \end{cases}.
%\end{equation}

%Then Specifically,  given that users' model  are  randomly distributed over the range$[r_1,r_2] $, and assume without loss of generality  having a uniform quantizer $Q_K_g \in \mathcal {Q}$  with $K_g$ quantization levels,  $\mathbf{ x}_i(k)$, for $k=1, \dots, m$   will be  uniformly quantized using $K_g$ levels, each of width $\Delta=\frac{2t}{K_g-1}$ to generate the quantized local model updates  $\bar {\mathbf{ x}}_i$. The output of the qunatizer $\bar {\mathbf{ x}}_i(k)$  will take a discrete  value from this range $\{-r, r_1+\Delta,r_1+2\Delta,  \dots, 0, \dots,  t-\Delta,  t\}$
\noindent\textbf{Step 4 (Encoding):} Following the quantization step, the set of users  $\mathcal {S}$   starts the encoding process on  $\{ \bar{\mathbf{  x}}_i(k)\}_{i \in \mathcal {S}} $, for $k=1, \dots, |\bar{\mathbf{  x}}_i|$,  by first mapping the  outputs of the quantizer  from the  $K_g$ real  values  that belongs to the discrete     range  $\{r_1, r_1+ \Delta_{K_g},r_1+2 \Delta_{K_g},   \dots,  r_2- \Delta_{K_g},  r_2\}$ to  integer values in this range   $[0,K_g-1]$. This mapping is performed   such that a real value 
 $  r_1$ maps   to $0$ and  $r_2$ maps to  $K_{g}-1$, etc.   The encoding process   is  completed    by allowing each  pair of users in $\mathcal {S}$  to use the pairwise random seeds  to    randomly  generate $0$-sum pairs of mask vectors to provide  the   privacy for  individual models. 
%,.Specifically, each pair of users $i, \; j \in \mathcal {S}$ first agree on a pairwise random seed $s_{i,j}$ by  using Diffie-Hellman key agreement, where $s_{i,j}$ is a function of  the public key   $s_j^{PK}$ of user $j$  and the private key    $s_i^{SK}$ of user $i$. In addition, user $i$ creates a private random seed $b_i$. The role of $b_i$ is to prevent the privacy breaches that may occur if user $i$ is only delayed instead of dropped (or declared as dropped by a malicious server), in which case the pairwise masks alone are not sufficient for privacy protection. 
The output vector of the encoder is given by
\begin{align}\label{encod}
\mathbf{ y}_{  \mathcal {S}, i
}=  \bar {\mathbf{x}}_i +  \text{PRG}(b_i)+\sum_{j:i<j}  \text{PRG}(s_{i,j})   -\sum_{j:i>j}  \text{PRG}(s_{j,i}) \text{ mod } R, 
\end{align}
where $\mathbf{ y}_{  \mathcal {S}, i
}$ is a vector of  $|\bar {\mathbf{ x}}_i|$ elements, and   $R=|\mathcal {S}|(K_g-1)+1$   to ensure that all possible aggregate vectors from the $|\mathcal {S}|$ users will be representable without overflow at the server. %For consistent,  the seeds $s_{j,i}$ in \eqref{encod} is generated by the public key of users $j$ and the private key of user $i$.  
PRG is a pseudo random generator  used to expand the different seeds  to  vectors    in $\mathbb {Z}_{R}$ to mask users' local models.

\noindent\textbf{Step 5 (Decoding):}  
From a subset of survived  users, the server collects either the  shares of private keys  the  belonging to  dropped users, or the shares of the private seed belonging to a surviving user (but not both). The server then   reconstructs  the private seed of each surviving user,  and    the pairwise seeds  $s_{i,j}$  of each dropped user  $i$. The server reconstructs $s_{i,j}$ by  combining the  reconstructed private key $s_i^{SK}$  with the corresponding available  public key at the server from user $s_j^{PK}$. Note that, the server hold all the public keys of all users. The server removes the masks of the dropped users  from the aggregate of the masked models. Finally, the server  computes the aggregated model
\begin{equation}
\mathbf{ x}_\mathcal{U}= \sum_{i\in \mathcal {U}} ( \mathbf{ y}_{  \mathcal {S}, i
} - \text{PRG}(b_i) )   -  \sum_{i\in \mathcal {D}}  \left(\sum_{j:i<j}  \text{PRG}(s_{i,j})- \sum_{j:i>j}  \text{PRG}(s_{j,i}) \right) \text{       mod } R 
=  \sum_{i\in \mathcal {U}} \bar {\mathbf{ x}}_i \mod R,
\end{equation}
where $\mathcal {U}$ and $\mathcal {D}$ represent the set of surviving and dropped users, respectively. The decoding process is completed by  mapping  the global   model   from  $\mathbb {Z}_{R}$ to the corresponding values in this discrete  set of real numbers $\{|\mathcal {U}|r_1, |\mathcal {U}|r_1+ \Delta_{K_g},|\mathcal {U}|r_1+2 \Delta_{K_g},  \dots,  |\mathcal {U}|r_2- \Delta_{K_g},  |\mathcal {U}|r_2\}$.

In the following, we discuss the importance of using the private mask $b_i$ in \eqref{encod} in preserving the privacy of the delayed model of user $i$.  According to the SecAg protocol,   the server will  consider any user, user $i$,  with   delayed model $y_{\mathcal{S},i}$  as a dropped user. Hence, according to the  decoding step,  the server will ask the set of survived users to get the shares of the private key $s_i^{SK}$ of the delayed user $i$. Getting the private key $s_i^{SK}$ allows the  server to reconstruct the set of  agreement keys $\{s_{i,j}\}_{j \in \mathcal{N}/i}$, and hence    remove the corresponding masks  from the aggregation of the masked models. Although the  server have  already    known   $\{s_{i,j}\}_{j \in \mathcal{N}/i}$, the privacy of the local model of node $i$ is still preserved  thanks to the private mask  PRG($b_i$)  as shown in \eqref{encod}. We  have  provided  a  simple  illustrative  example  for  SecAg  in  Appendix  \ref{ill}. 

\section{Problem Formulation}\label{sec-2}
We first describe  the  secure aggregation  with heterogeneous quantization  problem. After that, we explain  why   the conventional   SecAg protocol   can  not  be  applied  directly  to  our  problem. 

\subsection{System Model}
We  consider   a   FL    system  that consists of  a  central server  and a set $\mathcal {N}=\{1,. \dots, N\}$  of  $N$  mobile users  with heterogeneous  communication resources. These $N$ users  allow training a ML model locally   on their  local dataset, as described in the introduction.  
We   also consider having a set $\mathcal {Q}= \{Q_{K_0}, Q_{K_1}, \dots, Q_{K_{G-1}}\} $  of element-wise stochastic   quantizers, e.g., \cite{an2016distributed}, with $G$ different   levels      that can be  used  in this   system, where $K_g$ is  the  number of quantization levels  of  the quantizer $Q_{K_g}$, and  $K_0 < K_1< \dots< K_{G-1}$,  instead of having a single  quantizer as in SecAg.
In this problem, we assume that users are already clustered into $G$ different groups based on their communication resources. Each  user $i$ in group $g$ can quantize its model update $\mathbf{x}_i$ by  using quantizers from a pre-assigned set of quantizers with these levels $\mathcal {K}_g= \{K_0, K_1, \dots, K_g\} $ where  $K_g$ is the highest possible quantization levels that can be used by the users in group $g$ that is suitable for his transmission rate\footnote{The problem  of  the optimal clustering of the users based on their transmission rates  or    the optimal assignment of the quantizers to the users is not the main scope of our paper. Instead, our focus is to provide an approach that allows for doing secure aggregation when different quantizers can be utilized at different users, which is a challenging problem  as  we will show in Section III-B.}. 

 \noindent\textbf{Threat model:}   The server  is honest, in which it  honestly follows the  protocol as specified,   but it   can be curious and try  to extract any  useful information about the training data of the users from  their  received  models.  On the other hand,  users are curious and   can  only collude with each other, such that any colluding set of users only knows  the models from the  users in this set.  Furthermore,  $B$ users   out of the $N$ available  users  are    malicious and could  share false information during protocol execution, or send    malicious updates to the server.\\
At a high level, we want to design a scheme that achieves  1) Secure model aggregation,   where the server can only decode the aggregate model from all users,  while users are allowed to use different quantizers. 2)  Byzantine-resilience  and secure aggregation simultaneously.  We will formalize the objective in Section II-C. Now, we discuss  why SecAg   can  not  be  applied  directly  in  our  setting where users are using different quantizers \footnote{ The difficulties of applying SecAg in the  presence of  Byzantine users is described in  bullet two  in the introduction. }. 

\subsection{Challenges}\label{Challenges}
To describe the main challenge for applying   secure aggregation  protocols with additive masking (including SecAg)  to the case where users are using heterogeneous quantization, we   consider the following  simple example. In this example,  we first start by  describing  the case of    homogeneous  quantization.

\noindent \textbf{Example 1. } We consider having  two users, where user $i$ has an input  $\mathbf{x}_i \in \mathbb{R}$, and a centeral server,  which should only decode  the sum $\mathbf{x} = \mathbf{x}_1+\mathbf{x}_2$. User 1  is assigned these quantization levels, $\mathcal{K}_0 = \{2\}$, while user 2 is assigned  $\mathcal{K}_1 = \{2, 4\}$.   The  encoding processes for  the two users are given as follows:\\ 
\noindent a) \textit{Homogeneous quantization}: As a first step, each user quantizes its input  $\mathbf{x}_i$ by using the same  $K=2$ levels of quantization,  where we assume without loss of generality that the  output of the  quantizer  is denoted by  $\bar{\mathbf{x}}_i \in \{0, 1\}$, for $i =1,2$. The encoded messages $\mathbf{y}_1$  and  $\mathbf{y}_2$ from the two  users,  and the decoded message $\mathbf{x}$ at the server are given by
 \begin{equation}
  \mathbf{y}_1=\bar {\mathbf{x}}_1+ Z_{12} \; \text{mod} \; R, \; \; \; 
\mathbf{y}_2=\bar {\mathbf{x}}_2- Z_{12} \; \text{mod} \; R,  \; \; \;  
 \mathbf{x}=\mathbf{y}_1+\mathbf{y}_2 \; \text{mod} \;R = \bar {\mathbf{x}}_1 + \bar {\mathbf{x}}_2 \; \text{mod} \;R \label{22}, 
 \end{equation}
where the mask $Z_{12}$ is drawn uniformly at random from $[0,R)$. By working in the
space of integers  mod $R$ and sampling masks uniformly over $[0,R)$, this  guarantees that each user’s encoded message 
is indistinguishable from   its own input (mutual information $I(\bar{ \mathbf{x}}_i;\mathbf{y}_i) = 0 $).  The correct decoding in \eqref{22} is guaranteed  for the following two reasons.  First,  the two users and  the server are working in the same 
space of integers  mod $R$. Second, the   summation  and mod commute. Therefore,   the mask pairs will be cancelled out. We note that choosing $R = 3$ ensures that all possible outputs will be represented without any overflow. 

In our problem,  users   are supposed to   use different quantization levels in order to  compress their models to the size (in bits) which is  suitable   to their available communication resources. For instance, user 2 might want to quantize its model using the four levels of quantization which is suitable to its channel bandwidth.    In fact,   allowing  users to  transmit     different number of bits when using   SecAg  requires the size of the  space of  
integers   that users use  for the encoding, e.g., $R$ in \eqref{22},   to be different (the size of the masked  message  $\mathbf{y}_i$ in \eqref{22}  is $\ceil*{\log R}$ bits).   Hence, by using different modular at the users, we end up with incorrect decoding as  shown in the following  case. \\
\noindent b) \textit{Heterogeneous  quantization}: User  1 and user  2 (assuming user 2 has a higher bandwidth  than user 1)  quantize their  inputs  $\mathbf{x}_1$ and $\mathbf{x}_2$ by using    $K=2$  and $K=4$ levels of quantization, respectively.   We assume without loss of generality that the  output of the  two quantizers are    $\bar{\mathbf{x}}_1 \in \{0, 1\}$ and $\bar{\mathbf{x}}_2 \in \{0, \dots, 3\}$.  By further  assuming without loss of generality  that user 1 and user 2 are using mod $ 2$ and mod $4$ in \eqref{22}, respectively,  instead of the same  mod $R$, while having  $\bar{\mathbf{x}}_1 = 1$,   $\bar{\mathbf{x}}_2 = 3$  and $Z_{12}  = 1$, the decoded output  will be given by $\mathbf{x} = \mathbf{y}_1 +\mathbf{y}_2 = 2   $ instead of the true output $\mathbf{x} = 4$.  This confirms that having different modular at the users results in incorrect decoding.    Another issue for using different modular at the users   is that  the space of integer  in which   each user works on to  choose its masks at random will also      be different. Thus,  using SecAg's  approach for   generating the pairwise  masks   does not guarantee   having $0$-sum pairs of masks. For instance,  generating   $Z_{12}= 3$   at both  user 1 and 2 in Case (b) is not possible. The reason for that   $Z_{12}= 3$ only belongs to the space of integers that  node 2 uses to generate its  masks.

One method to cancel out  masks that belong to different spaces of integers is by using   modulus  $D$ at the server, where $D$ is an arbitrary integer, and let all users jointly choose a tuple of masks, whose sum mod $D$ equal to $0$, uniformly at random  from a set  of possible  tuples.  In this tuple of masks,  each mask for each user belongs to its space of integer. The main issue of this approach is that whenever wrapping  around occurs for the  transmitted masked model of any user, the masks will not be cancelled out  at the server side  and the aggregated model   will be distorted.  We will consider the following  example  for illustration.

\noindent \textbf{Example 2.} Assume having three users  with   the quantized model update vectors  $\bar {\mathbf{x}}_1 \in [0,1]^m$, $\bar {\mathbf{x}}_2 \in [0,2]^m$, and $\bar {\mathbf{x}}_3 \in [0,2]^m$, with dimension $m$. Without loss of generality,  we assume  the masks of user $1$, user $2$ and user $3$ take values randomly over $[0,1]^m$, $[0,5]^m$, and $[0,5]^m$, respectively. The transmitted masked models   and the sum of the  masked models at the server   are  given as follows
 \begin{align}
  \mathbf{y}_1=&\bar {\mathbf{x}}_1+ Z_1 \; \text{mod} \; 2, &
\mathbf{y}_2=&\bar {\mathbf{x}}_2+ Z_2 \; \text{mod} \; 6, \nonumber\\
\mathbf{y}_3=& \bar {\mathbf{x}}_3+ Z_3 \; \text{mod} \; 6, &
\mathbf{x}_{\{1,2,3\}}=&\mathbf{y}_1+\mathbf{y}_2+\mathbf{y}_3 \; \text{mod} \;6.
 \end{align}
 If the  users choose their tuple of  masks uniformly at random from the  set of tuples in TABLE \ref{TT0}, the mask  of each  user  becomes    uniformly distributed over its mask  range.  This guarantees user's model privacy in strong information-theoretic sense, i.e., $I(\mathbf{y}_i, \bar{\mathbf{x}}_i) = 0$, for $i = 1, 2, 3$. However, the main limitation  for this approach is that  no guarantee  for  correct decoding.  In particular,    once the masked models are added together, the masks will  not always   be canceled out, but it will only  be cancelled out when there is no   overflow happens for the transmitted masked model of any user. In other words,    the sum of users models will be distorted      whenever   an  overflow happens for the transmitted masked model of any user, which occurs with non-negligible probability.   For example, having  a tuple of masks $(Z_1(k), Z_2(k), Z_3(k)) = ( 1 , 1, 4)$, while the $k$-th element of the  model updates of  the  set of users   $(\bar {\mathbf{x}}_1(k), \bar {\mathbf{x}}_2(k), \bar {\mathbf{x}}_3(k)) = ( 1 , 0, 0)$,  an overflow will occur at user $1$, $\mathbf{y}_1=0$, and the sum will be $\mathbf{x}_{\{1,2,3\}}(k)= 5$  instead of being $\mathbf{x}_{\{1,2,3\}}(k)= 1$. 
 \begin{table}[]
 \caption{Tuples of masks that users could use in  Example 2. }
  \centering
\begin{tabular}{|l|l|l|l|l|l|l|l|l|l|l|l|l|}
\hline
$Z_1$ & $0$ & $0$ & $0$ & $0$ & $0$ & $0$ & $1$ & $1$ & $1$ & $1$ & $1$ & $1$ \\ \hline
$Z_2$ & $1$   & $2$   & $3$ & $4$ & $5$ & $0$ & $1$ & $2$ & $3$ & $4$ & $5$ & $0$ \\ \hline
$Z_3$ & $5$   & $4$   & $3$ & $2$ & $1$ & $0$ & $4$ & $3$ & $2$ & $1$ & $0$ & $5$ \\ \hline
\end{tabular}
\label{TT0}
\end{table}

To overcome the aforementioned  issues  associated with  incorrect decoding and 0-pairwise masks generation  in SecAg when having heterogeneous quantization,   SecAg  can leverage   multi-group structure.  In multi-group structure,  the set of user users  in group $S_g$, for $ g \in [G]$, where $G$ is the number of groups,      uses the same quantizer $Q_{K_g}$, which  has $K_g$ levels proportional to their  communication resources. After that,    each group  applies   the SecAg protocol   independently of all other groups. After that,    each group  applies   the SecAg protocol   independently of all other groups. However, in this strategy   the server would decode the aggregate of the  model updates from  each group  which implies   knowing  the average  gradient/model from each group.  Hence, this strategy  is not robust against  some attacks such as    membership-inference attack \cite{aaa,aaaa,8835245}  and gradient inversion attack  \cite{NEURIPS2019_60a6c400,geiping2020inverting,yin2021gradients}, specially when having a small group size.
In   gradient  inversion attack, the server  can  reconstruct   multiple images used in the training by a user  (subset of users)   from the  averaged gradients  corresponding to these images as  shown in \cite{geiping2020inverting,yin2021gradients}.  In   membership-inference attack,  the server    could  breach  users' privacy by inferring whether   a specific data point was      used in the training   by a certain subset of users or  not by using  the average  model from this targeted set of  users.   By letting    the server observe the average model from  a small group,  the  attack becomes much stronger and   the inferred information may  directly reveal the identity of the users  to whom the data belongs.  Our goal  is to  leverage the benefits of  grouping strategy while limiting the threat of  membership/inversion  attacks.   In particular, we want to address the following question \textit{“Can we design a grouping strategy that allows for secure model aggregation while using heterogeneous quantization    such that the server can not  unmask (decode) the entire  average  gradient/model  from any subset of users”?.  }

In this paper, we propose  HeteroSAg, a scheme which is  based on a specific segment grouping strategy. This segment grouping strategy    leverages  the multi-group structure  for performing     secure model aggregation  with     heterogeneous quantization  while preventing  the server from unmasking  (decoding) the entire  average  model  from any subset of users. At  a high level, our proposed  segment grouping strategy is based on     partitioning the edge users   into groups, and  dividing the local model updates of these users   into segments. Instead of applying  the secure  aggregation protocol  to the entire local model update vectors,  it is  applied   on segments with specific coordination between users, allowing  segments to  be quantized by different quantizers. More specifically, segments from different set of users are grouped  such that they are quantized  by the  same quantizer while being encoded and decoded together at the server independently  of all other segments. This is different from    SecAg where the entire local model updates from all users (or a subset of users in  the  multi-group structure)  are quantized  by the same  quantizer, while being  encoded  and decoded together. Furthermore, unlike   SecAg with  multi-group structure, where the server can decode the entire average model from each group, the   key  objective  of  our  segment grouping strategy    is   to limit the ability of a curious server from launching  inference/inversion attacks on an arbitrary subset of groups. This is achieved  in HeteroSAg by allowing the server to  only decode    a fraction of  at most $\frac{2}{G}$ segments  from the average gradient/model of any set of users which approaches  $0$ for sufficiently large number of  segments $G$.  The remaining segments from this average  model interfere with segments from the average models of some other  groups. 
%a hybrid model from any  subset of users instead of decoding the clear average model from these subset of users. This hybrid model   is composed of only a fraction of clear segments from  the   average  model  from  any subset of users, while the remaining segments in this hybrid model are segments from this  subset of users    interfere with segments from  the remaining set of users.
We quantify     the smallest fraction of segments that the server will not successfully decode from
the aggregated model from any set of users $\mathcal{S} \subsetneq \mathcal{N}$  by the inference robustness $\delta$. In the following subsection, we formally define the inference robustness and discuss  its implications.

\subsection{ Performance metric}
 
Let  $\bm {{\theta}}^p_i=[\bm {\theta}_i^0, \dots, \bm {\theta}_i^{G-1}]$  denotes the segmentation   of the  local model $\bm {\theta}_i$ of user $i$, and   $\bm {\bar{\theta}}^p_{\mathcal{S}}=[\bm {\bar{\theta}}_{\mathcal{S}}^0, \dots, \bm {\bar{\theta}}_{\mathcal{S}}^{G-1}]$ denotes   the segmentation   of the  average model $\bm {\bar{\theta}}_{\mathcal{S}} $ from the set  of users $\mathcal{S}$,  where $\bm {\theta}_i^l, \bm {\bar{\theta}}_{\mathcal{S}}^l \in \mathbb{R}^{\frac{m}{G}}$ for $l=0, \dots, G-1$,  $\ i \in [N]$,  and   $ \mathcal{S} \subsetneq  \mathcal{N}$. We define  $\mathcal{A}( \{\bm {\theta}^p_1, \dots,  \bm {\theta}^p_N \})$  to be an arbitrary 
  segment  grouping strategy   that leverages the multi-group structure for doing secure model aggregation on the segment level. This   strategy $\mathcal{A}$      groups each set of   segments from $\{\bm {\theta}^p_1, \dots,  \bm {\theta}^p_N \}$ together such that they are encoded and decoded together independently of all other segments. We define   $\mathcal{A}( \{\bm {\theta}^p_1, \dots,  \bm {\theta}^p_N \})$  to be feasible if it satisfies these three conditions 1) The server could only receive   a masked model $\bm {\tilde{\theta}}_i$ from user  $i \in \mathcal{N}$,  where the mutual information $I(\bm {\theta}_i;  \bm{\tilde{\theta}}_i) =0$. 2)  The server could only decode  at most a  fraction  of   $ \alpha_{\mathcal{S}}$ segments from the average model  $\bm {\bar{\theta}}^p_{\mathcal{S}}$, where $\mathcal{S} \subsetneq \mathcal{N}$,  while  each segment in the remaining $1-\alpha_{\mathcal{S}}$ fraction of segments   interferes with segments from the average model  $\bm {\bar{\theta}}^p_{\mathcal{S^*}}$ from other sets of users $ \mathcal{S}^*$,  where $ \mathcal{S}^* \subseteq \mathcal{S}^c$ such that $\mathcal{S} \cup \mathcal{S}^c =\mathcal{N}$. 3) The server could decode the average model $\bm {\theta}$ from all users. Now, we define our inference robustness metric.

\noindent  \textbf{Definition 1.} (Inference robustness $\delta$) \textit{  For a feasible segment  grouping strategy  $\mathcal{A}( \{\bm {\theta}^p_1, \dots,  \bm {\theta}^p_N \})$,  the inference robustness $\delta(\mathcal{A})$, where $\delta (\mathcal{A}) \in [0,1]$,   is given as follows: 
\begin{equation}
\label{infer}
\delta(\mathcal{A}) = \min \left\{  1-\alpha_{\mathcal{S}}: \mathcal{S}\subsetneq   \{ 1, \dots, N\} \right\}, 
\end{equation}
where $1-\alpha_{\mathcal{S}}$ is the  fraction of segments from $\bm {\bar{\theta}}^p_{\mathcal{S}}$ that    interferes with segments from the average models  $\bm {\bar{\theta}}^p_{\mathcal{S^*}}$,  where $ \mathcal{S}^* \subseteq \mathcal{S}^c$ such that $\mathcal{S} \cup \mathcal{S}^c =\mathcal{N}$. }

\begin{remark} The underlying objective of a good segment  grouping strategy  $\mathcal{A}( \{\bm {\theta}^p_1, \dots,  \bm {\theta}^p_N \})$ is to limit the ability of a curious server from launching  inference/inversion   attacks on an arbitrary subset of users $\mathcal{S} \subsetneq \mathcal{N}$, by allowing the server to decode only   a fraction  $ \alpha_{\mathcal{S}} \in [0,1)$ of segments from the average model  $\bm {\bar{\theta}}^p_{\mathcal{S}}$. This is  different from   the worst case scenario  where the server can decode the entire  target model $ \bm{\bar{ \theta}}_{\mathcal{S}}$.   The segments from the average model  $\bm {\bar{\theta}}^p_{\mathcal{S}}$ that interfere with segments from other users outside the set ${\mathcal{S}}$ can be viewed as  clear segments plus random noise, where the number of noisy segments  is determined    by  the inference robustness $\delta(\mathcal{A})$. The worst case inference robustness  for  HeteroSAg   is  $\delta  = \frac{G-2}{G}$,  which approaches one for sufficiently large number of segments.
\end{remark}

\section{The proposed HeteroSAg    } \label{sec-4} 
We first present  our    HeteroSAg,   and then    state     its theoretical  performance  guarantees. 
\subsection{HeteroSAg for heterogeneous quantization}\label{sec-4.1} 
HeteroSAg starts by letting the  set of  $N$ users   share their  keys and masks according to Step 1 in  Section \ref{sec-1.2}. Each  user  $i \in \mathcal{N}$   then uses  Step 2 to   secret shares its  masks with all  other  users.  For clarity and  ease of  analysis of  the proposed scheme,  we first    consider the case where users are already   clustered into  $G$ groups based on their communication resources,  each of which  has the same number of users,   $|\mathcal {S}_g|=\frac{N}{G}=n$, for $g \in [G]$, where $[G] :=\{  0, \dots, G-1\}$. The case of having a different number of users in each group is presented in the  Appendix \ref{HeteroSAg  s2}. Without loos of generality,   we consider users in higher groups have   communication resources   higher than users in lower groups.     Following the secret sharing step,  each  local  model update  vector   $\{\mathbf{ x}_i\}_{i \in \mathcal{N}}$   is equally partitioned into $G$ segments  such that the segmented model update  of user $i$ is given by $\mathbf{ x}_i^p=[\mathbf{ x}_i^0, \mathbf{ x}_i^1, \dots, \mathbf{ x}_i^{G-1}]^T$, where $\mathbf{ x}_i^l \in \mathbb{R}^{\frac{m}{G}}$ for $l \in [G]$.  Also, the  aggregated model update   $\mathbf{ x}_\mathcal{S}$ at the server from any set of users  $ \mathcal {S} \subsetneq  \mathcal {N}$ can be viewed as   a set of $G$ segments $\mathbf{ x}^p_\mathcal{S}=[\mathbf{ x}_\mathcal{S}^0, \mathbf{ x}_\mathcal{S}^1, \dots, \mathbf{ x}_\mathcal{S}^{G-1}]^T$. Finally, instead of  the direct implementing of  SecAg protocol    where  (1)  All the unsegmented  vectors $\{\mathbf{ x}_i\}_{i \in \mathcal {N}}$ are quantized  by the same quantizer, (2)   All the $N$ users  jointly   encode (mask)  the $N$ quantized vectors together,  and (3)  All the $N$ encoded model updates  will be decoded together at the server,    we   apply  the segment grouping strategy $\mathcal{A}_{\text{HeteroSAg}}( \{\mathbf{ x}_1^p, \dots ,\mathbf{ x}_N^p$\}) such that SecAg protocol is applied on the segment level where     (1) Different sets of segments are quantized by using  different  quantizers, (2) Different sets of users   jointly  encode   their quantized segments together independently of all other users, and  (3) The  jointly encoded  segments will be  also jointly decoded  at the server.   
 
\begin{algorithm}[H]\small
\SetAlgoLined
 \For{$g=0, \dots, G-2$}{
 \For{$r=0, \dots,  G-g-2$}{
 $l=2g+r$\;
 $\mathbf{B}( l \mod G,g)=\mathbf{B}(l  \mod G, g+r+1)= g$\;
 }
 }
The remaining entries of  $\mathbf{B} _{\text{ HeteroSAg }}$ will hold $*$.
 \caption{The SS Matrix $\mathbf{B}$ for HeteroSAg}
 \label{algo}
\end{algorithm}

 Each   set of segments  and its    corresponding  set of users that  jointly executes   SecAg  protocol together   according to the  segment grouping strategy $\mathcal{A}_{\text{HeteroSAg}}$ is   given by  the 
   $G\times G$  Segment Selection (SS)  matrix $\mathbf{B}$ produced by Algorithm 1.
   In this matrix and   as illustrated in  the  example   given in Figure.  \ref{matrix},  

\begin{figure}[h]
\centering
  \[\footnotesize
\myMatrix{ \mathbf{B}} = \begin{blockarray}{cccccc}
0& 1 & 2& 3 & 4 \\
\begin{block}{(ccccc)c}
0 & 0 &2&*&2&0\\
0 & *  &0& 3&3&1\\
0 &1  &1&0&*&2\\
0 &1 & *&1 &0&3\\
* &1 & 2&2&1 &4\\
\end{block}
\end{blockarray} 
\]
\captionof{figure}{ Segment selection matrix   $ \mathbf{B}$ for  $G=5$ groups.}
\label{matrix}
\label{fig:image1}
\end{figure}the label for each column represents the index of the group (index $g$ of the set of users $\mathcal{S}_g$, for $g \in [G]$),  where the  communication resources  of the set of users $ \mathcal {S}_g$  in group $g$ is smaller  than those  of the set of users  $\mathcal {S}_{g'}$,  where $g
 < g'$. On the other hand, the label of each row represents the index $l$ of the  segment $\mathbf{ x}_i^l$.     In this  matrix  having an entry $\mathbf{B} (l,g)=*$  means that the  set  of users $ \mathcal {S}  = \mathcal {S}_g$  will execute SecAg protocol   on the set   of segments  $\{\mathbf{ x}_i^{l}\}_{i \in \mathcal {S}_g}$. In other words, having $\mathbf{B} (l,g)=*$ means that  the  set  of users $  \mathcal {S}_g$    will  quantize  the  set of segments $\{\mathbf{ x}_i^{l}\}_{i \in \mathcal {S}_g}$  according to  Step 3 in Section \ref{sec-1.2} by using the  quantizer   $Q_{K_g}$,    and   jointly  encode the resulting  quantized  segments  together according to Step 4. At the server, these set of segments will be decoded together.  Similarly,   when $\mathbf{B} (l,g)=\mathbf{B} (l,g') = g $, this means that  the  set of users $ \mathcal {S}  = \mathcal {S}_g \cup {S}_{g'} $  corresponding to these  columns $g$ and $g'$, where $g<g'$,    will  quantize  the    set of segments $\{\mathbf{ x}_i^{l}\}_{i \in \mathcal {S}_g \cup {S}_{g'} }$ by using the quantizer  $Q_{K_g}$  and   jointly  encode the resulting quantized  segments.  At the server side, these set of segments will be decoded together. 
Finally,  the server  aggregates each   set of  decoded segments $\{\mathbf{ x}_\mathcal{S}^l\}_{\mathcal{S}\subset\{0, \dots, G-1\}}$, which results from different sets of users and belongs to the same segment level $l$,  together.   The server   concatenates these sets of aggregated segments,  which belong to these levels   $l \ \in [G]$,  to get the global update $\mathbf{ x}$. To  illustrate HeteroSAg and understand how its  inference robustness is measured,    we consider the following example.

\noindent \textbf{Example 3.}  We consider a system  which consists  of $N$ users, and  a set of $G=5$  quantizers  $\mathcal {Q}= \{Q_{K_0}, Q_{K_1}, Q_{K_2} , Q_{K_{3}} ,Q_{K_{4}}\}$, where  $K_0<\dots<K_4$. HeteroSAg execution     starts   by letting the $N$ users first  share their keys and masks with each other,  and then each user  secret shares its  masks with the other  users in the system. We consider having $G=5$ groups  with $n$ users in each group, where groups are arranged in ascending order based on the  communication resources of their users. The local  model  of each user,  $\mathbf{ x}_i$ for  $i \in \mathcal{N}$,  is equally partitioned into $G=5$ segments $\mathbf{ x}_i=[\mathbf{ x}_i^0, \mathbf{ x}_i^1, \mathbf{ x}_i^2, \mathbf{ x}_i^3, \mathbf{ x}_i^4]^T$, where $\mathbf{ x}_i^l \in \mathbb{R}^{\frac{m}{5}}$, for $l \in [5]$. 
  The SS matrix $\mathbf{B}$ that is used for managing the execution of  HeteroSAg      is given in  Figure.  \ref{matrix}.

\begin{figure*}
\centering
\includegraphics[width=\textwidth,height=8cm,keepaspectratio]{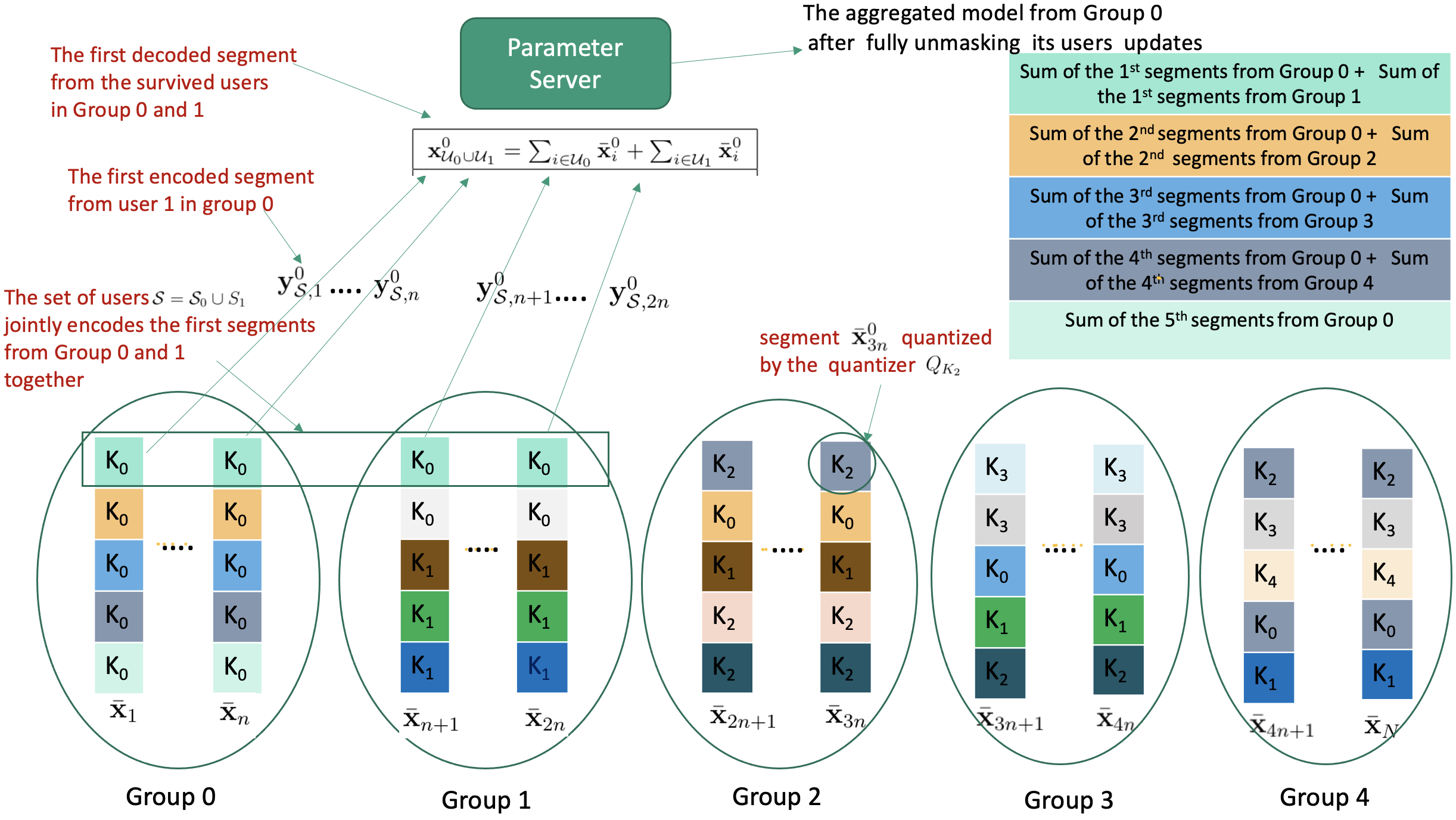}

\caption{ A system with $N$ users partitioned into $G=5$ groups, with $n$ users in  each group. Each user holds a quantized local model update    $\bar {\mathbf{ x}}_i$, $i \in [N]$. The segment selection and grouping is completed by using the SS matrix $\mathbf{B}$. }
\label{EX1}
\end{figure*}

To  further formalize the execution of HeteroSAg for  this  example, we consider  Figure. \ref{EX1}. In this  figure,      each  set of segments that executes the SecAg  together is  given the same color.   In particular,  the set of segments  $\{\mathbf{ x}_i^{0}\}_{i \in \mathcal {S}_0 \cup {S}_{1} }$ will be quantized by the quantizer  $Q_{K_0}$ according to   the third step in Section \ref{sec-1.2}. By using  the encoding   step in Section \ref{sec-1.2}, the output of the quantizer  $\{\bar { \mathbf{ x}}_i^{0}\}_{i  \in \mathcal {S}  }$, where $ \mathcal {S}= \mathcal {S}_0 \cup {S}_{1}$, will be first   mapped    from the   values  that belongs to its  discrete    range   to   integer values in this range   $[0,K_0-1]$.  By  generating the  random $0$-sum pairs of masks and  the individual masks,    the encoded  segment for each user  $i \in \mathcal {S}  $ will be given   as follows \begin{equation}
\mathbf{ y}^0_{  \mathcal {S}, i
}= \bar {\mathbf{ x}}_i^0 +  \text{PRG}(b_i)+\sum_{ j:i<j}  \text{PRG}(s_{i,j})- \sum_{j:i>j}  \text{PRG}(s_{j,i}) \text{       mod } R, 
\end{equation}

\noindent where $R=|\mathcal {S}|(K_0-1)+1$,   $|\mathcal {S}|=2n$, and   $j \in \mathcal {S}  $, while the PRG  is used to expand the different seeds  to  segments     in $\mathbb {Z}_{R}$.
The  server     collects  the  shares and    reconstructs the private seed of each surviving user, and the pairwise seeds of each dropped user. Then its uses the   PRG  along with the reconstructed seeds to expand them   to  segments     in $\mathbb {Z}_{R}$, where $R=|\mathcal {S}|(K_0-1)+1$,   $|\mathcal {S}|=2n$,  to be removed from the aggregate of the masked segments. The server then computes this   segment
\begin{align}
\mathbf{ x}^0_\mathcal{\mathcal{U}}=&   \sum_{i\in \mathcal{U}} ( \mathbf{ y}^0_{  \mathcal {S}, i
}- \text{PRG}(b_i) )  -  \sum_{i\in \mathcal{D}}  \left(\sum_{j:i<j}  \text{PRG}(s_{i,j})- \sum_{j:i>j}  \text{PRG}(s_{j,i}) \right) \mod  R \nonumber \\ 
=& \sum_{i\in \mathcal {U}_0} \bar {\mathbf{ x}}^0_i+\sum_{i\in \mathcal {U}_1} \bar {\mathbf{ x}}^0_i \mod R, 
\end{align}
   where the  set   $\mathcal {U}_i$    represents the set of survived users from $\mathcal {S}_i$, for $i = 0 ,1$.  
   
   The  aggregate model update from group  $0$  after fully unmasking its users' models  is  given by    Table \ref{T.2},  where   $\mathcal{U}_g \subseteq \mathcal {S}_g$  for $g\in [4] $, represents the set of survived users from group  $g$. According to Table \ref{T.2},   the  server will  decode    only the last segment from the aggregated model update  from  group 0, while the other segments from that group  interfere with  segments from some other groups \footnote{Decoding   a segment $\mathbf{ x}^l_{\mathcal{S}}$ from the  model update  $\mathbf{ x}_{\mathcal{S}}$ implies decoding the  segment   $\mathbf{ \theta }^l_{\mathcal{S}}$ from  the average  model  from that $\mathcal{S}$.   We note that   $\mathbf{ \theta }^l_{\mathcal{S}} =     \mathbf{ \theta }^l + \frac{1}{|\mathcal{S}|}\mathbf{ x}^l_{\mathcal{S} }$, where $ \mathbf{ \theta }^l$ is the $l$-th  segment from  the global model   $ \mathbf{ \theta }$.}. In particular, the first  segment from  group 0   (first row in Table \ref{T.2})  results from  the sum of the first set of segments  from the survived users in group 0 and group 1, as  these segments were encoded together, and hence must be decoded together.  More generally, the server   will   decode   only  one clear  segment from each individual group which is   corresponding to the index  denoted by  $*$   in the SS matrix $\mathbf{B}$. Also,  it can be easily seen that  the server will  not  decode  more than  $0.2  $ of clear   segments from the average model  from  any set of users $\mathcal {S} \subsetneq \mathcal {N}$,   and hence the  inference robustness     will be 
   $\delta (\mathcal{A}_{\text{HeteroSAg}})= \frac{4}{5} = 0.8$. 

\begin{table}[h]
\centering
\caption{The aggregated model update $\mathbf{x}_{\mathcal{S}_0}$     from  group $0$   after     fully  unmasking the  model updates   of its users \vspace{3mm}}
\begin{tabular}{|c|}
\hline
$\mathbf{ x}^0_{\mathcal{U}_0 \cup  \mathcal{U}_1}= \sum_{i\in \mathcal {U}_0} \bar {\mathbf{ x}}^0_i+\sum_{i\in \mathcal {U}_1} \bar {\mathbf{ x}}^0_i$ \\ \hline
$\mathbf{ x}^1_{\mathcal{U}_0 \cup  \mathcal{U}_2}= \sum_{i\in \mathcal {U}_0} \bar {\mathbf{ x}}^1_i+\sum_{i\in \mathcal {U}_2} \bar {\mathbf{ x}}^1_i$ \\ \hline
$\mathbf{ x}^2_{\mathcal{U}_0 \cup  \mathcal{U}_3}= \sum_{i\in \mathcal {U}_0} \bar {\mathbf{ x}}^2_i+\sum_{i\in \mathcal {U}_3} \bar {\mathbf{ x}}^2_i$ \\ \hline
$\mathbf{ x}^3_{\mathcal{U}_0 \cup  \mathcal{U}_4}= \sum_{i\in \mathcal {U}_0} \bar {\mathbf{ x}}^3_i+\sum_{i\in \mathcal {U}_4} \bar {\mathbf{ x}}^3_i$ \\ \hline
$\mathbf{ x}^4_{\mathcal{U}_0}= \sum_{i\in \mathcal {U}_0} \bar {\mathbf{ x}}^4_i$                                           \\ \hline
\end{tabular}
\label{T.2}
\end{table}

\begin{figure*}[t]
\centering
\includegraphics[width=13cm,height=8cm,keepaspectratio]{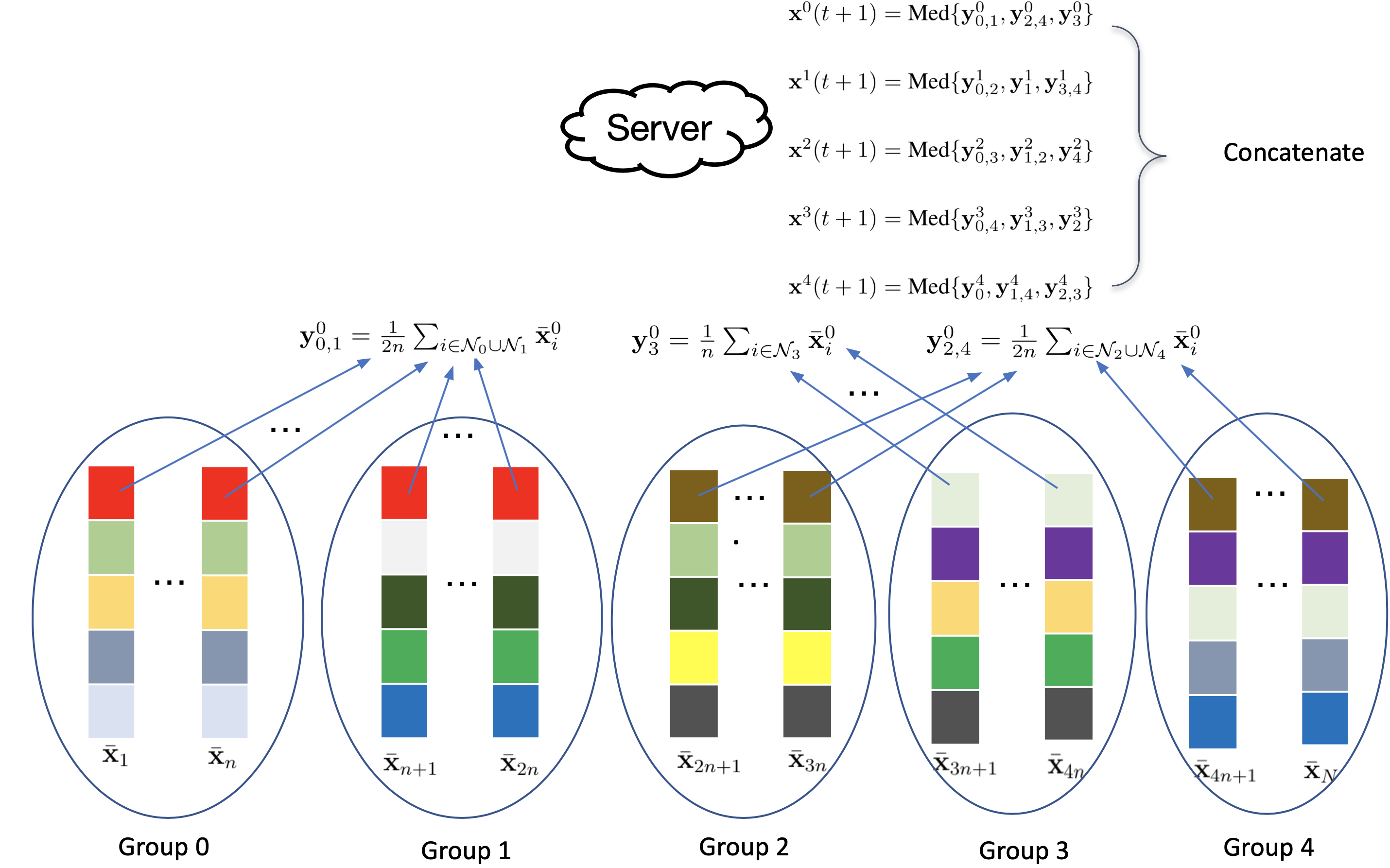}

\caption{  A system with $N$ users partitioned into $G=5$ groups, with $n$ users in  each group. Each user holds a  local model update    $\bar {\mathbf{ x}}_i$, $i \in [N]$. The segment selection and grouping is completed by using the SS matrix $\mathbf{B}$.  Segments with the same color will be encoded and decoded together independently of all other segments.   }
\label{Fig.3}
\end{figure*}
\subsection{HeteroSAg for  Byzantine-Resilience}\label{sec-4.2}
Now, we extend   the   segment grouping strategy $\mathcal{A}_{\text{HeteroSAg}}$ given by the SS matrix $\mathbf{B}$ generated by    Algorithm  \ref{algo}  to further  provide  Byzantine robustness   while achieving  secure model  aggregation simultaneously. This can be done be integrating $\mathcal{A}_{\text{HeteroSAg}}$  with   some  coordinate-wise defense  techniques, such as  coordinate-wise median (Median) \cite{By},  which have provable convergence guarantees. Integrating  Median with HeteroSAg is possible thanks to  the design of the SS matrix.  Particularly in the SS matrix, for a   given row $l$ out of the available $G$ rows  (segment index $l$ out of the $G$ segments indices),    the server  observes a set of unmasked segments.  Each unmasked segment  results from averaging  the segments that were encoded together.
 This is difference than the design  of the other secure aggregation protocols  which make the   server       only  receive the masked model of each user  and the aggregate of all users model updates.   Observing the  masked model from each user  makes the adaption of the  state-of-the-art  Byzantine-robust techniques  difficult to implement, as     these defense techniques are based on observing the individual clear model update  of each user to compare the updates from different users with each other and remove the outliers.  To further illustrate how the design of the SS matrix has  solved the aforementioned  limitation of the convention secure aggregation protocol,  we consider     the  following example. \\
 \textbf{Example 4}.   We consider a system  with  $N$ users, where users are divided equally among $G =5$  groups, and  each group has $n$ users,  as illustrated in Figure.  \ref{Fig.3}. We assume that node 1 in the first group is a Byzantine node.  The local model update of each node is divided equally into $G=5$ segments. Each set of  segments that  is  encoded and decoded together are given the same color as shown in Figure.  \ref{Fig.3}. The segments are grouped and colored according to the SS matrix $\mathbf{B}$  given in Figure.  \ref{matrix}. As can be seen from  Figure.  \ref{Fig.3},   the  server   decodes the  segments that were encoded   together, e.g.,
$\{ \mathbf{y}^0_{0,1} = \frac{1}{2n} \sum_{i \in \mathcal{N}_0 \cup \mathcal{N}_1 } \bar{\mathbf{x}}^0_i$, 
$\mathbf{y}^0_{2,4} = \frac{1}{2n}\sum_{i \in \mathcal{N}_2 \cup \mathcal{N}_4 } \bar{\mathbf{x}}^0_i$, $\mathbf{y}^0_{3} = \frac{1}{n} \sum_{i \in \mathcal{N}_3} \bar{\mathbf{x}}^0_i$\}. The  segments in this set  are   from the sets  of users in this tuple of sets $(\mathcal {S}_0 \cup \mathcal {S}_1, \mathcal {S}_2 \cup \mathcal {S}_4, \mathcal {S}_3)$, respectively,     and belong to the same  segment level $l=0$.    Since we have more than one segment in this set,   coordinate-wise median    can be applied as demonstrated  in Figure.   \ref{Fig.3}.

The coordinate-wise median   scheme  in  \cite{By} is presented for the case where the number of Byzantine users is less than half the total  number of users, i.e., the number of benign models is more than   the number of faulty models. To find the number of allowed Byzantine users in our setting while using  the segment grouping strategy $\mathcal{A}_{\text{HeteroSAg}}$,   we  consider the worst case  scenario  where Byzantine users are distributed uniformly  among the  groups. To make sure that each  set of  unmasked segments, which  belongs to the same level $l$, e.g., this set of unmasked segments $\{ \mathbf{y}^0_{0,1} = \frac{1}{2n} \sum_{i \in \mathcal{N}_0 \cup \mathcal{N}_1 } \bar{\mathbf{x}}^0_i$, 
$\mathbf{y}^0_{2,4} = \frac{1}{2n}\sum_{i \in \mathcal{N}_2 \cup \mathcal{N}_4 } \bar{\mathbf{x}}^0_i$, $\mathbf{y}^0_{3} = \frac{1}{n} \sum_{i \in \mathcal{N}_3} \bar{\mathbf{x}}^0_i$\},  contains benign  segments more than  faulty segments,  the number of Byzantine users, $B$, should be  $ B \leq \ceil*{0.25 G}-1$. The former result comes from the fact that having one Byzantine user in one group  makes all the segments of the average model from  this group  faulty.  Also  in   HeteroSAg,   we can see that some  segments belong to the aggregated model  from two groups. In particular, we can see the faulty model from  user 1 in group $0$ results in having the following faulty segments  $\{\mathbf{y}^0_{0,1}, \mathbf{y}^1_{0,2}, \mathbf{y}^2_{0,3}, \mathbf{y}^3_{0,4},\mathbf{y}^0_{0} \}$.  By taking this extreme case where    one Byzantine node has an impact on the segments from two groups, and  the number of  benign segments should be more than  the faulty segments within each set of decoded segments,  the number of Byzantine users, $B$, should be  $ B \leq \ceil*{0.25 G}-1$.  

We have included further  discussion for  the Byzantine robustness of HeteroSAg in Appendix \ref{Byz_extra}.

 \subsection{Theoretical guarantees of   HeteroSAg }\label{Assum}
We  state our main theoretical results.  The proofs of the theorems, propositions, and lemmas  are  presented in  the Appendix.
 \begin{theorem}(Inference robustness)\label{th1}
     For a FL  system with $N$ users clustered into $G$ groups, and the model update of each user is divided equally  into $G$ segments, the segment grouping strategy of     HeteroSAg    achieves    inference robustness  of $\delta(\mathcal{A}_{\text{HeteroSAg}})=\frac{G-2}{G} $ when the number of groups is even,  and $\delta(\mathcal{A}_{\text{HeteroSAg}})=\frac{G-1}{G} $   when the number of groups  is odd.  
\end{theorem}

\begin{remark}
Theorem \ref{th1} shows that we can achieve full inference robustness  for sufficiently large number of groups.  In particular, the maximum value for  inference robustness, $\delta$, is reached when $G=m$. In this case,  the server can decode   $\frac{2}{m} $ (or $\frac{1}{m}$  when $m$ is odd)  of the average model from  any set  of users,   which approaches zero for sufficiently large model size. In the Appendix,  we  show how we can further portion the users in each group to smaller sub-groups when the size of the set of quantizers $G$ is small  to increase $\delta$. 
\end{remark}
We derive the convergence  guarantees  of HeteroSAg under the following standard assumptions.  \\
\noindent\textbf{Assumption  1 } (Unbiasedness) \textit{The stochastic gradient  $\mathbf{x}_i^{(t)} = \mathbf{g}_i(\bm{\theta}^{(t)}) $   is an unbiased estimator  for the true gradient of the global  loss function in (1) such that  $\mathbb{E} [\mathbf{x}_i^{(t)}]  =\nabla F(\bm {\theta^{(t)}}$)}\\
 \noindent\textbf{Assumption  2 }(Smoothness) \textit{The  objective  function  $F(\bm {\theta})$  in (1) is convex, and  its  gradient is   $L$-Lipschitz that is  $||\nabla F(\bm {\theta}) - \nabla F(\bm {\theta'})|| \leq L|| \bm {\theta}- \bm {\theta'}||$,  for all  $\bm {\theta}, \bm {\theta'} \in \mathbb{R}^m$. }
 
 \noindent \textbf{Lemma 1} \textit{For any vector    $\mathbf{x}_i= [\mathbf{x}_i^0, \dots, \mathbf{x}_i ^{G-1}]  \in  \mathbb{R}^m$, where   $\mathbf{x}_i^l \in \mathbb{R}^{\frac{m}{G}}$   and its  values belong to this interval $[r_1,r_2]$, and by  letting  $\bar{\mathbf{x}}_i$ to be the quantization of $\mathbf{x}_i$,   we have that (i) $E[\bar{\mathbf{x}}_i] = \mathbf{x}_i$ (unbiasedness),  (ii) $E ||\bar{\mathbf{x}}_i - \mathbf{x}_i||_2^2 \leq \frac{m}{G}  \sum_{l=0}^{ G-1}   \frac{(\Delta_i^l)^2}{4} $ (bounded variance), where $\Delta_i^l$  is  the  quantization interval associated with the stochastic  quantizer $Q_{K_i^l}$  used to quantize the $l$-th  segment $\mathbf{x}_i^l$. (iii) $\mathbf{E}  || \bar {\mathbf{p}} - \mathbf{p} ||^2_2 \leq \sigma^2 $ (total quantization error), where $ \bar{\mathbf{p}} =  \frac{1}{N}\sum_{i=1}^{ N} \bar{\mathbf{x}}_i$, $\mathbf{p} =  \frac{1}{N} \sum_{i=1}^{ N} \mathbf{x}_i$ and $\sigma^2 = \frac{(r_2-r_1)^2}{4N^2} \frac{m}{G} \sum_{i=1}^{N}  \sum_{l=0}^{G-1}
\frac{1 }{(K_i^l-1)^2}$. Here,  $G$ is the number of segments and $K_i^l$ is the number of levels used to quantize the $l$-th segment of $\mathbf{x}_i$.}
\begin{theorem}(Convergence)\label{Th4}
Consider   FL  system  with  $N$ users,  each of which  has a local gradient vector  $\mathbf{ x}_i^{(t)}= \mathbf{g}_i(\bm{\theta}^{(t)}) \in \mathbb{R}^m$, such that  the elements of each     local gradient $\mathbf{ x}_i^{(t)}$, for $i=1, \dots, N$,  fall in  the range $[r_1,r_2]$.   Suppose the conditions in Assumptions 1-2
are satisfied. When each local gradient vector  is partitioned equally  into $G$   segments such that  segments are quantized  by using the  set   $\mathcal{Q}$ of $G$   quantizers according to    the  SS matrix $\mathbf{B}$,   and by using  constant step size $\eta =  1/L$,     HeteroSAg guarantees 
\begin{align}\label{error111}
\mathbb{E}\left[ F\left(\frac{1}{J} \sum_{t=1}^J \bm{\theta}^{(t)}\right)\right] -  F(\bm{\theta}^{*})  \leq \frac{||\bm{\theta}^0-\bm{\theta}^{*}||^2}{2\eta J} + \eta \sigma_{\text{HeteroSAg}}^2,
%\epsilon \leq \frac{(r_2-r_1)^2}{4N^2} \frac{m}{G} n  \sum_{g=0}^{G-1} \frac{2(G-g )-1}{(K_g-1)^2 }.
\end{align}
where $\sigma_{\text{HeteroSAg}} = \frac{(r_2-r_1)^2}{4N^2} \frac{m}{G} n \sum_{g=0}^{G-1} \frac{2(G-g )-1}{(K_g-1)^2} $, and   $\bm {\theta}^0$ is the  initial model. 
 \end{theorem}

\begin{remark} 
  HeteroSAg  has
 a convergence rate of $O(1/J)$. The  term  $\eta \sigma_{\text{HeteroSAg}}$
is a residual error in the training which can be reduced  by using an adaptive (decreasing) learning rate and  by using a set of  high level quantizers.
\end{remark}
 
\begin{remark}According to Theorem 1 and the bound  on the number of Byzantine nodes  given in  Section \ref{sec-4.2}, increasing the  number of groups by further partitioning each group out of the $G$ available  groups  equally into $L$ subgroups results in increasing      1)  The number of Byzantine to be tolerated  $ B \leq \ceil*{0.25 LG}-1$,  2) The inference robustness  $\delta = \frac{LG-2}{LG}$.  On the other hand, the residual error in Theorem 2 will not increase as stated  in  the next proposition. 
 \end{remark}

 \noindent \textbf{Proposition 1} (Quantization error) \textit{ Let $G$  be the number of  quantizers to be used in the system, and users are partitioned  equally into $G$ groups, by extra partitioning  each group equally into $L$ subgroups while using the segment grouping strategy of  HeteroSAg, the total quantization error  will not be changed and will  also be  given by $\sigma_{\text{HeteroSAg}}^2$. }

\begin{theorem} (Privacy leakage and dropout) For HeteroSAg  when  the number of users in each subgroup is given by $\bar n = \frac{N}{LG}$, where $LG$ is the total number of subgroups,    and  the dropout probability of each user is $p$, the probability of  privacy leakage, i.e., having  only  one  survived  user in  any  subgroup,   is given  by 
\begin{equation}\label{bound}
\mathbb {P}[ \text{Privacy leakage}] = \mathbb P(X  =1) =   \bar n (1-p) p^{\bar n-1},
\end{equation}
where having one survived user in any group implies that the server will be able to decode one clear segment  from  the model of that user, i.e., $I(\mathbf{x}^l_i; \mathbf{y}^l_i) \neq 0$, for a given segment $l$, where  $\mathbf{y}^l_i$ is the $l$-th encoded segment from user  $i$. 
 \end{theorem}

\begin{remark}The probability in \eqref{bound} approaches zero by either having a small   probability of dropout $p$, or by increasing the number of users in each subgroup. Therefore,   the number of users in each subgroup makes a trade-off between the  benefits of extra partitioning discussed in Remark 4 and   the privacy of users' models.  To further illustrate the impact of the subgroup size $\bar n$ on the 
 probability  in \eqref{bound}, we consider the following
  example. Let the total number of users in each subgroup to be $\bar n = 8$, and by considering $p = 0.1$,  a typical number for the  probability of dropout \cite{drop}, the probability in \eqref{bound} turns out to be $ 7.2 \times 10^{-7}$, which is negligible.    We further  note that      by using HeteroSAg,  the    dropout  rate of users   becomes  smaller.  The reason for  that  in  HeteroSAg  users  consider   their   transmission rates    when they  choose  their quantizers.   This decreases their   probabilities  of being delayed and hence  being considered dropped out by the server.
\end{remark}

 \noindent \textbf{Proposition 2} (HeteroSAg  communication and computation costs) \textit{ Each  user has a  computation cost of   $\mathcal{O}(N^2+  m \bar n)$  as opposite to  $\mathcal{O}(N^2+ m  N )$ for SecAg, where $m$, $N$ and $\bar n$ are the model size, total number of users and number of users in each group, respectively, and the same    communication cost $\mathcal{O}(N+m)$  as    SecAg. However, the bandwidth expansion, which measures the ratio between the size of the encoded model in bits to the size of the clear model, is much lower for HeteroSAg. The communication and the computation complexities at the server are the same as in SecAg.  }

 \begin{table}[h]
 \centering
 \caption{A Comparison between  SecAg \cite{cc} and our proposed HeteroSAg. Here,  $N$ is the total number of nodes, $m$ is the model size, $G$ is the number of groups, $n$ is the number of users in each group, and $K_g$  is the number of quantization levels.}
\begin{tabular}{lll}
\hline
                                  & SecAg                                                 & HeteroSAg                                                                              \\ \hline
Adaptive quantizers               & No                                                    & Yes                                                                                    \\
Communication complexity          & $\mathcal{O}(N+m)$                                    & $\mathcal{O}(N+m)$                                                                     \\
Computation complexity            & $\mathcal{O}(N^2+ m  N )$                             & $\mathcal{O}(N^2+  m  n)$                                                          \\
Inference robustness              & $1$                                                   & $\frac{G-2}{G}$                                                                          \\
Byzantine Robustness              & No                                                    & Yes                                                                                    \\
Quantization error bound          & $ \frac{(r_2-r_1)^2}{4N^2}  m (nG) \frac{1}{(K-1)^2}$ & $\frac{(r_2-r_1)^2}{4N^2} \frac{m}{G} n \sum_{g=0}^{G-1} \frac{2(G-g )-1}{(K_g-1)^2} $ \\
Probability of local model breach & $0$                                                   & $\to 0$                                                                                \\ \hline
\end{tabular}
\label{Table_x_sup}
\end{table}

In Table  \ref{Table_x_sup}, we give  a  comparison between   HeteroSAg and SecAg \cite{cc}. As we can observe  from  this table  that HeteroSAg is an adaptive algorithm that   allows users to use different quantizers to balance their communication load to their channel quality, while achieving secure model aggregation. This is different from  SecAg which requires all users to use the same quantizer to guarantee correct decoding as discussed in Section \ref{Challenges}. Additionally, HeteroSAg achieves Byzantine robustness, while SecAg fails in the presence of Byzantine nodes, as we will  demonstrate in Section \ref{Experiments}. Furthermore, the communication complexity of HeteroSAg is lower than SecAg, while both algorithms have the same communication complexity. Regarding the inference robustness, SecAg achieves inference robustness of  $1$,  meaning that the server will not decode any segment from the average model of  any subset of users. On the other hand, HeteroSAg achieves lower inference robustness. However, the inference robustness of HeteroSAg approaches $1$  for sufficiently large number of groups. The probability of local model breach when users dropped out in SecAg is $0$,   while in HeteroSAg this probability approaches $0$ when increasing   the number of users in each group (Theorem 3).

\section{Numerical Experiments}\label{Experiments}
We run two different  experiments  to show the performance gains achieved by  HeteroSAg. Experiment 1   highlights  the  benefits of using        heterogeneous quantization.   The second experiment is  to demonstrate   how the   secure aggregation strategy  of HeteroSAg can be effective   along with   coordinate-wise median   against   Byzantine users.  

\subsection{Experiment 1 (Heterogeneous quantization) }\label{Experiment 1}   We consider the setup of $N=25$ users,  where  users are  equally partitioned   into $G=5$ groups and each   model update vector  is equally  partitioned into $5$ segments.  We   consider    MNIST dataset \cite{MNIST} and  use a neural network with two fully connected layers. The details of the neural network is presented in Appendix \ref{model}.   For the data distribution, we sort the training data as per class, partition the sorted data into $N$ subsets, and assign each node one partition. We set the number of epochs to be $5$,  use a batch size of $240$, and constant  learning rate $0.03$.   We consider three different scenarios for the performance comparison based on the quantization scheme. The three scenarios  apply the same segment grouping strategy given by the SS matrix B in Figure.  \ref{matrix} in terms of the encoding  and decoding strategy (e.g., the first segment from group 0 and group 1 will be encoded together and decoded together at the server in the same way for  the three scenarios), while they are different in the quantization scheme.  \iffalse For the data distribution, we sort the training data as per class, partition the sorted data into $N$ subsets, and assign each user one partition. For running the experiment, we use learning rate of $0.03$,  set the number of epochs to be $5$, and use  $0.1$ of the local dataset as a mini-batch. \fi
%{\color{red} that state-of-the-art approaches have achieved a test accuracy of 96.5% [19] for CIFAR; nevertheless, the standard
%model we use is sufficient for our needs, as our goal is to
%evaluate our optimization method, not achieve the best possible accuracy on this task.}
%In particular,  we split the training data samples with the same
%label (from the same class) to $4$ disjoint subsets.  We
%then assign two  subsets of data samples from two consecutive classes to a different user.

\begin{figure}[]
  \centering
   \subfigure[Test accuracy for   MNIST dataset]{\includegraphics[scale=0.4]{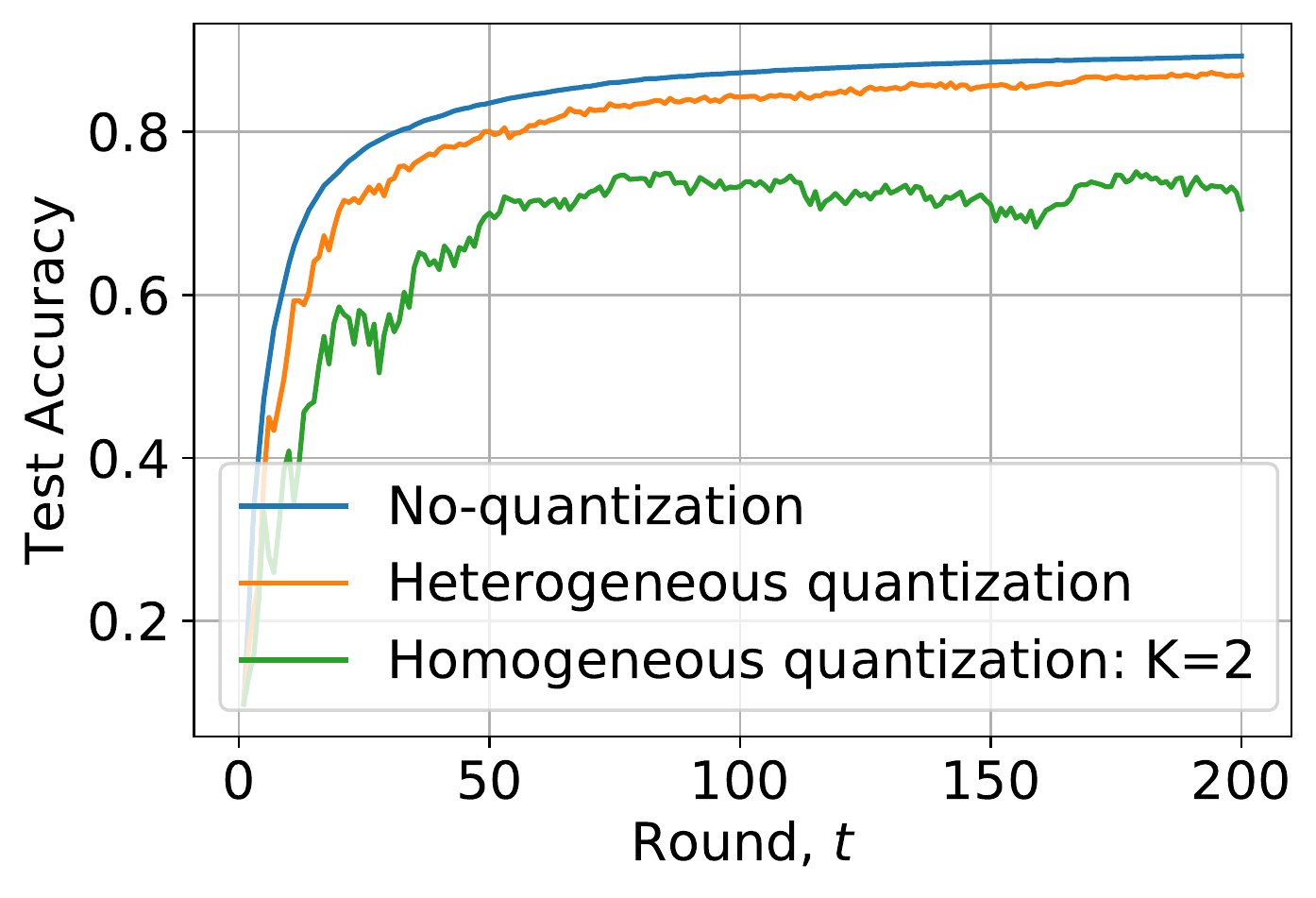}
  \label{2}}%
  \quad   \quad \quad   \quad \quad   
    \subfigure[Total communication time ]{\includegraphics[scale=0.4]{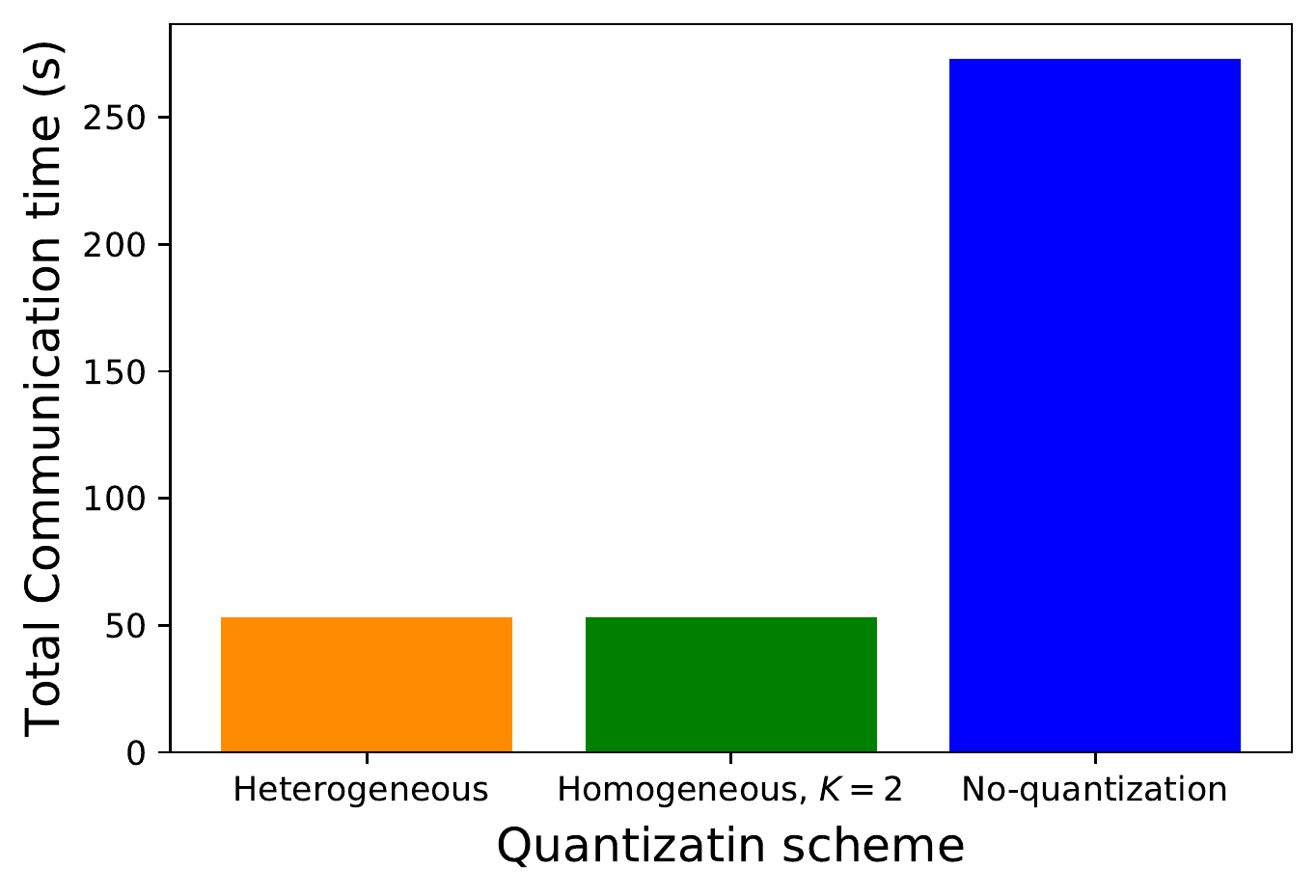}
  \label{1.11}}%
  \quad 
  \caption{  The performance of HeteroSAg   under  different quantization schemes for the  non-IID setting.}
  \label{fig-5}
 \end{figure}

\noindent \textbf{Quantization. }  We consider three scenarios based on the quantization scheme.\textit{Heterogeneous quantization}:  We consider a set $\mathcal{Q}$ of  $G=5$  quantizers with  these levels of quantization $(K_0, K_1, K_2, K_3, K_4)=(2, 6, 8, 10, 12)$, where using these quantizers follows the pattern given in Figure.  \ref{matrix}. \textit{Homogeneous quantization}:  All the  segments from  all users are   quantized by using  $K=2$ levels quantizer. \textit{No quantization}:  All  segments are represented in floating-point numbers  ($32$ bits)\footnote{Using HeteroSAg with no quantization is the same as using FedAvg with no quantization with respect to  the test accuracy. The   difference between the two schemes is that in HeteroSAg models are   encoded unlike FedAvg where clear  models are sent to the server. This  just  results in   a model with larger size  as described in details in   Appendix \ref{prpo}. We consider  HeteroSAg under different quantization schemes  for a fair comparison  regarding  the communication cost and time given in Figure.  \ref{1.11}.}.  We consider  group 0  as a straggler group which  includes    users with    limited  communication resources including low   transmission rates. In particular, we let each user  in group 0 to have   $1$Mb/s transmission rate while users in higher groups to have  more  than  $2$Mb/s, in order  to have a comparison between the three cases. 

We have run    the same experiment with  $N=100$ users.  The  results lead to the same conclusion, and can be found in  the Appendix \ref{I-2-1}.  The details of the  results in Figure.  \ref{1.11} are   given in Appendix \ref{H.1}. Additionally, we have further evaluate the performance of HeteroSAg using CIFAR10 dataset in Appendix \ref{I-2-2}.

\begin{figure}[h]
    \centering
    \subfigure[Illustrating the results for   MNIST]{\includegraphics[width=0.48\textwidth]{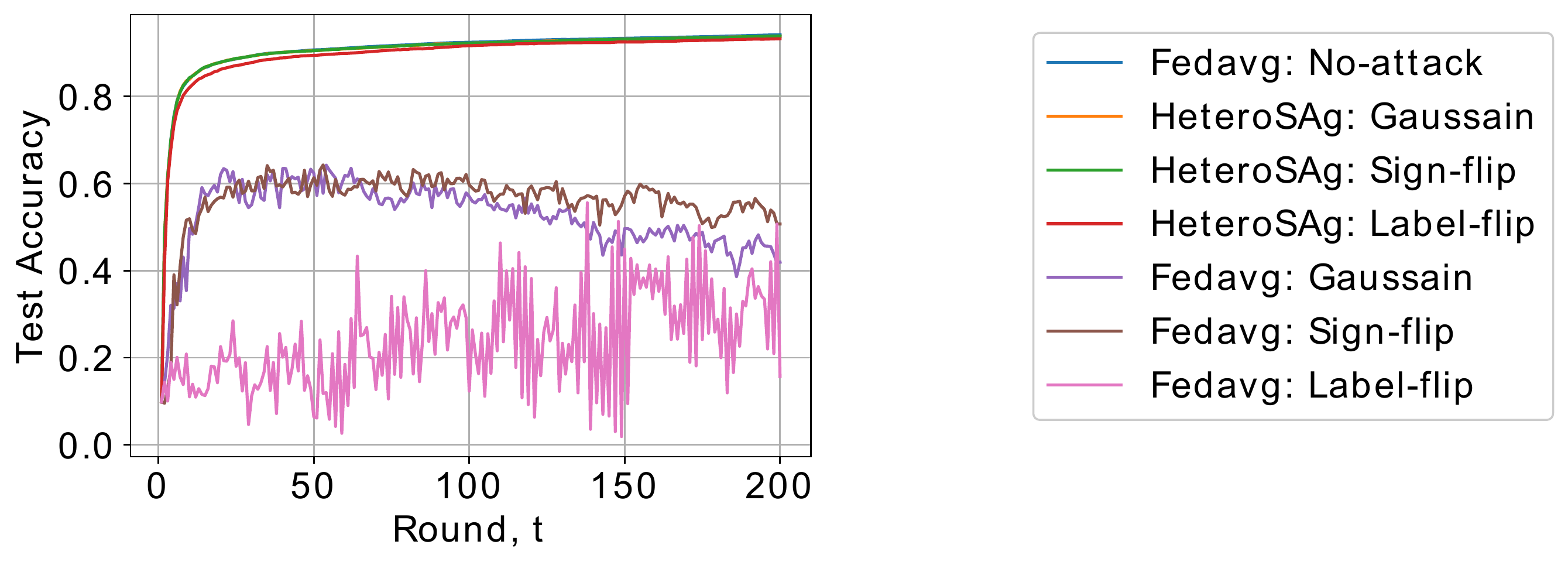}\label{1}} 
    \subfigure[Illustrating the results for CIFAR10]{\includegraphics[width=0.48\textwidth]{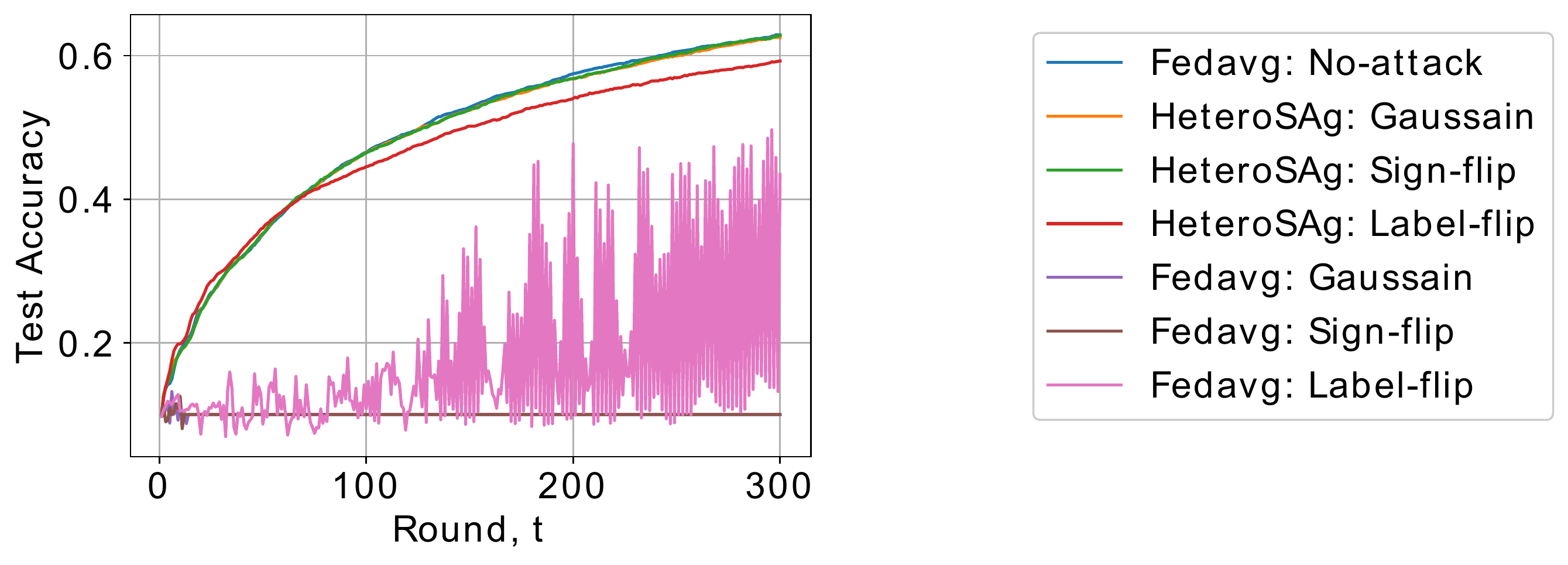}\label{3}}
    \caption{ The performance of HeteroSAg  and FedAvg  under three different attacks for the IID setting.}
    \label{fig:foobar}
\end{figure}

Figure.  \ref{2} illustrates   that HeteroSAg  with  heterogeneous quantization  achieves accuracy close to the  baseline (no-quantization).  Additionally,  after $t=200$ rounds   of communication   with the server, the total  communication time  when using heterogeneous  is   less the case with no quantization by a factor of $5.2 \times$ according Figure.  \ref{1}.  Furthermore,   HeteroSAg with heterogeneous quantization    maintains  superior performance over  the  case of  homogeneous  quantization   with  $K=2$ levels  with more than  $15 \%$ improvement in test accuracy,  while the  communication time is  the same for the both settings. This confirms our motivation that  by adapting the  quantization levels to the transmission rates of the  users,  we can achieve  high  accuracy with    small  training  time.

\subsection{Experiment 2 (Byzantine robustness)} \label{main_Byz}
We show  how the   secure aggregation strategy  of HeteroSAg can be effective   along with  the coordinate-wise median  against   Byzantine users.  For running the experiment, 
  we   consider a setup of $N=300$ users, in which $B=18$ of them are Byzantines.  We consider   the IID setting,  where we 
randomly split the training data samples to $N=300$ disjoint subsets, and assign each subset to a distinct user. We use a learning rate of $0.06$,  set the number of epochs to be $1$, and use  batch size  40. (model details are presented in Section \ref{model}). The performance of HeteroSAg in the non-IID setting is presented in Appendix  \ref{Add_Byz}.  \\
\textbf{Scheme.} We consider two schemes: HeteroSAg with $G=75$ groups and $n=4$ users in each group along with coordinate-wise median, and  FedAvg implemented with secure aggregation \cite{FedAvg}.\\
\textbf{Attack model.} We assume that the  Byzantine users are distributed over $18$ groups. We note that since the focus here is the presence of Byzantine users  while doing  secure aggregation,  where users are sending masked model to the server, Byzantine users can sends any faulty model with extreme values without being individually decoded and hence  filtered out.   \textit{Gaussian Attack}: Each Byzantine user replaces its model parameters with entries drawn from a Gaussian distribution with mean 0 and standard distribution $\sigma = 5$.  \textit{Sign-flip}: Each Byzantine user multiplies its model updates by $-5$. \textit{Label-flip}: Each Byzantine user subtract 9 from its labeled data, and then multiplies its resulting model update by $30$.   

As we can see in  Figure. \ref{fig:foobar}, HeteroSAg with coordinate-wise median is robust to the three attacks  and gives performance almost the same as  the case with no Byzantine users. On the other hand, in the presence of these  attacks,  FedAvg scheme gives very low performance.  As a final remark,  HeteroSAg   achieves both  privacy for the  users' local models  and   Byzantine robustness, simultaneously. This is different from  the SecAg  protocol  that  only achieves  model privacy, and different from   the naive coordinate-wise median \cite{Byz}  that solely  achieves  Byzantine robustness.

%\[\footnotesize
%myMatrix{\mathbf{B} _{\text{HeteroSAg }}} = \begin{pmatrix}
%K_0  & K_0 &K_2&K_3&K_4\\
%K_0 &K_1&K_1&K_3&K_4\\
%K_0 &K_1 &K_2&K_2&K_4\\
%K_0 &K_1  &K_2&K_3&K_3\\
%K_0&K_1  &K_2 &K_3&K_0\\
%\end{pmatrix} \; \myMatrix{\mathbf{B} _{\text{ HeteroSAg }}} = \begin{pmatrix}
%K_0  & K_0 &K_2&K_3&K_2\\
%K_0 &K_1&K_0&K_3&K_3\\
%K_0 &K_1 &K_1&K_0&K_4\\
%K_0 &K_1  &K_2&K_1&K_0\\
%K_0&K_1  &K_2 &K_2&K_1\\
%\end{pmatrix},
%\] 
%where $\mathbf{B} (l,g) =K_k$ means that the $l$-th segment of group $g$ will  be quantized by using  $K_k$ level quantizer.

\section{Conclusion}
We propose  HeteroSAg, a scheme    that allows secure aggregation with heterogeneous quantization.
This enables the users to adjust their quantization proportional to their communication resources, which can provide a substantial     better trade-off between  the  accuracy  of training and the communication time. 
We show that the proposed strategy used in  HeteroSAg can be utilized to  mitigate Byzantine users.     Finally, we demonstrate that   HeteroSAg    can  significantly  reduce  the bandwidth expansion of the sate-of-the-art secure aggregation protocol.

\bibliographystyle{IEEEtran}
\bibliography{ref}
\begin{appendices}

 \section{Illustrative example for SecAg} \label{ill}
In this simple example, we  illustrate SecAg protocol. We consider a  secure aggregation problem in FL,  where there are  $N=3$ nodes with drop-out resiliency  $D = 1 $.  Node $i \in \{1,2,3\}$ holds a  local model update vector   $\mathbf{x}_i \in \mathbb{R}^m$. In the following, we present the steps for executing the  SecAg protocol. 

\textbf{Step 1: Sharing keys and masks.} User 1 and User 2 agree on pairwise random seed $s_{1,2}$. User 1 and User 3 agree on pairwise random seed $s_{1,3}$.  User 2 and User 3 agree on pairwise random seed $s_{2,3}$. Each one of these pairwise seeds is a function of the  public key and the private keys  of  paried users (more details is given in Step 1 in Section II).  In addition, user $i \in \{1,2,3\}$ creates a private random seed $b_i$.

\textbf{Step 2: Sharing keys and masks.} Each user $i \in \{1,2,3\}$ secret shares $b_i$ and the private key $s_i^{SK}$ with the other users via Shamir's secret sharing. The threshold for secret sharing is set to $2$.

\textbf{Step 3: Quantizing.} user $i \in \{1,2,3\}$  quantizes its model $\mathbf{x}_i$ using $K$ levels of quantization according to equation \eqref{quanti}. The output of the quantizer  $\bar {\mathbf{  x}}_i(k)=Q_{K}(\mathbf{  x}_i(k))$   takes a discrete  value from this range $\{r_1, r_1+ \Delta_{K},r_1+2 \Delta_{K}, \dots,  r_2- \Delta_{K},  r_2\}$, where $\Delta_K = \frac{r_2-r_1}{K-1}$ is the quantization interval.  The quantized model of each node will be mapped from $\{r_1, r_1+ \Delta_{K},r_1+2 \Delta_{K}, \dots,  r_2- \Delta_{K},  r_2\}$ to the integer range $[0, K-1]$. 

\textbf{Step 4: Masking.}

To provide privacy for each individual model, user  $i \in \{1,2,3\}$, masks its model $\bar{\mathbf{x}}_i$ as follows:

\begin{equation}\label{ex33}
\mathbf{y}_1 = \bar{\mathbf{x}}_1 + n_1+ z_{1,2} + z_{1,3} \mod R, \quad  \mathbf{y}_2 = \bar{\mathbf{x}}_2 + n_2+ z_{2,3} - z_{1,2}  \mod R, \\ \quad  \mathbf{y}_3 = \bar{\mathbf{x}}_3 + n_3- z_{1,3} - z_{2,3} \mod R, 
\end{equation}
 where $n_i = \text{PRG}(b_i)$ and $z_{i,j} = \text{PRG} (s_{i,j})$ are the random masks generated by a pseudo random number generator. Here, $R = 3 (K-1)+1  $ to ensure that all possible aggregate vectors from the three users will be representable without any overflow.  After that, user $i \in \{1,2,3\}$ sends its masked model $\mathbf{y}_i  $ to the server.
 
 \textbf{Step 5: Decoding.} This phase for the aggregate-model recovery. Suppose that user $1$ drops in the previous phase. The goal of the server is to compute the aggregate of the models $\bar{\mathbf{x}}_2 + \bar{\mathbf{x}}_3$.  The aggregated model at the server from the survived users (user 2 and user 3) is given as follows 
 
\begin{equation}
\label{ex-555}
\mathbf{x}_{2,3} = \bar{\mathbf{x}}_2 +  \bar{\mathbf{x}}_3 + (  n_2+ n_3) - z_{1,2} - z_{1,3} \mod R.
\end{equation}
 Hence, the server needs to reconstruct masks $  n_2$,  $n_3$,  $z_{1,2}$, and  $z_{1,3}$ to recover $\bar{\mathbf{x}}_2 + \bar{\mathbf{x}}_3$. To do that, the server has to collect two shares for each of $b_2$, $b_3$ and $s_i^{SK}$ from the two survived users. Therefore, the server can reconstruct the missing masks and remove them from \eqref{ex-555}. Note that, if node $1$ is delayed while  the server has already reconstructed     $z_{1,2}$, and  $z_{1,3}$, the local   model  of node $1$ is still protected by the private mask $b_i$ in \eqref{ex33}.  
 
\section{Proof of Theorem 1}
First, we  state  four main  properties of the  SS matrix $\mathbf{B}$ for  HeteroSAg. These properties  will be used to prove the   inference robustness of HeteroSAg.
 \subsection{Main properties for the SS matrix of  HeteroSAg }\textbf{Property 1.}\textit{  Each column in the SS matrix contains only one   $*$. This  implies that  each group of  users independently of the other groups executes the secure aggregation on only one segment from its  model update. }

\noindent\textbf{Property 2.}\textit{  Any two  distinct columns $g$ and $g'$ in the SS matrix have at most   one row with the same number, where we do not   consider the symbol  $*$ as a number.  This implies that each two groups  of users  corresponding to these columns  independently of the other groups execute the secure aggregation on only one segment from their   model updates. }

\noindent\textbf{Property 3.}\textit{  For the case where the number of groups $G$ is odd, each row in the SS matrix  contains only one  $*$. On the other hand, when the number of groups is even, only a  pair of $*$'s can be found in the odd rows (the indices of the  rows in the SS matrix  started from  $0$). A pair of $*$'s  for  a given row belongs to one pair of groups from this set $\{(g,g+\frac{G}{2}): g = 0, \dots \frac{G}{2}-1\}$ as shown in Figure. \ref{matrix_even}. }
\begin{figure}[h]
\centering
    \[\footnotesize
\myMatrix{ \mathbf{B}} = \begin{blockarray}{cccccc}
\begin{block}{(cccccc)}
0 & 0  &2&3 &3&2\\
0 & *  &0&3 &*&3\\
0 &1  &1&0 &4&4\\
0 &1  &*&1 &0& *\\
0 &1  &2&2 &1&0\\
* &1 &2& * &2&1\\
\end{block},
\end{blockarray} 
\]
\caption{ Matrix   $ \mathbf{B}$ for  $G=6$.}
\label{matrix_even}
\end{figure}

\noindent\textbf{Property 4.}\textit{ We say that we have a pair when    having  two  equal numbers.  In the SS matrix,  if row $i$ contains only   pairs  of numbers from the   set of columns  $\mathcal{S} \subset  \{ 0, \dots, G-1\}$ where $|\mathcal{S}|=2n$, for $n = 2,  \dots, \frac{G-i}{2} $,  where $i =2$ for even number of groups and $i =1$ for odd number of groups, each row in  the   remaining set of rows $\{0,\dots, G-1\}/i$ corresponding to  the  set of columns $\mathcal{S} $ will have  at least two unpaired   numbers. For example, in  the first row of  the SS matrix given in  Figure.  2,  we have these two pairs $\left(\mathbf{B}(0,0), \mathbf{B}(0,1) \right)$    and $\left(\mathbf{B}(0,2), \mathbf{B}(0,4) \right)$, where $\mathbf{B}(0,0) =  \mathbf{B}(0,1) =0  $  and   $\mathbf{B}(0,2) =  \mathbf{B}(0,4) =2 $.  On the other hand,  for  these column indices  $   \mathcal{S}= \{0, 1, 2, 4\}$, each row in the set of     remaining rows $\{1, \dots, 3\}$   does  not contain numbers to be paired.}

 \subsection{Proof of Theorem 1}
According to the SS matrix $\mathbf{B} $, we have these 4 cases: 

\textbf{Case 1: Single group.}  For a single group  $\mathcal{S} \in  \{ 0, \dots, G-1\}$,  the server can only decode  the  segment denoted by   $*$   in the SS matrix from the model update $\mathbf{x}_{\mathcal{S}}$  by  using Property 1. 
 
\textbf{Case 2: A Pair of  groups.}  From   any   pair  of groups $\mathcal{S} \subset  \{ 0, \dots, G-1\}$,  where $|\mathcal{S}|=2$, the server can  successfully decode at  most one segment from the local model update $\mathbf{x}_{\mathcal{S}}$  when the number of groups is odd. On the other hand, when the total  number of groups is even,   the server can  decode  at most two segments. The former results come from: First,  Property 1 and Property 2 show that for any  pair of groups $\mathcal{S} \subset  \{ 0, \dots, G-1\}$,  where $|\mathcal{S}|=2$,   the server can decode    one segment from the model update  $\mathbf{x}_{\mathcal{S}}$. This segment  is the segment that is  jointly  encoded by the set of groups in  $\mathcal{S}$.  Second, Property 3  states that  when the total  number of groups $G$ is odd, the segments denoted by $*$'s in any   pair of  groups $\mathcal{S}$ do  not get  aligned   in the same row, but interfere with segments from other groups. On the other hand, when the total number of groups $G$ is even, the segments denoted by $*$'s   might be aligned together  on the same row, and hence make the server able to decode another segment from the   model update  $\mathbf{x}_{\mathcal{S}}$ according to Property 3.

\textbf{Case 3:  Even number of    groups.}    From   any  even number of    groups   $\mathcal{S} \subset  \{ 0, \dots, G-1\}$,  where $|\mathcal{S}|=2n$ for $n = 2,  \dots, \frac{G-i}{2} $  with  $i =2$ when $G$ is even  and $i =1$ when $G$ is odd,  the server can only decode one segment    from the model update $\mathbf{x}_{\mathcal{S}}$ according to Property 4.  This segment  results from   the sum of the decoded  segments  from each   pair of groups in $\mathcal{S}$, while the segment denoted by  $*$  from each group in the  set $\mathcal{S}$  interferes with a segment  from another group  according to Property 3.

\textbf{Case 4:  Odd number of    groups.} For  an  odd  number of   groups in the  set   $\mathcal{S} \subsetneq  \{ 0, \dots, G-1\}$,  where $|\mathcal{S}|=2n+1$  for   $n = 1,  \dots, \frac{G-i}{2} $ with  $i =4$ when $G$ is even and $i =2$ when $G$ is odd,   the server can not decode any segment from the  model update $\mathbf{x}_{\mathcal{S}}$.  The reason for that   for any  given row in the SS matrix there is at least two segments to not   be paired.  These four cases complete the proof.

\section{Proof of Lemma 1}

 i) (Unbiasedness) One can easily prove that $E[\bar{\mathbf{x}}_i] = \mathbf{x}_i$.\\
 ii)(Variance bound)
 \begin{align}\label{error111}
\mathbf{E}   || \bar{\mathbf{x}}_i - \mathbf{x}_i||_2^2\overset {a}{=}& \sum_{l=0}^{ G-1} \sum_{k=1}^{ \frac{m}{G}}  \mathbf{E} ( \bar{\mathbf{x}}_i^l(k) - \mathbf{x}^l_i(k))^2 
\overset {b}{=} \sum_{l=0}^{ G-1} \sum_{k=1}^{ \frac{m}{G}}  \left( T(l+1) -\mathbf{x}_i^l(k)\right) \left(\mathbf{x}_i^l(k)-T(l) \right) \nonumber\\
\overset {c}{\leq} & \sum_{l=0}^{ G-1} \sum_{k=1}^{ \frac{m}{G}}  \frac{(T(l+1)-T(l))^2}{4}
=\frac{m}{G}  \sum_{l=0}^{ G-1}   \frac{(\Delta_i^l)^2}{4}, 
\end{align}
where  $(a)$ follows from the fact that  the random quantization is IID  over elements of the vector $\mathbf{x}_i$,  (b)  from the variance of  the quantizer in \eqref{quanti}, and  (c)  from  the  bound  in \cite{an2016distributed}, which  sates that having  $x$ such that   $a\leq x \leq b$, this implies $(b-x)(x-a)\leq \frac{(b-a)^2}{4}$.\\
iii) (Total quantization error)
 \begin{align}\label{error11}
\mathbf{E}  || \bar {\mathbf{p}} - \mathbf{p} ||^2_2 = &\mathbf{E}   \left\|  \frac{1}{N}\sum_{i=1}^{ N} \bar{\mathbf{x}}_i - \frac{1}{N} \sum_{i=1}^{ N} \mathbf{x}_i\right\|^2_2
= \frac{1}{N^2} \mathbf{E}   \left\| \sum_{i=1}^{ N} \bar{\mathbf{x}}_i - \mathbf{x}_i\right\|^2_2
\overset {d}{=}\frac{1}{N^2}\sum_{i=1}^{ N} \mathbf{E}   || \bar{\mathbf{x}}_i - \mathbf{x}_i||_2^2 \nonumber  \\  \overset {e} \leq& \frac{(r_2-r_1)^2}{4N^2} \frac{m}{G} \sum_{i=1}^
{N}  \sum_{l=0}^{G-1}
\frac{1 }{(K_i^l-1)^2 } = \sigma^2.
\end{align}
where  $(d)$ follows from the fact that  the random quantization is IID  over the $N$ local gradients $\{\mathbf{x}_i\}_{i =1}^N$, and 
 (e)  from \eqref{error111}. We note that $\sigma_{\text{HeteroSAg}}^2$ (total quantization error when using HeteroSAg)  in Theorem 2  can be derived from \eqref{error11} by    counting the number of segments that  is quantized by each quantizer. According to the SS matrix  $\mathbf{B}$, each user $i \in \mathcal {S}_g$   in group $g$, for  $0\leq g \leq  G-1$  uses quantizer $Q_{K_g}$ to quantize $G-g$ segments,  and the remaining $g$ segments are  quantized by the set of quantizers $\{Q_{K_0}, Q_{K_1}, \dots, Q_{K_{g-1}}\}$, with one segment for each quantizer.    Hence,   the total number of segments used  quantizer $Q_{K_g}$, where $0\leq g \leq  G-1$, is  given by $(2(G-g)-1)n$.

 \section{Proof of Theorem 2}
 From the $L$-Lipschitz
continuity of $\nabla F(\bm{\theta})$, we have
\begin{equation}\label{lemma}
 F(\bm{\theta}^{(t+1)}) \leq  F(\bm{\theta}^{(t)}) + \langle\, \nabla F(\bm{\theta}^{(t)}), \bm{\theta}^{(t+1)} -\bm{\theta}^{(t)}\rangle\, + \frac{L}{2} ||\bm{\theta}^{(t+1)} -\bm{\theta}^{(t)}||^2  \overset{a}=   F(\bm{\theta}^{(t)}) - \eta  \langle\,  \nabla F(\bm{\theta}^{(t)}), \bar{\mathbf{p}}^{(t)}\rangle\, + \frac{L\eta^2}{2} ||\bar{\mathbf{p}}^{(t)}||^2 ), \nonumber
\end{equation}
where $\bar{\mathbf{p}}^{(t)}=  \frac{1}{N}\sum_{i=1}^{ N} \bar{\mathbf{x}}_i$,  and  $\bar{\mathbf{x}}_i = \bar {\mathbf{g}}_i(\bm{\theta}^{(t)})$ is the quantized  local gradient at node $i$. We  used  this relation  $\bm{\theta}^{(t+1)} = \bm{\theta}^{(t)} - \eta \bar{\mathbf{p}}^{(t)}$ to get (a).  By taking the expectation with respect to the quantization noise and data sampling randomness,
\begin{align} 
\mathbb{E} \left[F(\bm{\theta}^{(t+1)}) \right] \overset {a}\leq & F(\bm{\theta}^{(t)}) - \eta  || \nabla F(\bm{\theta}^{(t)})||^2 + \frac{L\eta^2}{2} \left( || \nabla F(\bm{\theta}^{(t)})||^2  +  \sigma_{\text{HeteroSAg}}^2\right) \nonumber \\
  \overset {b} \leq &  F(\bm{\theta}^{(t)}) - \frac{\eta}{2} || \nabla F(\bm{\theta}^{(t)}||^2 + \frac{\eta}{2} \sigma_{\text{HeteroSAg}}^2 \nonumber \\ \overset {c}  \leq &  F(\bm{\theta}^{*}) + \langle\, \nabla F(\bm{\theta}^{(t)}), \bm{\theta}^{(t)} - \bm{\theta}^{*} \rangle\, - \frac{\eta}{2} || \nabla F(\bm{\theta}^{(t)})||^2 + \frac{\eta}{2} \sigma_{\text{HeteroSAg}}^2 \nonumber \\   = &  F(\bm{\theta}^{*}) + \langle\, \mathbb{E} [\bar{\mathbf{p}}^{(t)}], \bm{\theta}^{(t)} - \bm{\theta}^{*}\rangle\, - \frac{\eta}{2}|| \mathbb{E} [\bar{\mathbf{p}}^{(t)}]||^2 + \frac{\eta}{2}  \sigma_{\text{HeteroSAg}}^2 \nonumber \\  \leq &  F(\bm{\theta}^{*}) + \langle\, \mathbb{E} [\bar{\mathbf{p}}^{(t)}], \bm{\theta}^{(t)} - \bm{\theta}^{*}\rangle\, - \frac{\eta}{2}\mathbb{E}||  \bar{\mathbf{p}}^{(t)}||^2 + \eta \sigma_{\text{HeteroSAg}}^2 \nonumber \\ 
= &  F(\bm{\theta}^{*}) +  \mathbb{E} \left[ \langle\,  \bar{\mathbf{p}}^{(t)}, \bm{\theta}^{(t)} - \bm{\theta}^{*}\rangle\, - \frac{\eta}{2}||\bar{\mathbf{p}}^{(t)}||^2 \right]+ \eta  \sigma_{\text{HeteroSAg}}^2 \nonumber \\ = &  F(\bm{\theta}^{*}) + \frac{1}{2\eta}    \left(  \mathbb{E} ||\bm{\theta}^{(t)} - \bm{\theta}^{*}||^2 - \mathbb{E} ||\bm{\theta}^{(t+1)} - \bm{\theta}^{*}||^2 \right)+ \eta  \sigma_{\text{HeteroSAg}}^2
\end{align}
where (a) follows from that $\mathbb{E}||\bar{\mathbf{p}}^{(t)}||^2 =  \mathbb{E}  ||\bar{\mathbf{p}}^{(t)} - \mathbb{E} [ \bar{\mathbf{p}}^{(t)} ]||^2 + || \mathbb{E} [ \bar{\mathbf{p}}^{(t)} ]||^2$, where  $\mathbb{E} [ \bar {\mathbf{p}}^{(t)} ] = \mathbb{E} [ {\mathbf{x}_i}^{(t)} ]=  \nabla F(\bm{\theta}^{(t)}) $ according to Lemma 1-(i) and  Assumption 1, and $\mathbb{E}  ||\bar{\mathbf{p}}^{(t)} - \mathbb{E} [\bar {\mathbf{p}}^{(t)}]||^2 \leq \sigma_{\text{HeteroSAg}}^2$ according to Lemma 1-(iii), where  $\mathbb{E} [ \bar {\mathbf{p}}^{(t)}] =\mathbf{p}^{(t)} $ with respect to the quantization error.  Furthermore, (b) follows from using $\eta\leq \frac{1}{L}$,  and (c) from the convexity of $F(.)$.  By  summing the above equations for  $t = 0 \dots, J-1$ 
\begin{equation}\label{we}
 \sum_{t=0}^{J-1}  \left(\mathbb{E}\left[ F(\bm{\theta}^{(t+1)})\right] -  F(\bm{\theta}^{*})\right)   \leq  \frac{1}{2\eta}  ( \mathbb{E}  || \bm{\theta}^0-\bm{\theta}^{*}||^2 -  \mathbb{E} || \bm{\theta}^J-\bm{\theta}^{*}||^2) + \eta J \sigma_{\text{HeteroSAg}}^2 \leq  \frac{||\bm{\theta}^0-\bm{\theta}^{*}||^2}{2\eta } + \eta  J \sigma_{\text{HeteroSAg}}^2.
%\epsilon \leq \frac{(r_2-r_1)^2}{4N^2} \frac{m}{G} n  \sum_{g=0}^{G-1} \frac{2(G-g )-1}{(K_g-1)^2 }.
\end{equation}
 By using the convexity of $F(.)$, 
\begin{equation}
 \mathbb{E}\left[ F\left(\frac{1}{J} \sum_{t=1}^J \bm{\theta}^{(t)}\right)\right] -  F(\bm{\theta}^{*}) \leq  \frac{1}{J} \sum_{t=0}^{J-1}  \left(\mathbb{E}\left[ F(\bm{\theta}^{(t+1)})\right] -  F(\bm{\theta}^{*})\right)   \leq  \frac{||\bm{\theta}^0- \bm{\theta}^{*}||^2}{2\eta J } + \eta \sigma_{\text{HeteroSAg}}^2.
\end{equation}

\section{Proof of Theorem 3}
We recall that  the  model update $\mathbf{x}_i$ of user $i$, is partitioned    into $Z$ segments  when using  HeteroSAg, where $Z$ is the total number of subgroups. The partitioned model is denoted by $\mathbf{ x}_i=[\mathbf{ x}_i^0, \mathbf{ x}_i^1, \dots, \mathbf{ x}_i^{Z-1}]^T$, where $\mathbf{ x}_i^l \in \mathbb{R}^{\frac{m}{Z}}$. To guarantee   information theoretic  privacy for  the model update $\mathbf{x}_i$,   we should have  $I(\mathbf{x}^l_i; \mathbf{y}^l_i) = 0$, for $l= 0, \dots, Z-1$, where  $\mathbf{y}^l_i$ is the $l$-th encoded segment from user  $i$.  To achieve this  information theoretic  privacy for each segment in the model update, the server should not  be able to decode any  individual segment  $\mathbf{x}^l_i$
when recovering all  pairwise keys of  dropped users and the private  keys of the survived users.  Each segment  from user  $i$ in a subgroup $g$ is jointly encoded with either the remaining  $\bar n-1$ users from this subgroup,  or  $2\bar n-1$, users from its subgroup and from an additional  subgroup,  where $\bar n$  is the  total number  the number of users in each subgroup. Therefore,   the number  of survived users  in each   subgroup  can not be  one.  We assume that each user has a dropout probability  $p \in [0, 1]$. 
By using the fact that the number of survived users in each subgroup $X$ follows a binomial distribution with
parameters $ \bar n$, which is the number of users in each subgroup,    and $1-p$, Theorem 3 can be proven. 

\section{Proof of Proposition 3}\label{prpo}
 \noindent \textbf{Computation cost:} $\mathcal{O}(N^2+ \bar n m)$, where $m$, $N$ and $\bar n$ are the model size, total number of users and number of users in each group, respectively. Each user  computation can be broken up as (1) Performing the $2N$ key agreements, which
takes $\mathcal{O}(N)$ time, (2) Creating t-out-of-N Shamir secret shares of the private key of $s_i^{SK}$ and  $b_i$, which is order $\mathcal{O}(N^2)$ (3) Generating  the model masks  according to   \eqref{encod}  for all neighbors  which takes $\mathcal{O}( \bar n m)$ time in total. The former result   comes from the fact that each  element in the model update of any user  in HeteroSAg is masked by   either $\bar n-1$  masks or    $2 \bar n-1$ masks, unlike   SecAg   where the whole vector is masked by $N-1$ 0-sum pairwise masks. Therefore, for the case where the number of users in each group $\bar n = \log N$, the computation cost becomes $\mathcal{O}(N^2+ m \log N )$,  as opposite to  $\mathcal{O}(N^2+ m  N )$. 

\noindent \textbf{User communication  complexity:} $\mathcal{O}(N+m)$ The communication complexity is the same as the secure aggregation protocol; however,  the actual number of transmitted bits per user in  HeteroSAg  is lower.  In particular, HeteroSAg gives lower  per user  communication cost compared to SecAg.  Specifically,  having a set of $|\mathcal {S}|$ users executes the secure aggregation  protocol together    on the set of segments $\{\mathbf{ x}_i^{l}\}_{i \in \mathcal {S}}$, the actual number of transmitted bits from each user $i \in \mathcal {S}$ is  given by $R= | \mathbf{ x}_i^{l} |\log (|\mathcal {S}| (K_g-1)+1)$ according to \eqref{encod},  where  $| \mathbf{ x}_i^{l} |$ gives us the number of elements in this segment, and $K_g$ is the number of quantizer levels. On the other hand,     just sending the quantized  segments in  clear without any encoding  results in  $ | \mathbf{ x}_i^{l} | \log  (K_g)$ bits. This gives us an expression for what is called the   bandwidth  expansion  factor with respect to segments $ \frac{ \ceil[\big]{\log (|\mathcal {S}| (K_g-1)+1)}}{ \ceil[\big]{\log  (K_g)}}$, while  ignoring   the cost of sharing keys and masks and other cryptographic aspects of the protocol\footnote{The costs of sharing keys and masks in   HeteroSAg  are the same as SecAg, so we do not consider them in the evaluation.}.  In fact, the majority of  the bandwidth expansion
for the additive masking  in  SecAg comes from the number of users that execute the protocol together. In our proposed segment grouping strategy,   all segments  are   executed by either $\bar n$ or $2 \bar n$ users.  On the other hand,   SecAg    besides the fact that   the  local model of all users are quantized   by using the   same quantizer, even if they have different communication resources,  all the $N$ users execute the secure aggregation protocol together. This implies much larger bandwidth expansion factor than our  HeteroSAg.   
%In most cases the size of the model parameter $m$ is larger than the number of users $N$. 

In order to further illustrate   how   HeteroSAg   reduces  the bandwidth expansion of the SecAg protocol, we give the following  numerical explanations.  We assume having $N=2^{10}$ users, a number    used to evaluate this metric in  \cite{cc, bonawitz2019federated},   and assume    without loss of generality that  only one  quantizer to be used by the users.  When the  partitioning step results in $\bar n=8$ users in each subgroup, and when using  $K=2^{16}$  quantization levels, the  bandwidth  expansion factor   becomes $ 1.25 \times$   instead of being $ 1.625 \times$ for SecAg. For a single bit quantization,     the expansion factor is significantly  reduced from  $ 11 \times$  to  $4\times$. 
%The communication, computation, and storage cost of the server  are the same as the secure aggregation in \cite{cc}.

\section{ HeteroSAg for  Heterogeneous group Size }\label{HeteroSAg  s2}
We have considered  the  case  of uniform group sizes  for  HeteroSAg,   where clustering users results in the same number of users in each group, in Section \ref{sec-4} in the main submission. In this section,  we consider  a more general scenario,    where  instead of assuming  that the set of $N$ users are divided equally on the $G$  groups, where each group has $n$ users, we assume the case where the number of  users in each group is  different. 

\subsection{Execution of HeteroSAg     for heterogeneous group size}
Similar  to  HeteroSAg in Section \ref{sec-4.1},  key agreement and secret sharing are executed   according to Step 1 and    Step 2 in Section \ref{sec-1.2}. Here,  the  size of the set of users  in group  $g$ is denoted    by $|\mathcal {S}_g|=n_g$, for $g\in [G]$, where $\sum_{g=0}^{G-1} n_g=N$, and $G$ is the number of possible quantizers given in the set  $\mathcal{Q}$. 
 The second extension  for HeteroSAg  is that we allow  further  partitioning of the groups into smaller subgroups  when  the number of users in each  group is large.  Extra partitioning  results in   the benefits given in Remark 4, and in   decreasing  the expansion factor discussed in Section \ref{prpo}, which measures the ratio between the size of the masked model in bits to the size of the clear model.   Extra partitioning   is  achieved by dividing each  set of users $\mathcal {S}_g$, for $g\in [G]$,    into $L_g$ subsets (subgroups), $\mathcal {S}^d_g$, for $d=0, \dots, L_g-1$, such that each subgroup has the same number of users $\bar n$.  
 %The number of users that jointly executes the secure aggregation protocol together on the $l$-th  segment from user   $i \in \mathcal{S}_g$     is  $ |\mathcal {S}^l| \in \{\bar n, 2 \bar n\}$,  and hence by using \eqref{encod},  the total number of transmitted bits from that user  is given by  $\sum_{l=0}^{G-1}  | \mathbf{ x}_i^{l} |\log (|\mathcal {S}^{l}| (K_{g,l}-1)+1)$, where $K_{g,l}$ is the level of quantizers used to quantize the $l$-th segment from user  $i \in \mathcal{S}_g$.  Therefore, the   difference in the transmitted bits from different users      depends  only on  the level of quantizations and independent of the size of each group.
 Following the clustering step and the extra partition of the groups,     each    model  update vector   $\{\mathbf{ x}_i\}_{i \in [N]}$    is equally partitioned into $Z$ segments $\mathbf{ x}_i=[\mathbf{ x}_i^0, \mathbf{ x}_i^1, \dots, \mathbf{ x}_i^{Z-1}]^T$, where $\mathbf{ x}_i^l \in \mathbb{R}^{\frac{m}{Z}}$ and $Z=\sum_{g=0}^{G-1}L_g$,  for $l\in [Z]$.  Also, we should  have $Z\leq m$, and for sufficiently large     $N> m$, we might   restrict the number of subgroups to equal the size of the model parameter  $Z=m$, which means that     each segment of the local model update is  just one element.

     The segment grouping strategy $\mathcal{A} _{\text{HeteroSAg}}$ is given by the  SS matrix  $\mathbf{B}$ with   dimensions  $Z \times Z$ according to Algorithm  \ref{5}. As shown  in the example SS matrix  given  in  Figure.  \ref{matrix_even_}, 
     
 \begin{figure}[h]
\centering
    \[\myMatrix{\mathbf{B} ^e}= \footnotesize
\begin{blockarray}{cccccc}
(0,0) & (1,0) & (1,1)& (2,0) & (2,1) \\
\begin{block}{(ccccc)c}
(0, 0)  & (0,0)  &(1, 1)& * &(1, 1)&0\\
(0, 0) & *  &(0, 0)&(2, 0) &(2, 0)&1\\
(0, 0) &(1, 0) &(1, 0)&(0, 0) & *&2\\
(0, 0) &(1, 0)  &*&(1, 0)&(0, 0)&3\\
*&(1, 0)  &(1, 1) &(1, 1)&(1,0)&4\\
\end{block}
\end{blockarray}
 \]
\caption{ Matrix $\mathbf{B} ^e$ for  $G=3$ groups  with number of subgroups in each group  $L_0=1$, $L_1=2$,  and $L_2=2$, respectively.    }
\label{matrix_even_}
\end{figure} 
 each column is indexed by  two indices $(g,d)$  representing the   set of users $\mathcal {S}_g^d$, for $g\in [G]$ and   $d=0, \dots, L_g-1$,  while each  row $l$, for  $l=0, \dots, Z-1$,  represents the index of the   segment. Similar to the description   in Section \ref{sec-4.1},    having an entry $\mathbf{B} ^e(l, (g,d))=*$ means that the   set of users $\mathcal {S}=\mathcal {S}_g^d$ will quantize the  set of segments $\{\mathbf{ x}_i^{l}\}_{i \in \mathcal {S}_g^d}$   by the quantizer  $Q_{K_g}$ and   encode  them  together, while at the server side these set of segment will be  decoded together.     When $\mathbf{B} ^e(l,(g,d))=\mathbf{B} ^e(l,(g',d')) =(g,d) $, where $g\leq g'$, this means that  the  set of users  $S=\mathcal {S}_g^d \cup {S}_{g'}^{d'}$, corresponding to these  columns $(g,d)$ and $(g',d')$,   will  quantize  the  set of segments $\{\mathbf{ x}_i^{l}\}_{i \in \mathcal {S}_g^d \cup {S}_{g'}^{d'} }$ by using the quantizer  $Q_{K_g}$ and encode the output of the quantizer   together, while at the server side these set of segments will be decoded together. 

 \begin{algorithm}[H]\footnotesize
\SetAlgoLined
Define: $Z_{g-1}=\sum _{l=0}^{g-1} L_l$ and  $Z_{g-1} =0 $  when $g=0$, and $\mathds{1}$ is the indicator function\;
 \For{$g=0, \dots, G-1$}{
 \For{$d=0, \dots, L_g-1-\mathds{1}_{g= G-1}$ }{
 $i=0$ and $s=0$ \;
 \For{$r=0, \dots,  Z-Z_{g-1} -d-2$}{
 $m=2(Z_{g-1}+d)+r$  \;
  \eIf{$(d+r+1) \mod \sum_{l=0}^i L_{g+l}=0$ }{
  $i=i+1$ and  $s=0$  \; 
  $\mathbf{B} ^e((  m \mod Z,(g,d))=\mathbf{B} ^e(m  \mod Z,  ( \; (g+i), s ) )= (g,d)$}
  {$s=s+1$ \; 
 $\mathbf{B} ^e((  m \mod Z,(g,d))=\mathbf{B} ^e(m \mod Z, ( \; (g+i), s ))= (g,d)$}
   }
  
   }
 }
The remaining entries of Matrix   $\mathbf{B}^e$ will hold $*$
 \caption{Segment Selection matrix $\mathbf{B} ^e$ for  HeteroSAg}
 \label{5}
\end{algorithm}

 Now, we give the   theoretical guarantees of  HeteroSAg  for heterogeneous group size.
 
\noindent \textbf{Theorem 4.} (Inference robustness) \textit{
     For a FL  system with   $N$ users clustered into $Z$ subgroups, and the model update of each node is divided equally  into $Z$ segments,  HeteroSAg      achieves  an  inference robustness  $\delta (\mathcal{A}_{\text{HeteroSAg}}) = \frac{Z-2}{Z}  $,  when the number of subgroups is even,  and $\delta (\mathcal{A}_{\text{HeteroSAg}}) = \frac{Z-1}{Z}$, when   the number of subgroups   is odd, where $Z=\sum_{g=0}^{G-1}L_g$,  for $l=0, \dots, Z-1$, and $L_g$ is the number of subgroups in group $g$.}

\noindent \textbf{Lemma 2} (Quantization error bound) \textit{
Let  $L_g$ to be  the number of subgroups in group $g$, and  $Z_{g-1}=\sum _{l=0}^{g-1} L_l$  represent the sum of subgroups  of  group  $0$  to group $g-1$, such that $Z_{g-1} =0 $ when $g=0$. Additionally, let   $Z=\sum_{g=0}^{G-1}L_g$ to be the total number of subgroups. 
For a set of vector     $ \{\mathbf{ x}_i \in \mathbb{R}^m\}_{i=1}^N$, such that  the elements of each    vector   $\mathbf{ x}_i$, for $i=1, \dots, N$,  take value from   this interval  $[r_1,r_2]$, and each  vector  is partitioned into $Z$  equal segments, the quantization error bound $\sigma_{\text{HeteroSAg}+}$  when using the quantizers in $\mathcal{Q}$ along with the SS  matrix $\mathbf{B} ^e$ is given by 
$\sigma_{\text{HeteroSAg}+} =  \frac{(r_2-r_1)^2}{4N^2} \frac{m}{Z} \bar n  \sum_{g=0}^{G-1} \frac{\sum_{j=0}^{L_g-1}( 2(Z-Z_{g-1}-j)-1)}{(K_g-1)^2 }$.}\\
The proofs of  Theorem 4 and Lemma 2 can be derived similarly to the proofs of    Theorem 1 and Lemma 1, respectively.   The convergence rate is the same as  in Theorem 2  with replacing  $\sigma_{\text{HeteroSAg}}$ with $\sigma_{\text{HeteroSAg}+}$.

\section{Proof of Proposition 1}
From Lemma 2, when having $G$ groups and  each group is partitioned equally   into $L$  subgroups each of which has a size of     $\bar n = \frac{N}{LG}$ users and using these two results        $Z_{g-1}= gL $ and  $Z=GL$,   the quantization error bound will be the same as $\sigma_{\text{HeteroSAg}}$ given in Theorem 2.  This means that    extra partitioning of each group   does not change the quantization error.

\section{Byzantine robustness of HeteroSAg}\label{Byz_extra}
We further discuss the intuition behind the success of HeteroSAg in mitigating the Byzantine nodes in the following remark 

\begin{remark}[Byzantine robustness of HeteroSAg]
In this remark, we further motivate the reason behind the success of HeteroSAg in mitigating Byzantine nodes.  The reason for the success of  coordinate-wise median (Median) algorithm  \cite{By} in mitigating the Byzantine nodes in the  the IID setting is the same reason behind the success of HeteroSAg when it is integrated with coordinate-wise median. In particular,   the  success of Median      is guaranteed since   the model updates from all benign  users are similar to each other \cite{By}, where the similarity increases as the data at the users become more IID. Therefore,   taking  the  median over each coordinate across the model update of all users  ensures  that we get a representative model for all  the benign models while ignoring the outliers from each coordinate. For the same reason,  integrating median with   HeteroSAg  can provide  Byzantine robustness. In particular, unlike the case where   each  coordinate   represents    one element from the local model of each user (e.g., the k-th element $\bar{\mathbf{x}}_i(k)$ of the local model of node i) in the naive coordinate-wise median algorithm. In HeteroSAg, each coordinate becomes representing  the average of a set of elements from the local models of some users (e.g., the k-th element $\mathbf{y}^0_{0,1}(k)$  of the segment $\mathbf{y}^0_{0,1}$ given in Example 3, where  $ \mathbf{y}^0_{0,1}(k) = \frac{1}{2n} \sum_{i \in \mathcal{N}_0 \cup \mathcal{N}_1 } \bar{\mathbf{x}}^0_i(k)$). Similarly,   the average of a set of elements  from the local  models of  some benign nodes is   a reasonably good representative of those elements.   Therefore,  applying the median along the new coordinates  will   guarantee that we get a representative model of the benign models while ignoring the outliers from each coordinate. The outliers of each coordinate appear when having at least one faulty model contribute to the  average element in that coordinate. For instance,    $\mathbf{y}^0_{0,1}(k)$ will be faulty if at least one of these elements $\{ \bar{\mathbf{x}}^0_i(k),  i \in \mathcal{N}_0 \cup \mathcal{N}_1 \}$ is faulty.

 \end{remark}

\section{Complete Experimental Results for Section \ref{Experiments} }
\subsection{Evaluating  the results in Figure \ref{1.11}} \label{H.1}
\begin{table}[h]
\caption{  User communication cost and the total communication time in Experiment 1. }
\begin{center}
\begin{adjustbox}{width=.7\textwidth}
\begin{tabular}{|c|c|c|l|}
\hline
Quantization                                                                                                               & Group & \begin{tabular}[c]{@{}c@{}}User communication\\ cost (Mb)\end{tabular} & \multicolumn{1}{c|}{\begin{tabular}[c]{@{}c@{}}Communication \\ time (s)\end{tabular}} \\ \hline
\multirow{5}{*}{\begin{tabular}[c]{@{}c@{}}Heterogeneous \\ $(K_0, K_1, K_2, K_3, K_4)$\\ $=(2, 6, 8, 10, 12)$\end{tabular}} & 0     & 53                                                                    &                                                                                        \\
                                                                                                                           & 1     & 87                                                                    &                                                                                        \\
                                                                                                                           & 2     & 90                                                                    & \multicolumn{1}{c|}{53}                                                               \\
                                                                                                                           & 3     & 97                                                                    &                                                                                        \\
                                                                                                                           & 4     & 101                                                                    &                                                                                        \\ \hline
\begin{tabular}[c]{@{}c@{}}Homogeneous\\ $K=2$\end{tabular}                                                                & -     & 53                                                                    & \multicolumn{1}{c|}{53}                                                               \\ \hline
No-quantization                                                                                                            & -     & 279                                                                   & \multicolumn{1}{c|}{279}                                                              \\ \hline
\end{tabular}
\label{table_time}
\end{adjustbox}
\end{center}
\end{table}

The total communication time of the three heterogeneous scheme given in Figure. \ref{1.11} can be derived from the results in Table \ref{table_time}. The transmission rate of the  users  in group 0  is    $1$Mb/s,  while users in higher groups have transmission rate more than $2$Mb/s,   as given in Section \ref{Experiments}. The communication cost in (Mb)   per user $i$ in group $g$ is given by summing the size of the masked model sent by node $i$ and the size of the global model  received  from the server (in Mb).  The model size for the fully connected neural network considered for Experiment 1  is $ 79510$ elements. In HeteroSAg, the encoding is done on the segment level, and the size of the encoded segment $\mathbf{x}_i^l$ from  node $i$ is given by $|\mathbf{x}_i^l|  \ceil[\big]{\log (|\mathcal {S}| (K^l_g-1)+1)}$, where  $|\mathcal {S}| $ is the number of users who jointly encode this segment,  $K^l_g$ is the number of quantization levels used for quantizing $\mathbf{x}_i^l$. The former result is given according to the encoding step in  \eqref{encod}. By using the previous formula along with the segment grouping given in Figure.  \ref{matrix},  and the quantizers from the three different scenarios that we are considering (heterogeneous quantization with levels $(K_0, K_1, K_2, K_3, K_4)=(2, 6, 8, 10, 12)$, homogeneous quantization $K=2$, and  no quantization, i.e.,  $K =2^{32}$), the per-user communication cost after $t=200$ rounds can be evaluated.  The cost of sharing keys and masks is the same for the three scenarios, therefore,   we do not consider that in  the calculations of the communication cost. The communication time can simply be  computed by dividing the per-user  communication cost by the corresponding  transmission rate.
\subsection{ Additional experiment  (Heterogeneous quantization)} \label{H.2}

\subsubsection{MNIST dataset}\label{I-2-1}
\begin{figure}[h]
  \centering
   \subfigure[Test accuracy for    MNIST dataset]{\includegraphics[scale=0.4]{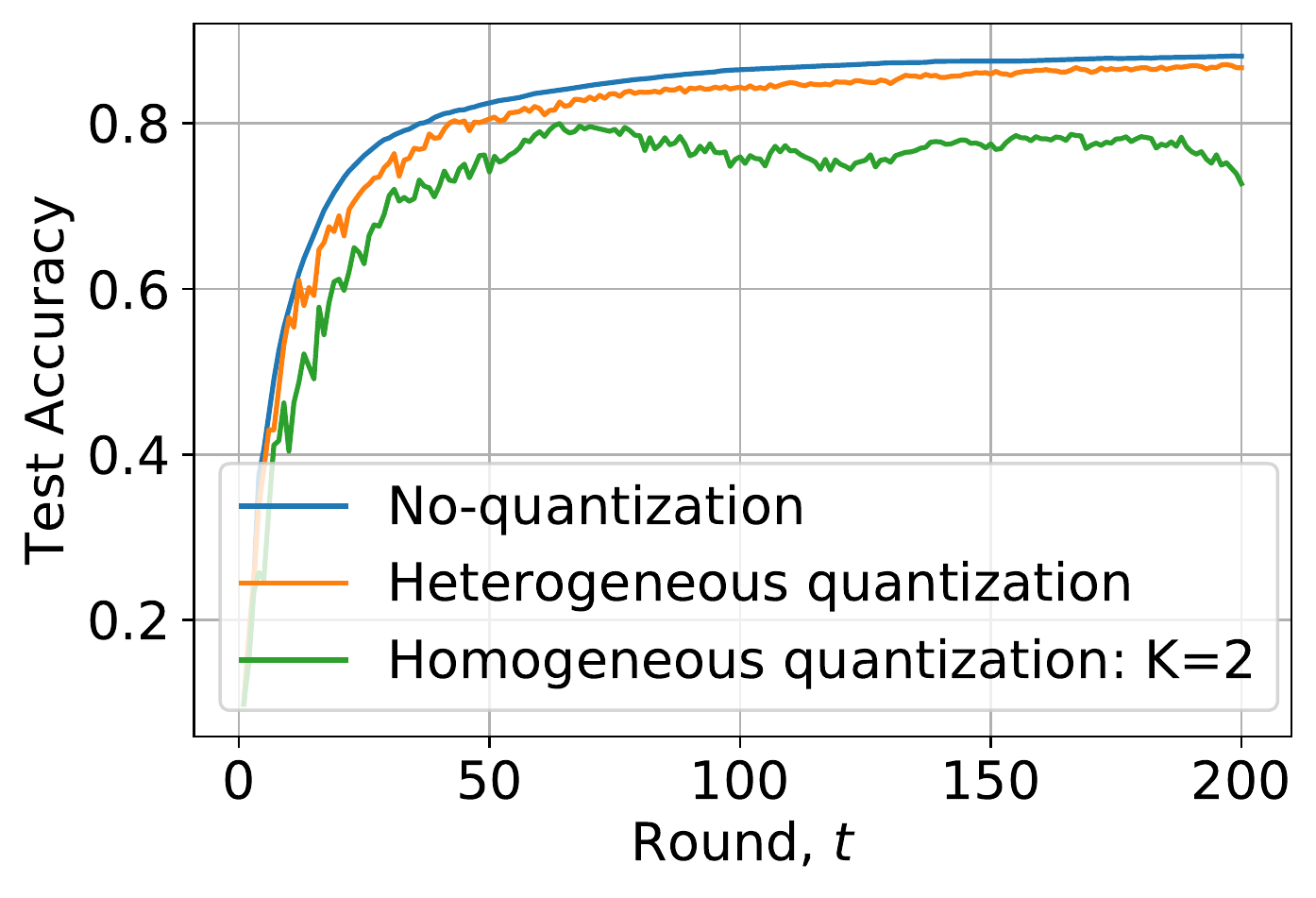}
  \label{21}}%
  \quad   \quad \quad   \quad \quad   
    \subfigure[Total communication time ]{\includegraphics[scale=0.4]{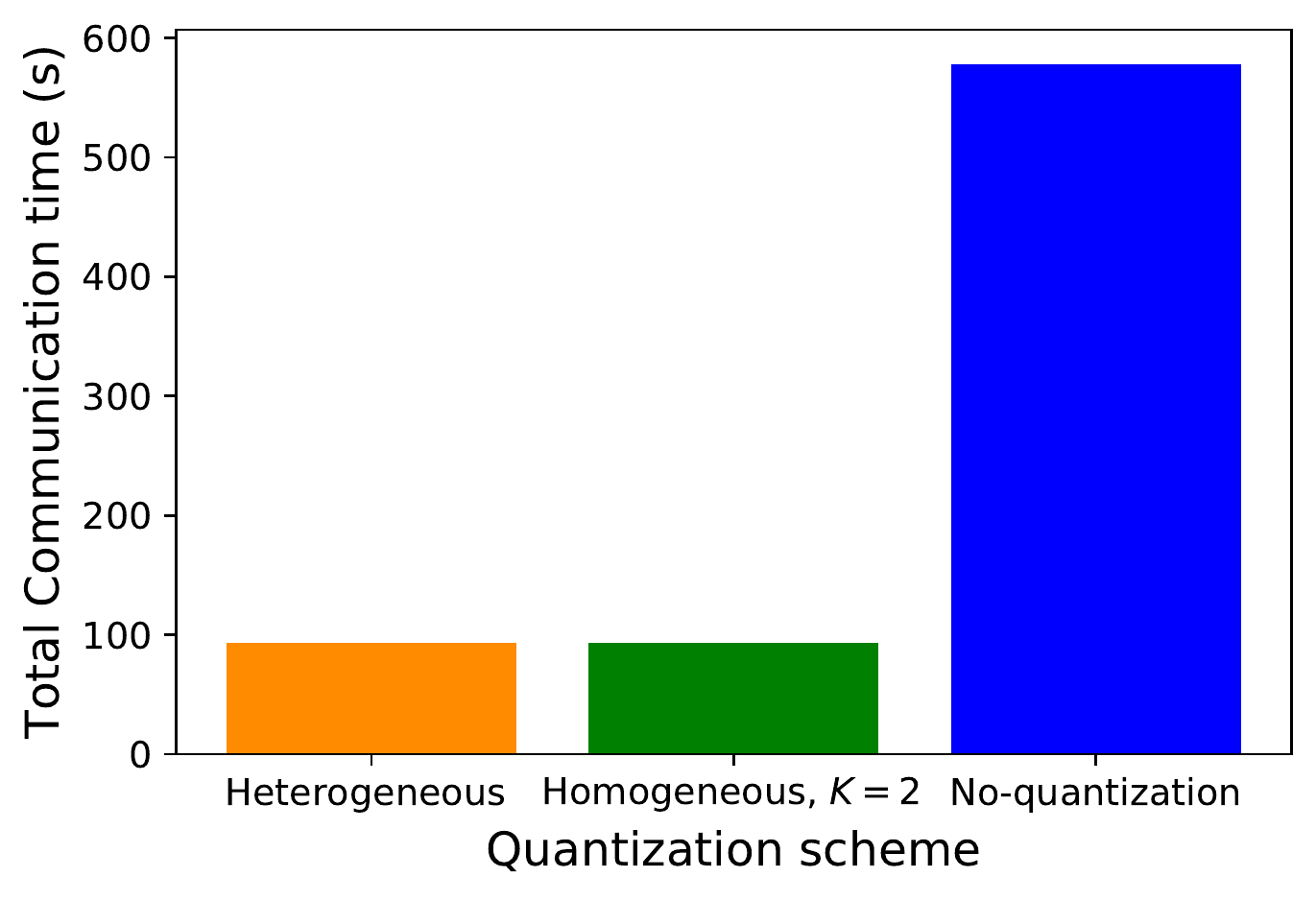}
  \label{1.113}}%
  \quad 
  \caption{The performance of HeteroSAg   under  different quantization schemes for the  non-IID setting ($N = 100$).}
  \label{fig9}
 \end{figure}

We consider  the same setup given for  Experiment 1 in Section \ref{Experiment 1}, while setting $N = 100$ users with $n = 20 $  in each group of the $G=5 $ groups for running the experiment in  Figure \ref{fig9}.  In this experiment, we use a batch size of $60$.

 \subsubsection{CIFAR10 dataset}\label{I-2-2}
 
 \textbf{Data distribution  and Hyperparameters}
  We set the total number of users to $N=100$.  We use a fixed learning rate  of $ 0.02$ for the first $150$ rounds, $ t \leq 150$, and then  gradually decrease the  learning rate according to $\frac{0.02}{(1 + 0.02 t)}$, for $t>150$.  We set  the batch size for each user to be $20\%$ of its local data. We consider epoch training, where the number of epochs is $5$.  We use CIFAR10 dataset with non-IID data distribution. In particular, we use  the  generic non-IID synthesis method based
on the Dirichlet distribution with parameter $\alpha$ proposed in \cite{hsu2019measuring}. In this method, increasing $\alpha$ makes the data more IID at the users. On the other hand, decreasing $\alpha$ makes each user have very few samples from some random classes. We implement this method using FedML library \cite{he2020fedml}.

  \begin{figure}[h]
  \centering
   \subfigure [Label distribution with $\alpha = 0.7$] {\includegraphics[scale=0.33]{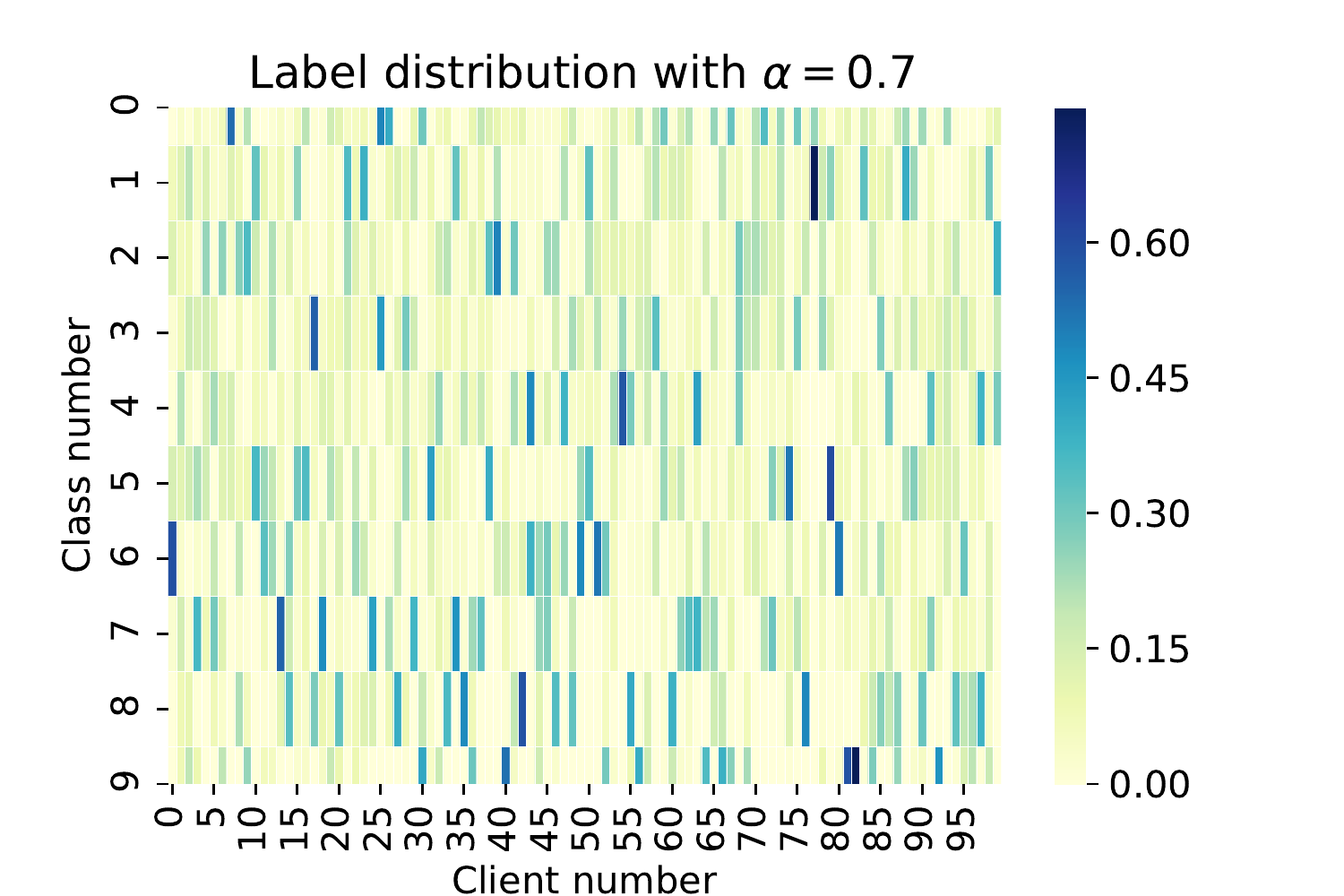}
  \label{cc-1}}%\quad
 \subfigure[Label distribution with $\alpha = 1$]{\includegraphics[scale=0.33]{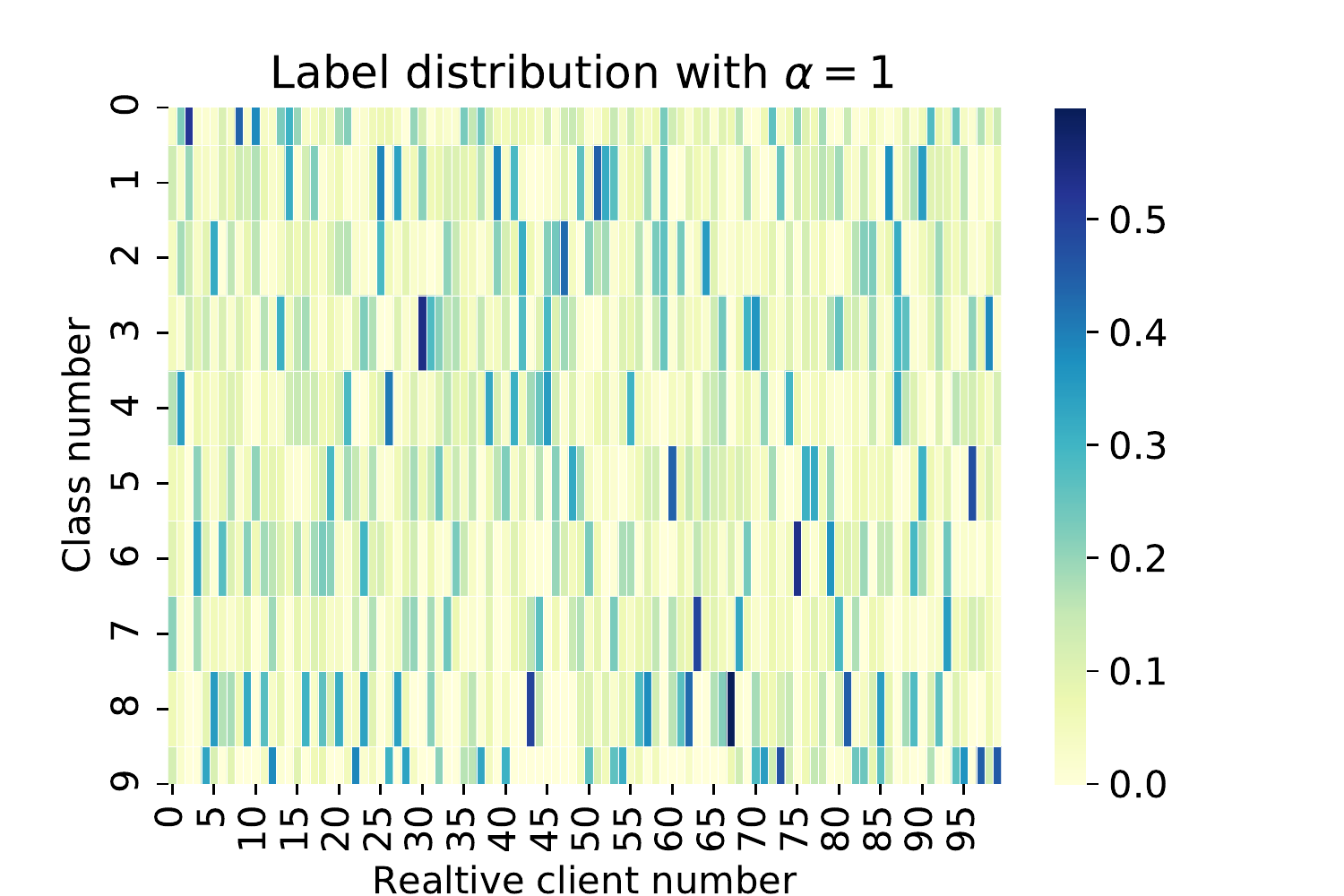}
  \label{cc-2}}%\quad
  \subfigure[Label distribution with $\alpha = 10$]{\includegraphics[scale=0.33]{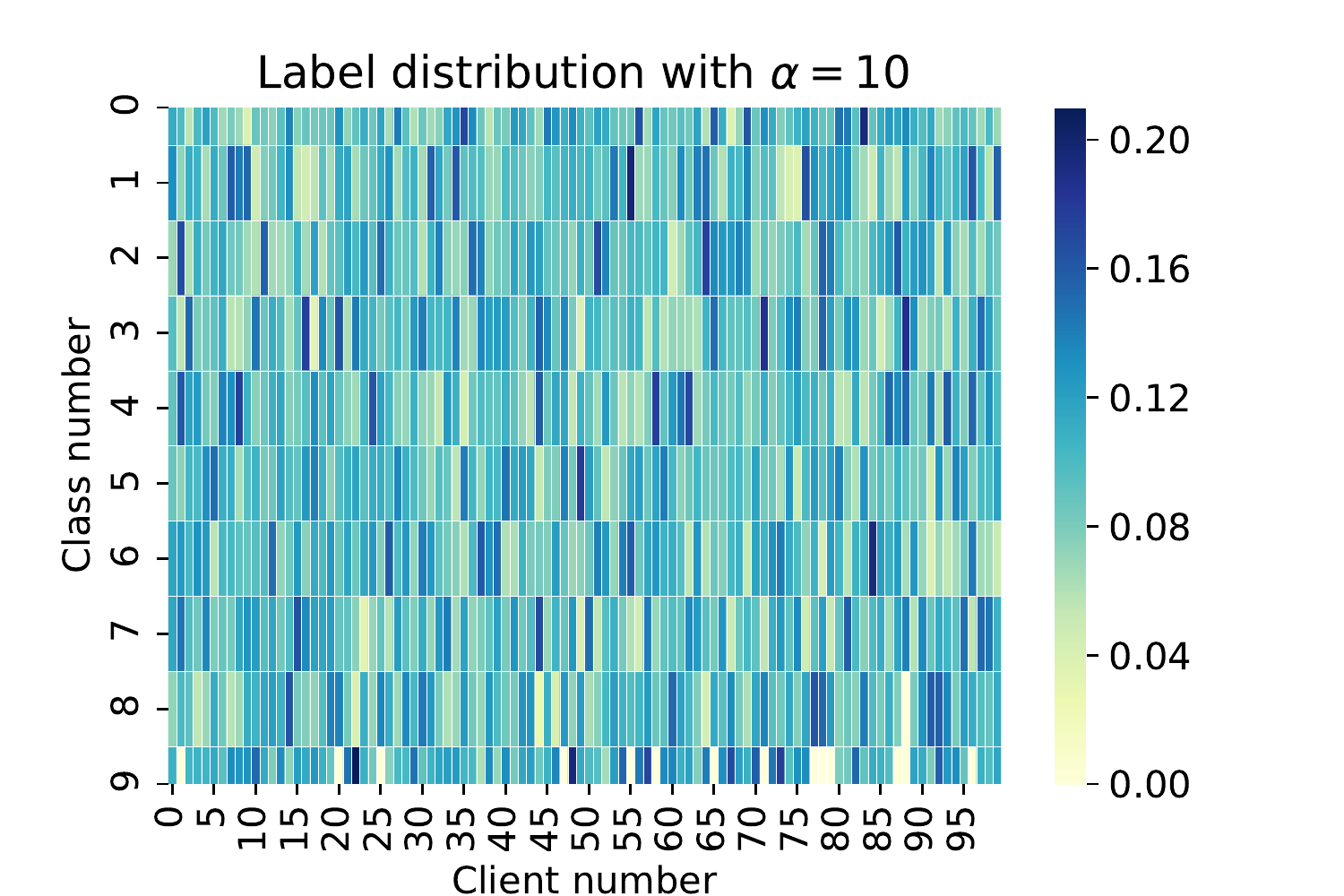}
  \label{cc-3}}
  \caption{The label distribution   over the    $N=100$ users using   Dirichlet distribution with  three different $\alpha$ parameters. }
  \label{fig10}
 \end{figure}

\begin{figure}[t]
  \centering
  \subfigure [Data distribution with $\alpha = 0.7$] {\includegraphics[scale=0.33]{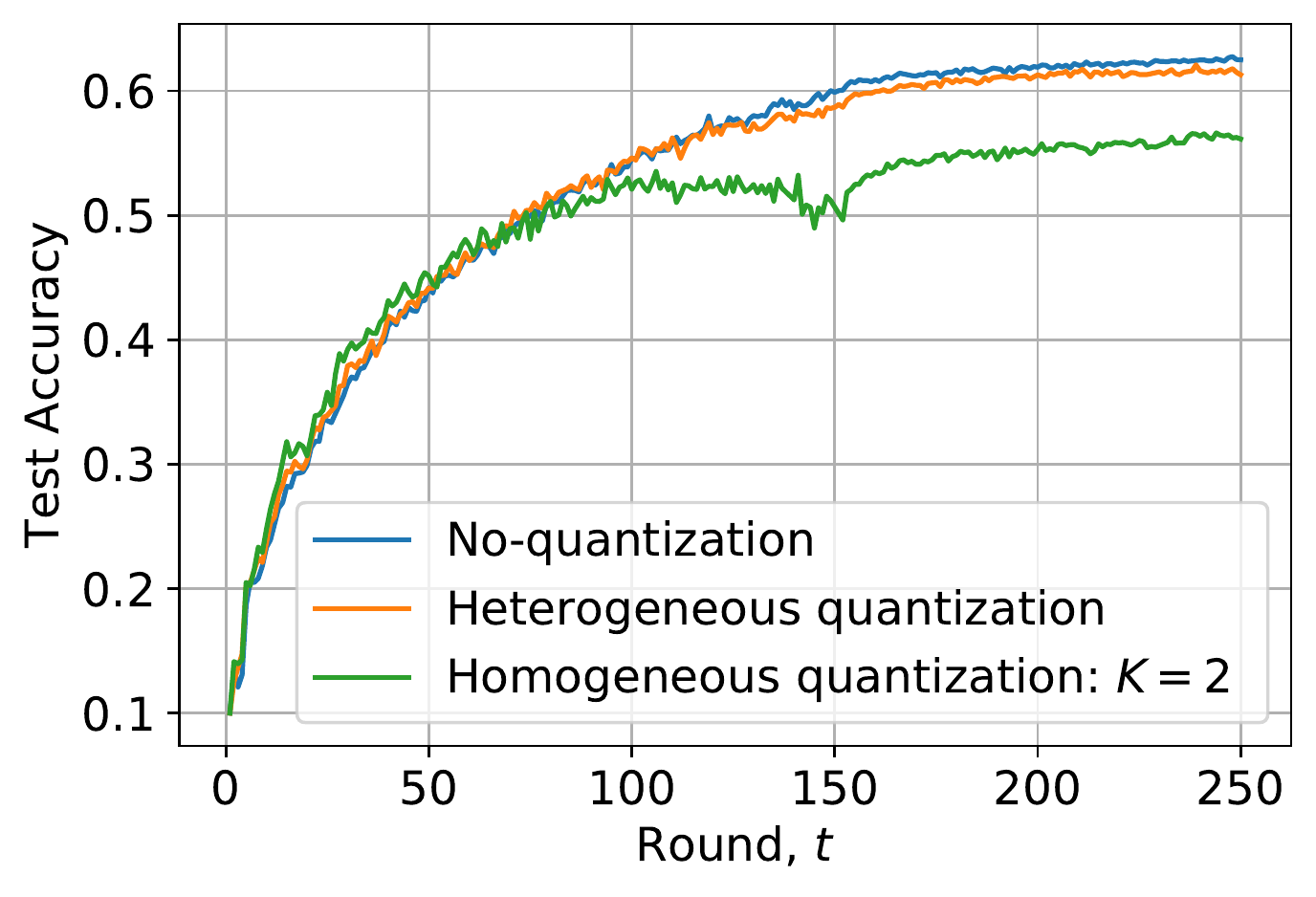}
  \label{c-1}}%\quad
 \subfigure[Data distribution with $\alpha = 1$]{\includegraphics[scale=0.33]{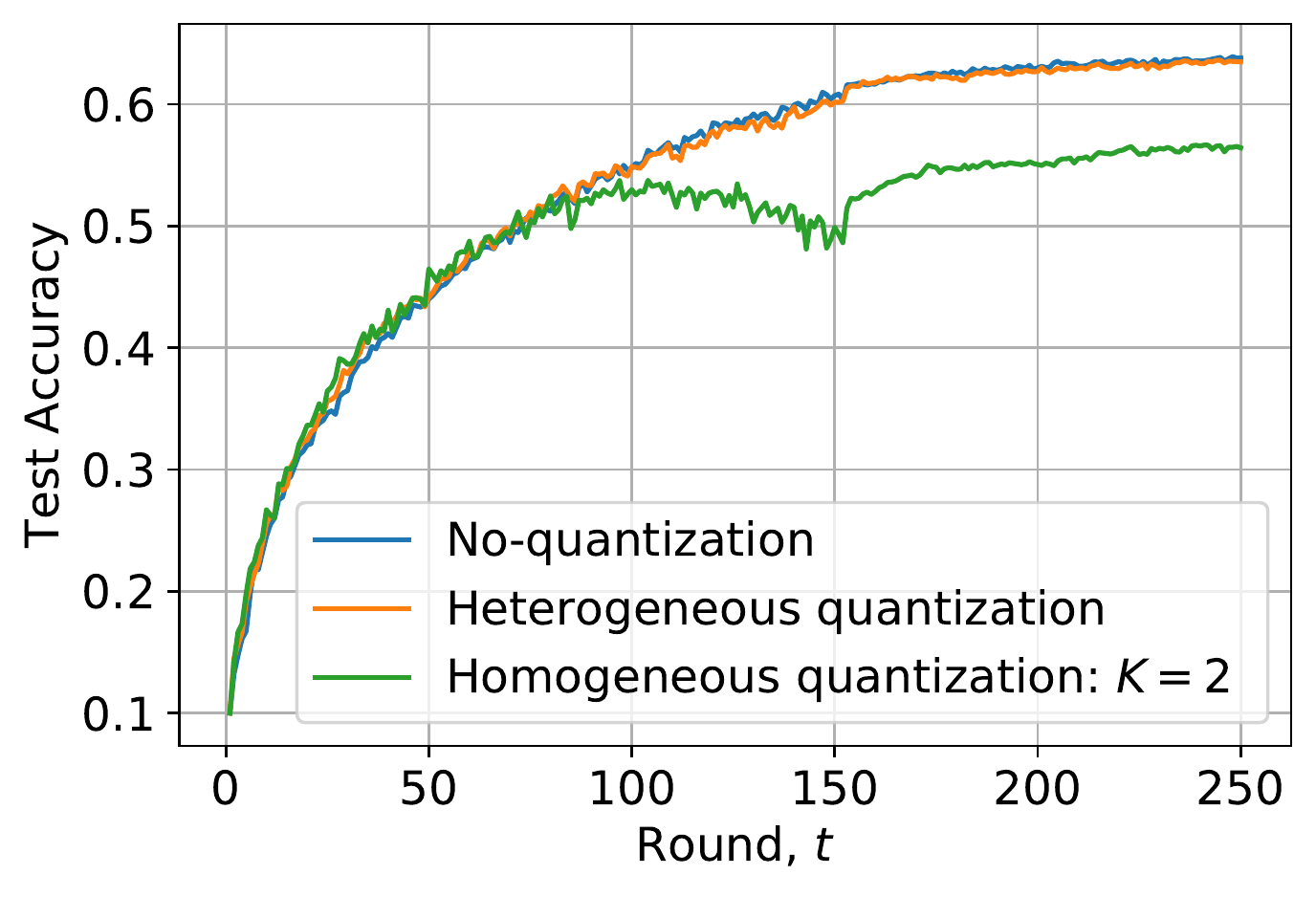}
  \label{c-2}}%\quad
  \subfigure[Data distribution with $\alpha = 10$]{\includegraphics[scale=0.33]{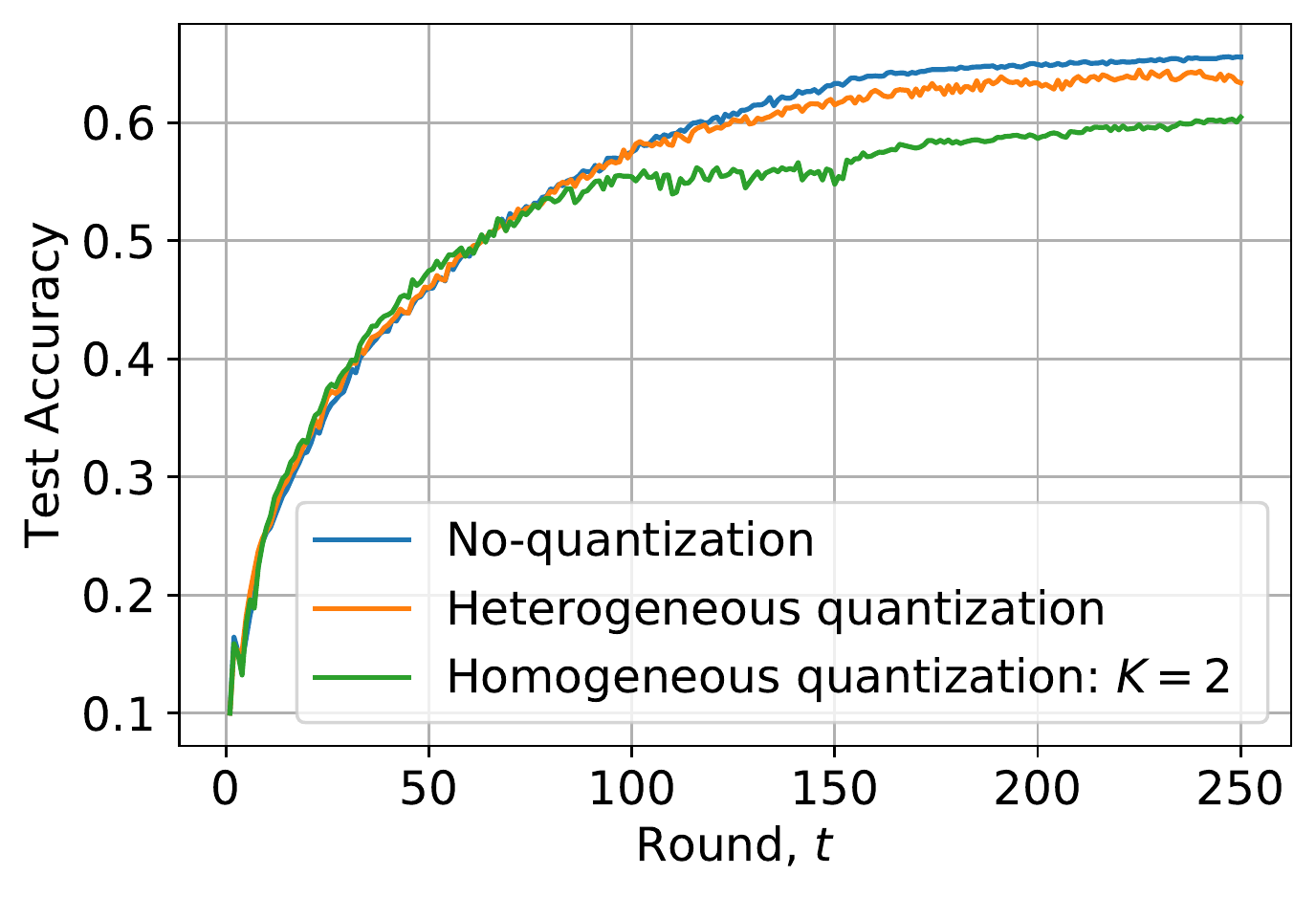}
  \label{c-3}}
  \subfigure[Total communication time]{\includegraphics[scale=0.46]{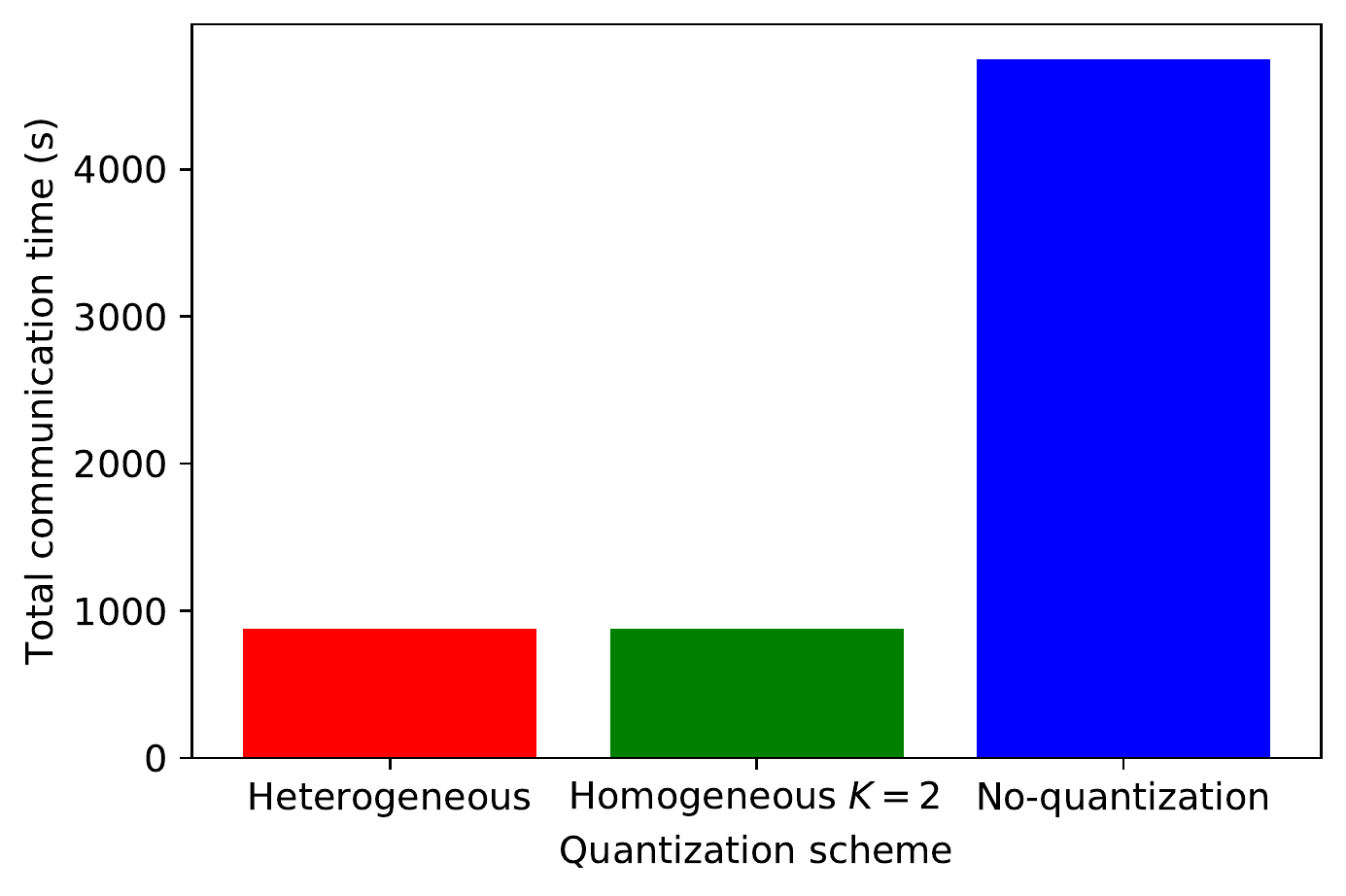}
  \label{c-4}}
  \caption{The performance of HeteroSAg and Fedavg under Gaussian attack and three different data distribution. }
  \label{fig-11}
 \end{figure}

\textbf{Results} Figure \ref{fig10} illustrates the label distribution  for  $N=100$ users with  Dirichlet distribution with different $\alpha$ parameter. 
Using this set of  distributions given in Figure \ref{fig10}, we  evaluate the performance of HeteroSAg using the same quantization schemes given in Section \ref{Experiment 1}. Similar to the performance of HeteroSAg presented in Figure \ref{fig-5} and Figure \ref{fig9} for the  MNIST dataset,   Figure        \ref{fig-11} demonstrates that      HeteroSAg  for CIFAR10 dataset still achieves  higher accuracy  than the Homogeneous quantization with $K=2$ over the three different data distribution settings in Figure \ref{fig10}. Additionally,  HeteroSAg  gives a comparable test accuracy to the   baseline (no-quantization).  The high test accuracy  of HeteroSAg  over the case of homogenous quantization  with $K=2$  is achieved  at no extra communication time as illustrated in Figure \ref{c-4}.   On the other hand,  after $t=250$ rounds   of communication   with the server, the total  communication time  when using heterogeneous  is   less the baseline case with no quantization by a factor of $5.4 \times$. 

%Additionally,  after $t=200$ rounds   of communication   with the server, the total  communication time  when using heterogeneous  is   less the case with no quantization by a factor of $5.2 \times$ according Figure.  \ref{1}.  Furthermore,   HeteroSAg with heterogeneous quantization    maintains  superior performance over  the  case of  homogeneous  quantization   with  $K=2$ levels  with more than  $15 \%$ improvement in test accuracy,  while the  communication time is  the same for the both settings. This confirms our motivation that  by adapting the  quantization levels to the transmission rates of the  users,  we can achieve  high  accuracy with    small  training  time.   

 \subsection{ Additional experiment  (Byzantine robustness)} \label{Add_Byz}
 
 In the following  set of experiments, we further  demonstrate the performance of HeteroSAg under Byzantine attacks using  non-IID data setting. 
 
 \textbf{Dataset and Hyperparameters}
 Similar to the setting given in Section \ref{main_Byz}, we set the total number of users to $N=300$, in which $B = 18$ of them are Byzantines. We use a fixed learning rate  of $ 0.02$, and setting the batch size for each user to be $20\%$ of its local data. We consider epoch training, where the number of epochs is $4$.  We use CIFAR10 dataset with non-IID data distribution. Here, we also  generate the non-IID data distribution using 
 Dirichlet distribution with parameter $\alpha$. 

\begin{figure}[t]
  \centering
  \subfigure{\includegraphics[scale=0.33]{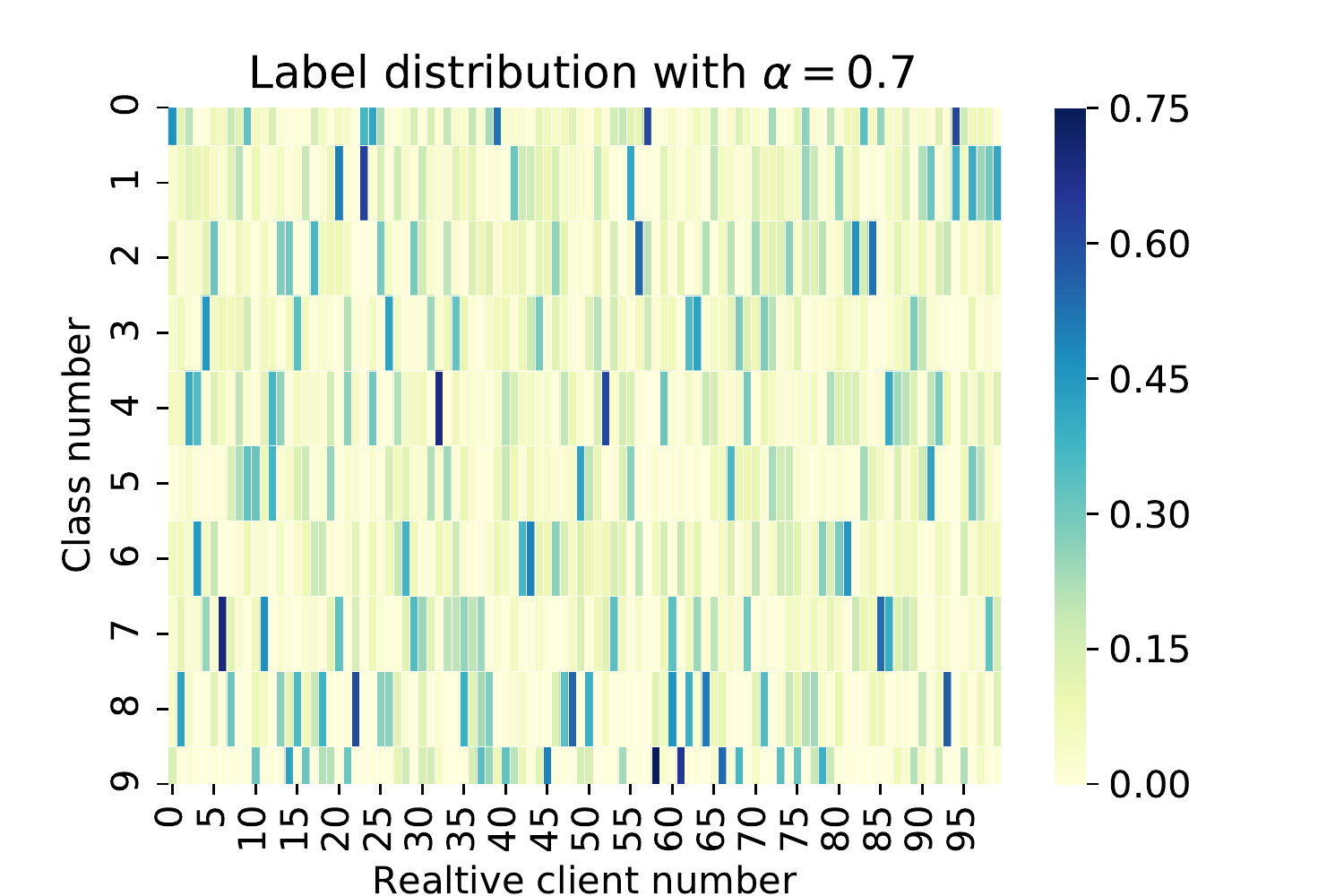}
  \label{a-1}}%\quad
 \subfigure{\includegraphics[scale=0.33]{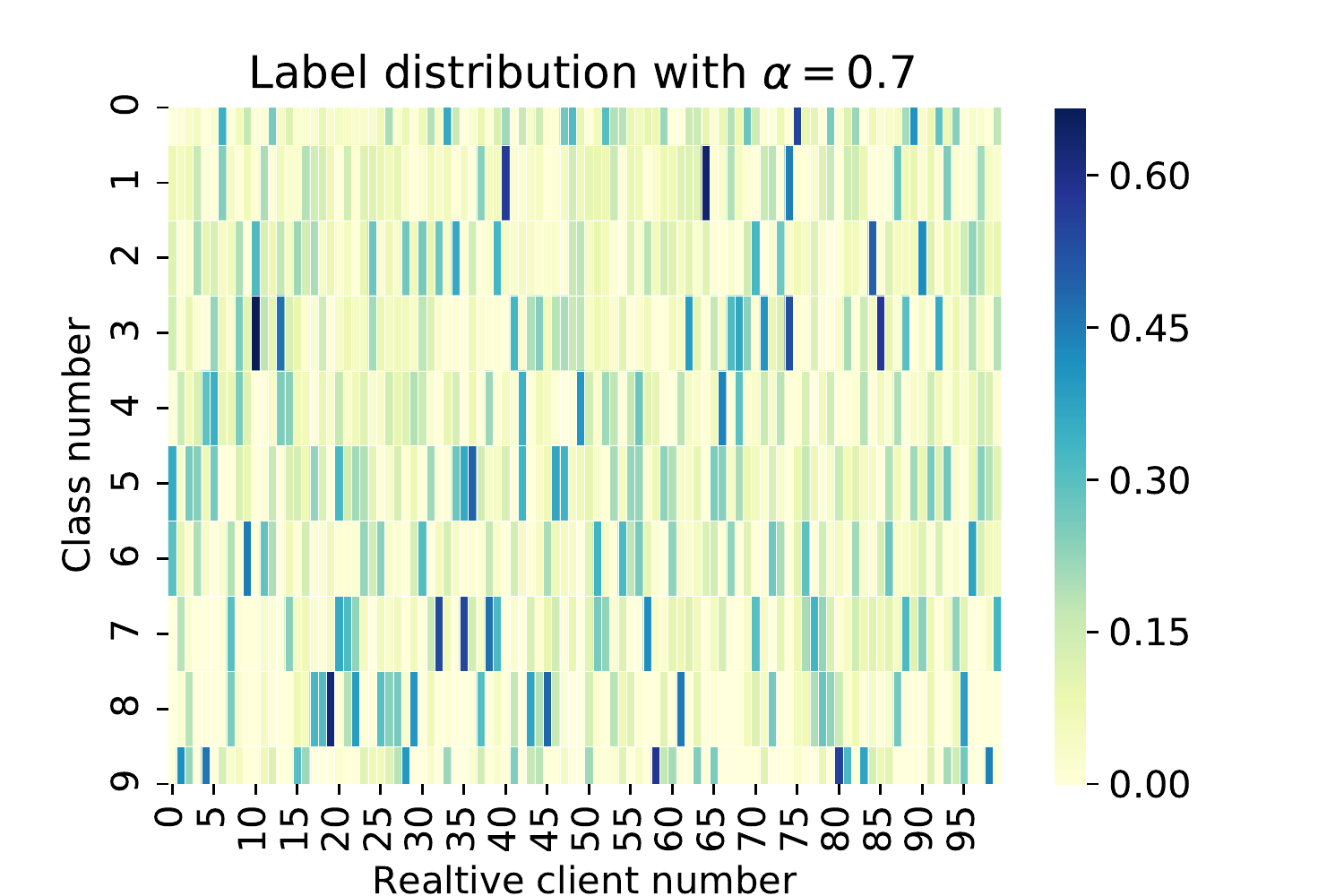}
  \label{a-2}}%\quad
  \subfigure{\includegraphics[scale=0.33]{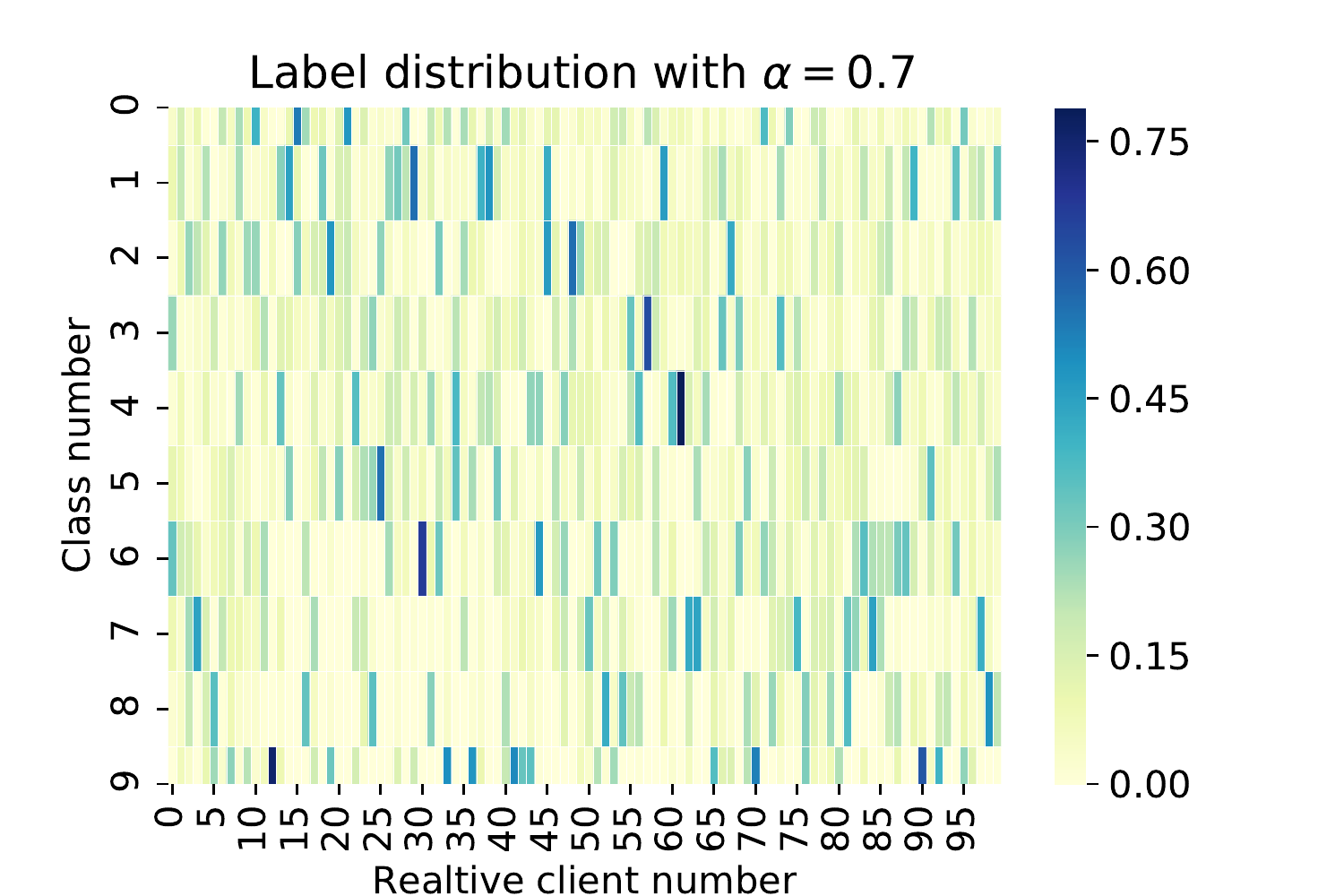}
  \label{a-3}}

   \subfigure{\includegraphics[scale=0.33]{alpha_1_cifar10heatmap_first_100.pdf}
  \label{b-1}}%\quad
 \subfigure{\includegraphics[scale=0.33]{alpha_1_cifar10heatmap_first_100.pdf}
  \label{b-2}}%\quad
  \subfigure{\includegraphics[scale=0.33]{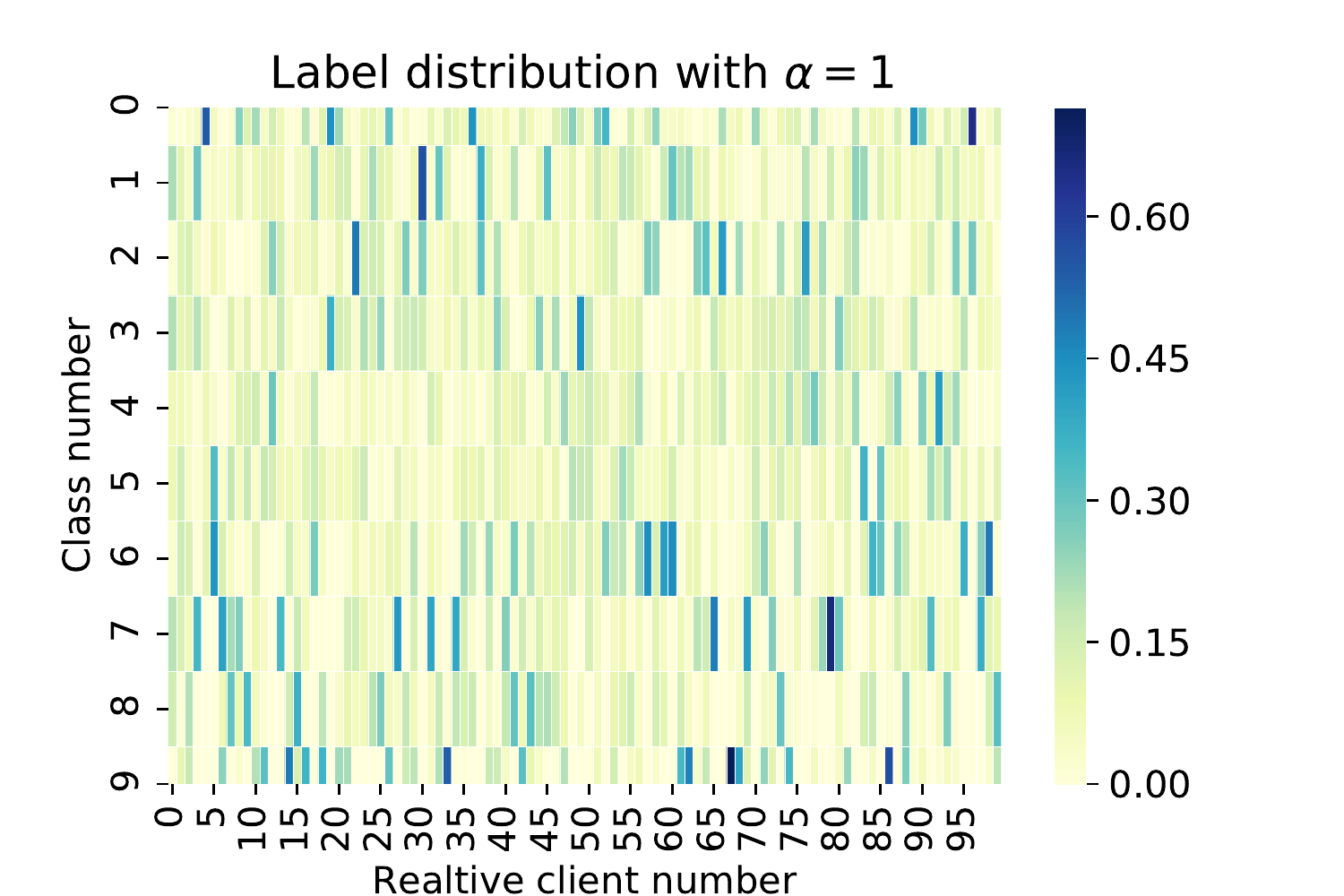}
  \label{b-3}}
  
   \subfigure [First $100$ users] {\includegraphics[scale=0.33]{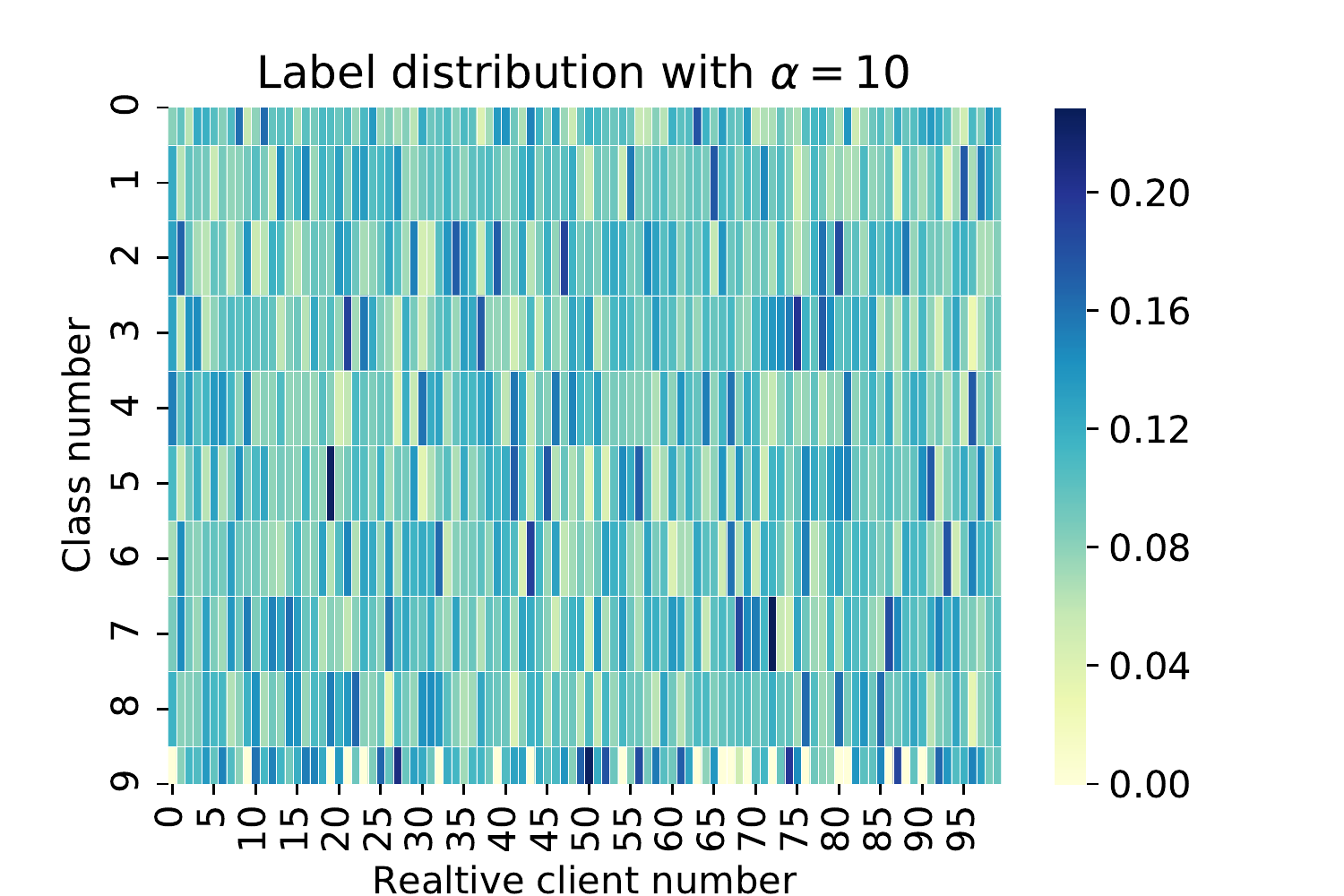}
  \label{c-1}}%\quad
 \subfigure[Second $100$ users]{\includegraphics[scale=0.33]{alpha_10_cifar10heatmap_first_100.pdf}
  \label{c-2}}%\quad
  \subfigure[Third $100$ users]{\includegraphics[scale=0.33]{alpha_10_cifar10heatmap_first_100.pdf}
  \label{c-3}}
  \caption{The label distribution   among   $N=300$ users using   Dirichlet distribution with different $\alpha$ parameters. }
  \label{Byz_dis}
 \end{figure}

\begin{figure}[t]
  \centering
  \subfigure [Data distribution with $\alpha = 0.7$] {\includegraphics[scale=0.33]{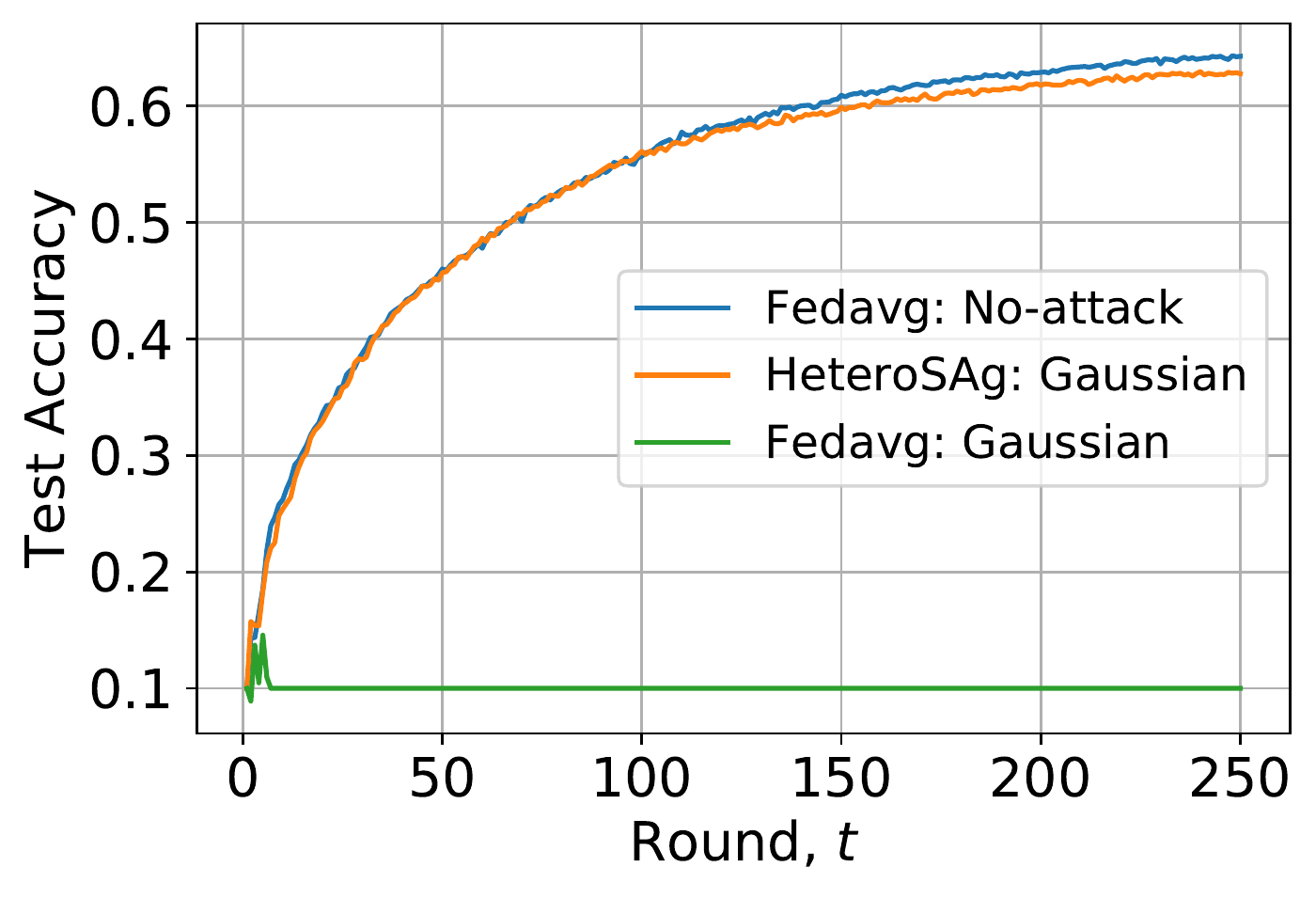}
  \label{c-1}}%\quad
 \subfigure[Data distribution with $\alpha = 1$]{\includegraphics[scale=0.33]{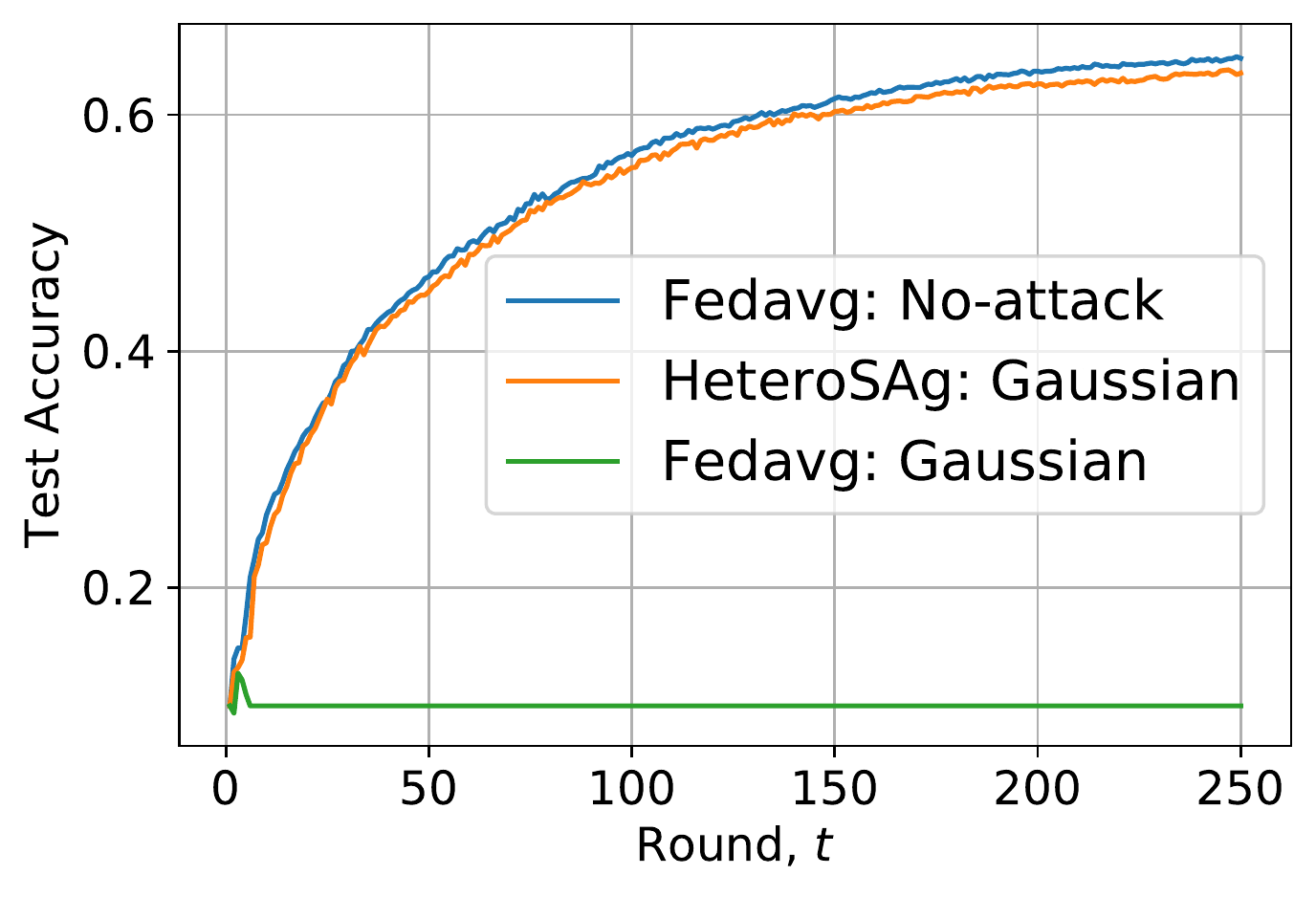}
  \label{c-2}}%\quad
  \subfigure[Data distribution with $\alpha = 10$]{\includegraphics[scale=0.33]{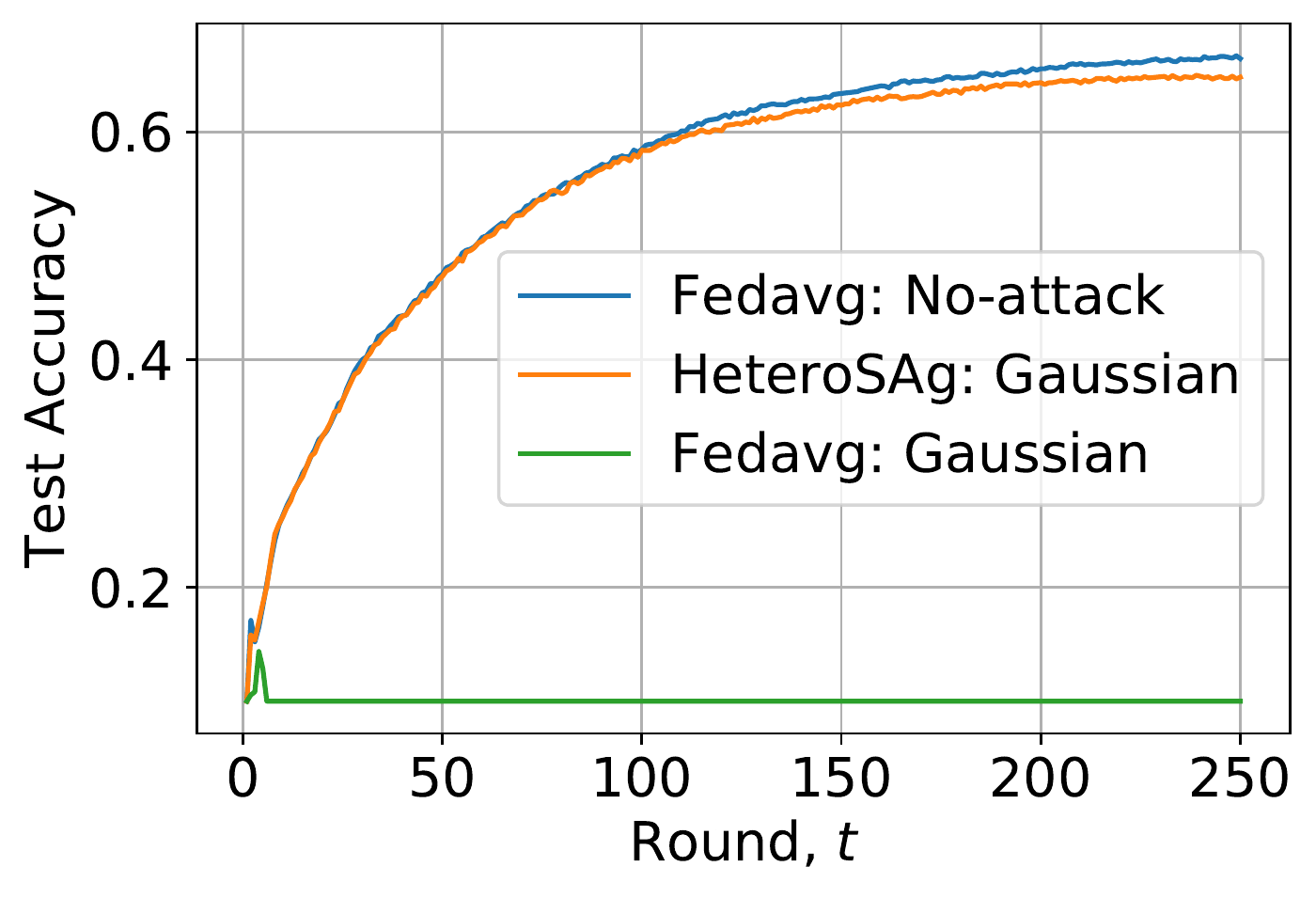}
  \label{c-3}}
  \subfigure [Data distribution with $\alpha = 0.7$] {\includegraphics[scale=0.33]{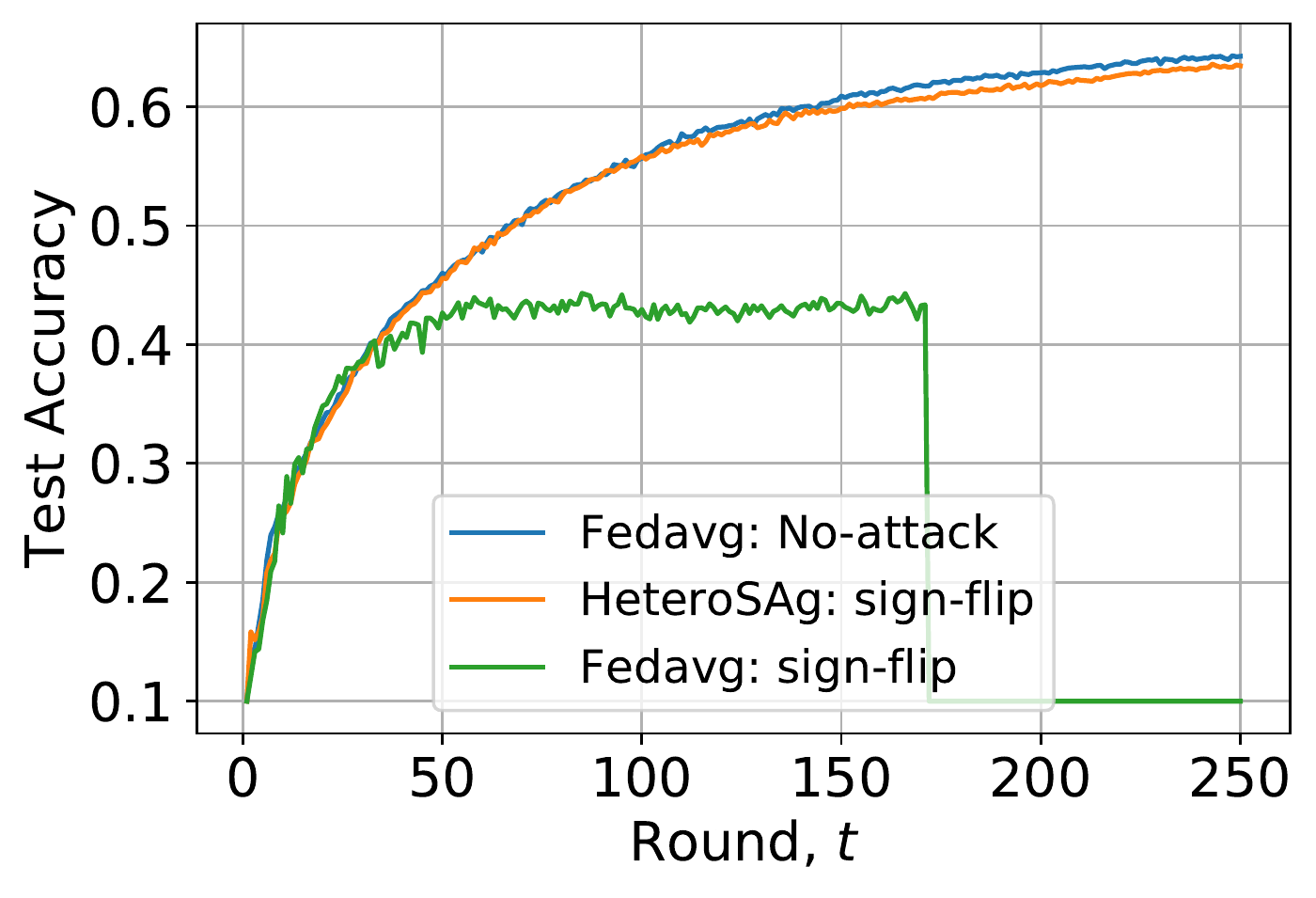}
  \label{c-1}}%\quad
 \subfigure[Data distribution with $\alpha = 1$]{\includegraphics[scale=0.33]{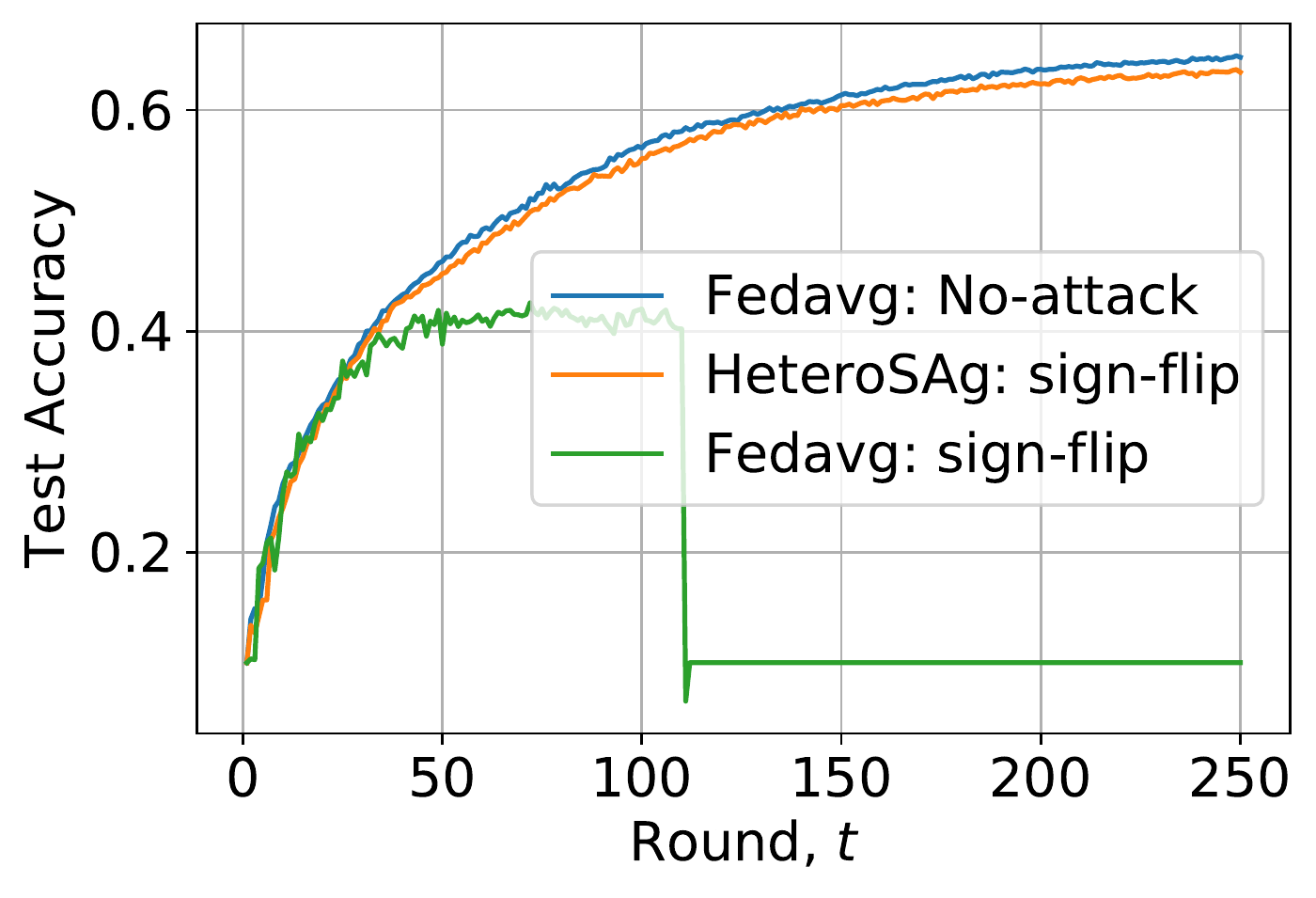}
  \label{c-2}}%\quad
  \subfigure[Data distribution with $\alpha = 10$]{\includegraphics[scale=0.33]{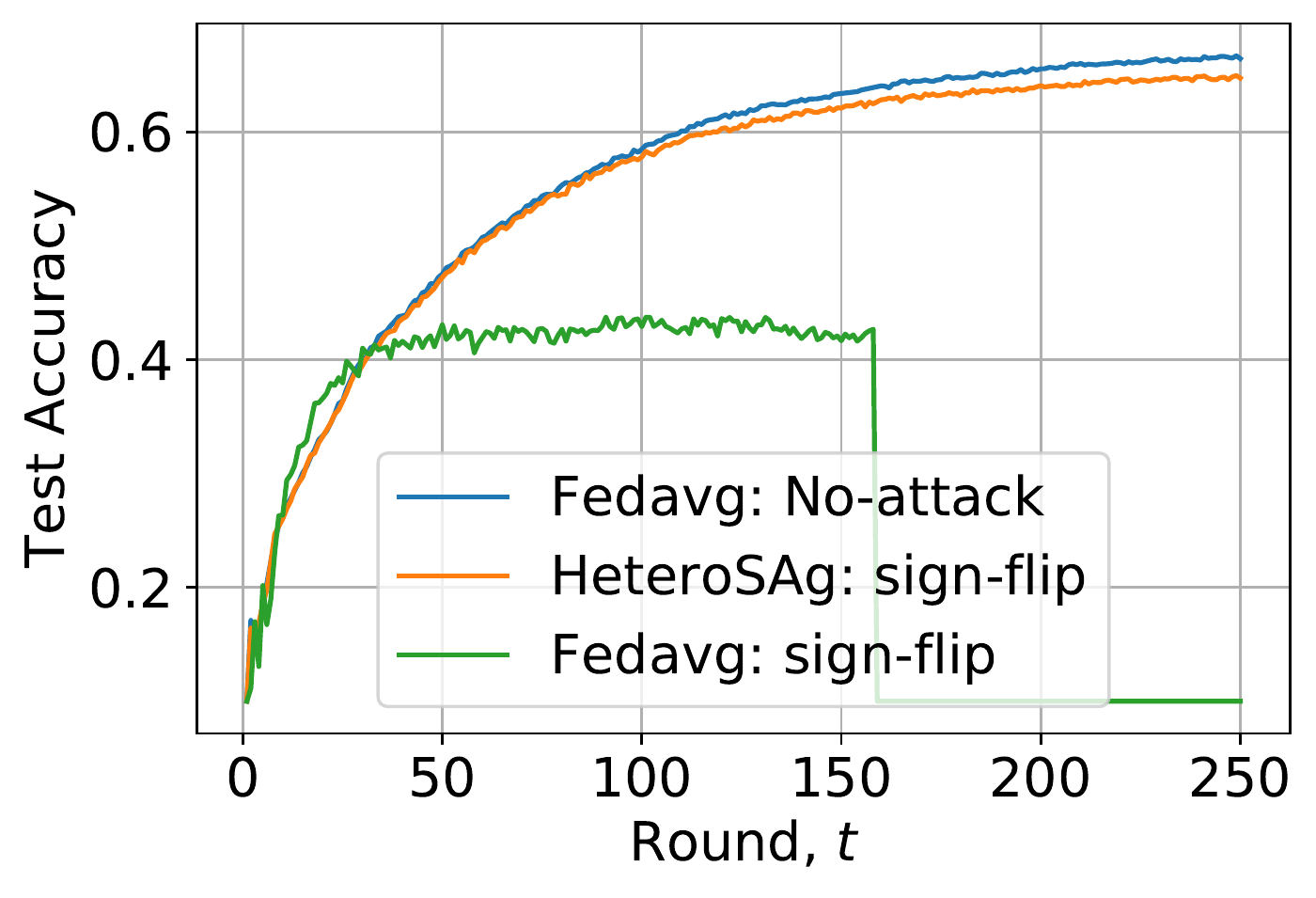}
  \label{c-3}}
  \caption{The performance of HeteroSAg and Fedavg under  Gaussian and sign-flip attacks with three different data distributions. }
  \label{fig_last}
 \end{figure}

\textbf{Results} 
Figure \ref{Byz_dis} gives   the label distribution  for  $N=300$ users with  Dirichlet distribution with three  different $\alpha$ parameters. 
Using this set of  distributions, we  evaluate the performance of HeteroSAg.  As we can see in Figure  \ref{fig_last}, HeteroSAg with coordinate-wise median is robust to Gaussian attack and the sign flip attack  while  giving  performance almost the same as the case with no Byzantine users. On the other hand, in the presence of these attacks, FedAvg scheme gives very low performance.

\subsection{Models}\label{model}
We provide the details of the neural network architectures used in our experiments. For MNIST, we use a model with two fully connected layers, and the details are provided in Table \ref{tab:1}. The first  fully connected layers is followed by ReLU, while softmax is used at the output of the last layer.

\begin{table}[htb!]
\caption{Details of the parameters in the architecture of the neural network used in our MNIST experiments.}\label{tab:1}
\centering
\vspace{5pt}
\begin{tabular}{|l|l|}
\hline
\textbf{Parameter} & \textbf{Shape}\\ \hline
fc1& $784\times100$\\  \hline
fc2& $100\times10$\\ \hline
\end{tabular}
\end{table}

\begin{table}[ht]
\caption{Details of the parameters in the architecture of the neural network used in our CIFAR10 experiments.}\label{tab:2}
\centering
\begin{tabular}{|l|l|}
\hline
\textbf{Parameter} & \textbf{Shape}\\ \hline
conv1& $3\times16\times3\times3$ \\ \hline
conv2& $16\times64\times4\times4$ \\ \hline
fc1& $64\times384$\\ \hline
fc2& $384\times192$\\ \hline
fc3& $192\times10$\\ \hline
\end{tabular}
\end{table}

For CIFAR10, we consider a neural network with two convolutional layers, and three fully connected layers, and the specific details of these layers are provided in Table \ref{tab:2}. ReLU and maxpool is applied on the convolutional layers. The first maxpool has a kernel size $3\times3$ and a stride of $3$ and the second maxpool has a kernel size of $4\times4$ and a stride of $4$. Each of the first two fully connected layers is followed by ReLU, while softmax is used at the output of the third one fully connected layer.

We initialize all biases to $0$. Furthermore, for weights in convolutional layers, we use Glorot uniform initializer, while for weights in fully connected layers, we use the default Pytorch initialization.
 \end{appendices}

\end{document}